\documentclass{aa}
\usepackage{graphicx}
\usepackage{lscape}
\usepackage{subfigure}
\usepackage{natbib}
\usepackage[varg]{txfonts}
\usepackage{longtable}
\usepackage{booktabs}
\usepackage[normalem]{ulem} 
\usepackage{epstopdf}
\usepackage{xcolor}
\usepackage{soul}
\usepackage{makecell}
\usepackage{multicol,tabularx,capt-of}
\usepackage{placeins}
\usepackage{scalerel}
\usepackage{tikz}

\usetikzlibrary{svg.path}

\defcitealias{Sanchez-Saez21a}{SS21a}

\definecolor{orcidlogocol}{HTML}{A6CE39}
\tikzset{
  orcidlogo/.pic={
    \fill[orcidlogocol] svg{M256,128c0,70.7-57.3,128-128,128C57.3,256,0,198.7,0,128C0,57.3,57.3,0,128,0C198.7,0,256,57.3,256,128z};
    \fill[white] svg{M86.3,186.2H70.9V79.1h15.4v48.4V186.2z}
                 svg{M108.9,79.1h41.6c39.6,0,57,28.3,57,53.6c0,27.5-21.5,53.6-56.8,53.6h-41.8V79.1z M124.3,172.4h24.5c34.9,0,42.9-26.5,42.9-39.7c0-21.5-13.7-39.7-43.7-39.7h-23.7V172.4z}
                 svg{M88.7,56.8c0,5.5-4.5,10.1-10.1,10.1c-5.6,0-10.1-4.6-10.1-10.1c0-5.6,4.5-10.1,10.1-10.1C84.2,46.7,88.7,51.3,88.7,56.8z};
  }
}

\newcommand\orcidicon[1]{\href{https://orcid.org/#1}{\mbox{\scalerel*{
\begin{tikzpicture}[yscale=-1,transform shape]
\pic{orcidlogo};
\end{tikzpicture}
}{|}}}}

\usepackage{hyperref} 
\hypersetup{
    colorlinks=true,
    citecolor=blue,
    linkcolor=blue,
    urlcolor=blue,
    }
\makeatletter
\renewcommand*\aa@pageof{, page \thepage{} of \pageref*{LastPage}}
\makeatother

\bibpunct{(}{)}{;}{a}{}{,} 
\usepackage{amstext} 
\definecolor{cadmiumred}{rgb}{0.89, 0.0, 0.13}

\definecolor{ste}{rgb}{0., 0.26, 0.15}

\begin{document} 
   \title{Persistent and occasional: searching for the variable population of the ZTF/4MOST sky using ZTF data release 11\thanks{Tables containing the classifications and features for the ZTF $g$ and $r$ bands, the labeled set, and the master catalog used to create the labeled set are available in electronic form at the CDS via anonymous ftp to cdsarc.cds.unistra.fr (130.79.128.5)
or via \url{https://cdsarc.cds.unistra.fr/cgi-bin/qcat?J/A+A/}. Individual catalogs per class and band, as well as the labeled set catalogs, can be downloaded at Zenodo via \href{https://doi.org/10.5281/zenodo.7826044}{10.5281/zenodo.7826045}.}}

   \author{
   P. S\'anchez-S\'aez\inst{1,2\orcidicon{0000-0003-0820-4692}},
   J. Arredondo\inst{3\orcidicon{0000-0002-2045-7134}},
   A. Bayo\inst{1,4\orcidicon{0000-0001-7868-7031}}, 
   P. Ar\'evalo\inst{4,5\orcidicon{0000-0001-8503-9809}},
   F. E. Bauer\inst{6,7,2,8\orcidicon{0000-0002-8686-8737}},
   G. Cabrera-Vives\inst{2,9,10\orcidicon{0000-0002-2720-7218}},
   M. Catelan\inst{6,2,7\orcidicon{0000-0001-6003-8877}},
   P. Coppi\inst{11\orcidicon{0000-0001-9604-2325}},
   P. A. Est\'evez\inst{2,12\orcidicon{0000-0001-9164-4722}},
   F. F\"orster\inst{13,2,14,15\orcidicon{0000-0003-3459-2270}},
   L. Hern\'andez-Garc\'ia\inst{2,4\orcidicon{0000-0002-8606-6961}},
   P. Huijse\inst{15,2\orcidicon{0000-0003-3541-1697}},
   R. Kurtev\inst{4,2\orcidicon{0000-0002-9740-9974}},
   P. Lira\inst{16,5\orcidicon{0000-0003-1523-9164}},
   A. M. Mu\~noz Arancibia\inst{2,14\orcidicon{0000-0002-8722-516X}},
   and G. Pignata\inst{17,2\orcidicon{0000-0003-0006-0188}}
   }

   \titlerunning{Searching for the variable population of the ZTF/4MOST sky using ZTF data release 11}
   \authorrunning{P. S\'anchez-S\'aez et al.}

\institute{
European Southern Observatory, Karl-Schwarzschild-Strasse 2, 85748 Garching bei München, Germany
\\e-mail: pasanchezsaez@gmail.com, paula.sanchezsaez@eso.org
\and
Millennium Institute of Astrophysics (MAS), Nuncio Monse\~nor Sotero Sanz 100, Of. 104, Providencia, Santiago, Chile
\and
Departamento de Ingeniería Informática, Universidad de Santiago de Chile, Av. Ecuador 3659, Santiago, Chile
\and
Instituto de F\'isica y Astronom\'ia, Facultad de Ciencias,Universidad de Valpara\'iso, Gran Breta\~na No. 1111, Playa Ancha, Valpara\'iso, Chile
\and
Millennium Nucleus on Transversal Research and Technology to Explore Supermassive Black Holes (TITANS)
\and
Instituto de Astrof{\'{\i}}sica, Facultad de F{\'{i}}sica, Pontificia Universidad Cat{\'{o}}lica de Chile, Casilla 306, Santiago 22, Chile
\and
Centro de Astroingenier{\'{\i}}a, Pontificia Universidad Cat{\'{o}}lica de Chile, Av. Vicu\~{n}a Mackenna 4860, 7820436 Macul, Santiago, Chile
\and 
Space Science Institute, 4750 Walnut Street, Suite 205, Boulder, Colorado 80301
\and
Department of Computer Science, Universidad de Concepci\'on, Chile
\and
Data Science Unit, Universidad de Concepción, Edmundo Larenas 310, Concepci\'on, Chile
\and
Department of Astronomy, Yale University, P.O. Box 208101, New Haven, CT 06520-8101, USA
\and
Department of Electrical Engineering, Universidad de Chile, Av. Tupper 2007, Santiago 8320000, Chile
\and 
Data and Artificial Intelligence Initiative (ID\&IA), University of Chile, Santiago, Chile
\and
Center for Mathematical Modeling (CMM), Universidad de Chile, Beauchef 851, Santiago 8320000, Chile
\and
Instituto de Informática, Facultad de Ciencias de la Ingeniería, Universidad Austral de Chile, General Lagos 2086, Valdivia, Chile
\and
Departamento de Astronom\'ia, Universidad de Chile, Casilla 36D, Santiago, Chile
\and 
Departamento de Ciencias Fis\'icas, Universidad Andres Bello, Avda. Republica 252, Santiago, Chile
}

   \date{}
  \abstract
   {}
   {We present a  variability, color and morphology based classifier, designed to identify multiple classes of transients, persistently variable, and non-variable sources, from the Zwicky Transient Facility (ZTF) Data Release 11 (DR11) light curves of extended and point sources. The main motivation to develop this model was to identify active galactic nuclei (AGN) at different redshift ranges to be observed by the 4MOST Chilean AGN/Galaxy Evolution Survey (ChANGES). Still, it serves as a more general time-domain astronomy study. }
   {The model uses nine colors computed from CatWISE and PanSTARRS1 (PS1), a morphology score from PS1, and 61 single-band variability features computed from the ZTF DR11 $g$ and $r$ light curves. We trained two versions of the model, one for each ZTF band, since ZTF DR11 treats independently the light curves observed in a particular combination of field, filter, and CCD-quadrant. We used a hierarchical local classifier per parent node approach, where each node was composed of a balanced random forest model. We adopted a 17-class taxonomy, including non-variable stars and galaxies, three transients (SNIa, SN-other, and CV/Nova), five classes of stochastic variables (lowz-AGN, midz-AGN, highz-AGN, Blazar, and YSO), and seven classes of periodic variables (LPV, EA, EB/EW, DSCT, RRL, CEP, and Periodic-other). }
   {The macro averaged precision, recall and F1-score are 0.61, 0.75, and 0.62 for the $g$-band model, and 0.60, 0.74, and 0.61, for the $r$-band model. When grouping the four AGN classes (lowz-AGN, midz-AGN, highz-AGN, and Blazar) into one single class, its precision, recall, and F1-score are 1.00, 0.95, and 0.97, respectively, for both the $g$ and $r$ bands. This demonstrates the good performance of the model classifying AGN candidates. We applied the model to all the sources in the ZTF/4MOST overlapping sky ($-28\leq \text{dec} \leq8.5$), avoiding ZTF fields covering the Galactic bulge ($|gal\_b|\leq9$ and $gal\_l\leq50$). This area includes 86,576,577 light curves in the $g$-band and 140,409,824 in the $r$-band, with 20 or more observations, and with an average magnitude in the corresponding band lower than 20.5. Only 0.73\% of the $g$-band light curves and 2.62\% of the $r$-band light curves were classified as stochastic, periodic, or transient with high probability ($P_{init}\geq0.9$). Even though the metrics obtained for both models are similar, we found that, in general, more reliable results are obtained when using the $g$-band model. Using the latter, we identified 384,242 AGN candidates (including low-, mid-, and high-redshift AGNs and Blazars), 287,156 of which have $P_{init}\geq0.9$. }
   {}

   \maketitle

\section{Introduction}\label{section:intro}

We are approaching an era in astronomy in which data sets are too large to process efficiently with traditional tools and methods. In the last decades, a majority of astronomers have used tools that require only the computing power of a simple laptop to obtain insights from a given data set. However, with the advent of surveys such as the Zwicky Transient Facility (ZTF; \citealt{Bellm19}) and the upcoming Vera Rubin Observatory Legacy Survey of Space and Time (LSST; \citealt{LSST}), we need to develop new tools to extract the most out of the data and manipulate the huge samples observed by these surveys. Recently, several community brokers have begun to process the ZTF public alert stream, including the Automatic Learning for the Rapid Classification of Events (ALeRCE; \citealt{Forster21}), Alert Management, Photometry and Evaluation of Lightcurves (AMPEL \citealt{Nordin19}), Arizona-NOAO Temporal Analysis and Response to Events System (ANTARES \citealt{Narayan18}), Fink \citep{Moller21Fink}, and LASAIR \citep{Smith19}. These brokers make use of different statistical and machine learning techniques to process the large amounts of data generated by the ZTF survey, which amount to over 300,000 alerts per night.\footnote{ZTF produces an alert when there is a 5$\sigma$ detection in the difference image of a particular object. This detection can be produced by the flux variations of persistently variable and transient objects, as well as by moving targets, or bogus detections.} 

The use of alert streams is optimal for studies of transient objects, such as supernovae (SNe), tidal disruption events (TDE), or kilonovae, as demonstrated by several works (e.g., \citealt{vanVelzen21,Sanchez-Saez21a,Carrasco-Davis21,Leoni22, Miranda22}). However, no information is provided by this stream when sources do not vary with amplitudes large enough to produce an alert, or it can suffer from strong miscalibrations due to changes in the reference images used to generate the difference images. On the other hand, the study of persistent and/or low/amplitude variable objects can benefit from denser sampling on their light curves, where all epochs are considered without an absolute threshold in their relative variations. This is the case of the ZTF data release (ZTF DR) light curves, which are constructed from the point spread function (PSF) photometry over all the ZTF science images, for objects detected in the ZTF reference images (for details see \citealt{Masci19}). For extended sources, ZTF DR light curves are quite sensitive to seeing variations, and can show spurious variations, thus, they are not optimal for variability studies. Active galactic nuclei (AGNs), young stellar objects (YSOs), and binary stars are among the persistent variable objects whose analysis benefits the most from the use of DR light curves, as their alert light curves can be very incomplete and biased. An example of this is the work presented by \cite{Sanchez-Saez21b}, where ZTF DR5 light curves were used to identify changing-state AGN (CSAGNs; sources that change their classification as type 1 or type 2 AGNs) candidates and found that several of their promising CSAGN candidates present only a few or even no alerts.

\cite{Chen20} presents one of the first attempts to classify ZTF DR light curves. They used ZTF DR2 light curves to classify 781,602 periodic variable stars into 11 classes. They selected variable stars from the DR2 light curves using the period (computed using the Lomb-Scargle periodogram; \citealt{Lomb76,Scargle82}) and the variability amplitude, as well as other statistics, and applied feature cuts to classify their candidates, with a misclassification rate of 2\%. \cite{vanRoestel21} presented the ZTF project light curve classifier, which was designed to classify all persistent point sources in the ZTF DR time series. They presented two models, one using variability features and a gradient-boosted decision tree classifier, and another using deep neural networks applied to magnitude–time histograms (\textit{dmdt}; \citealt{Mahabal17}), as well as variability features. To test their selection technique, they applied their models to 10 pairs of ZTF fields, obtaining precision and recall scores larger than 0.8 in most of the classes included in their taxonomy. More recently, \cite{Aleo22} used ZTF DR4 light curves to search for missed transients in the ZTF alert stream, using variability features computed with the \texttt{light-curve} package \citep{Malanchev21}, and applying a k-D tree algorithm \citep{Bentley75}. From this model, they were able to identify 11 missing transients.

In this work, we present a ZTF DR light curve classifier that has been designed to work with the PSF light curves of both extended and point sources, constructed from their ZTF science images. As previously mentioned, these light curves are not optimal for variability studies of extended sources, but in this work, we demonstrate that they can still be used to separate non-variable and variable extended sources, as well as to identify transients whose host is detected directly in the science images. The model separates sources into 17 different classes of persistently variable objects, transient events, and non-variable galaxies and stars. The main purpose of this classifier is to identify AGN candidates at different redshifts to be observed by the 4MOST instrument \citep{deJong19} as part of the 4MOST Chilean AGN/Galaxy Evolution Survey (ChANGES; F. Bauer et al., in prep.). To this end, we focus on the classification of objects located in the region of overlap between ZTF and 4MOST skies (ZTF/4MOST sky), excluding the Milky Way bulge. The total size of the ZTF DR11 light curves in this area amounts to 139.5 GB in the $g$-band and 484.5 GB in the $r$-band, including 86,576,577 and 140,409,824 light curves with 20 or more observations and an average magnitude lower than 20.5, respectively.

\begin{figure*}[tbhp]
    \centering
    \includegraphics[width=\linewidth]{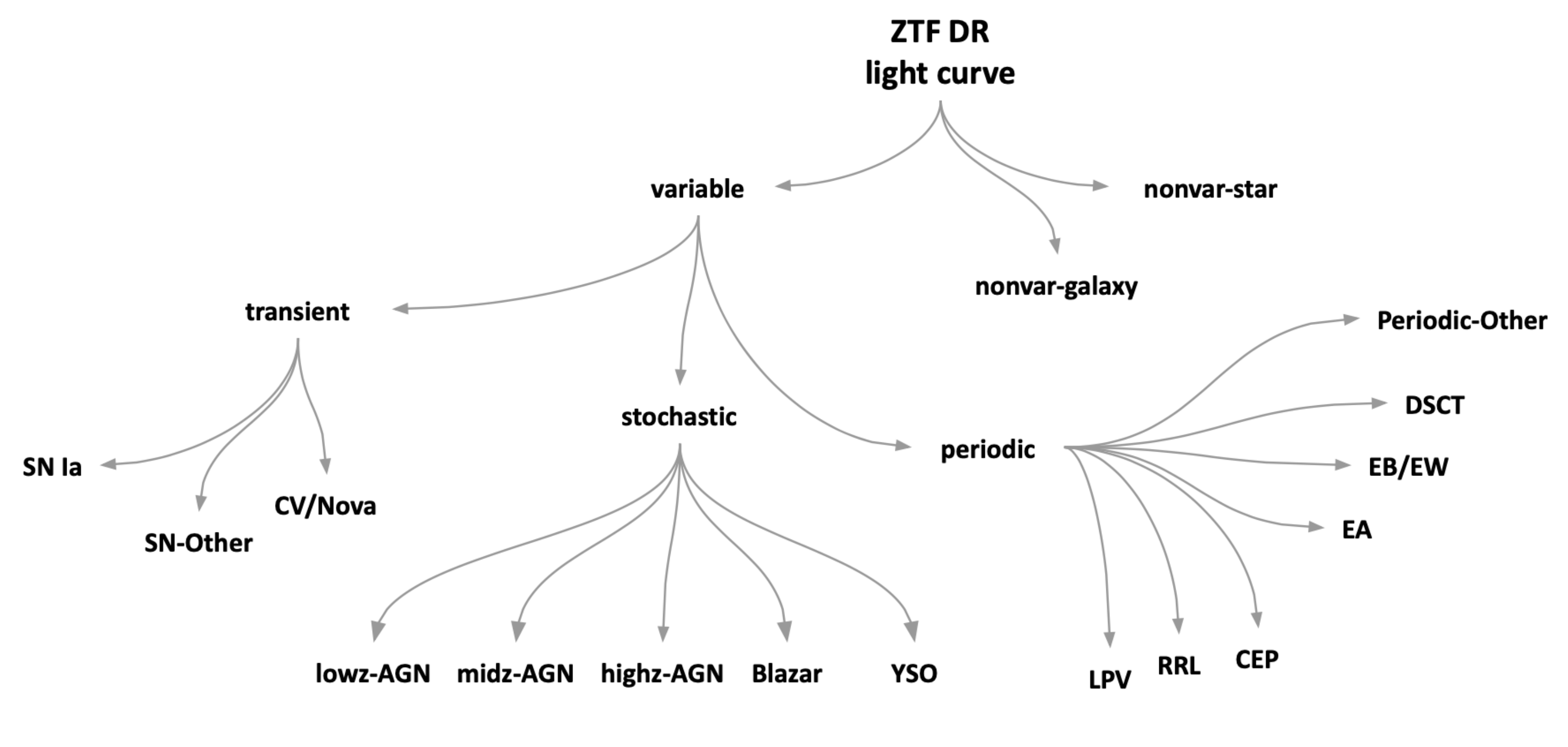}
    \caption{Hierarchical taxonomy used in this work. See Section \ref{section:taxonomy} for details\label{figure:taxonomy}}
\end{figure*}

The classifier presented here uses 61 variability features computed using the ZTF DR11 \citep{Masci19,Bellm19} light curves, nine colors (when available) obtained from PanSTARRS1 (PS1; \citealt{PS1}) and CatWISE \citep{Eisenhardt20,Marocco21}, and one morphology score from PS1 \citep{Tachibana18}. The classifier is inspired by the ALeRCE broker alert light curve classifier \cite{Sanchez-Saez21a} (hereafter \citetalias{Sanchez-Saez21a}). However, several modifications have been made to this original model, as we are dealing now with ZTF DR light curves: due to the large number of sources considered in this work, and the presence of non-variable classes, we changed the taxonomy tree (see Figure \ref{figure:taxonomy}) and used a fully hierarchical approach for the classifier; we excluded the non-detection features presented in \citetalias{Sanchez-Saez21a}; each ZTF light curve with unique DR ID is treated independently (implicitly meaning there are no multi-band features); and, in order to have an estimate of the redshift of our AGN candidates, we separated the AGN classes in three broad redshift bins. 

The paper is organized as follows. In Section \ref{section:data} we describe the data used for this work. In Section \ref{section:taxonomy_lb} we describe the taxonomy considered by our model, and the construction of the labeled set (LS). In Section \ref{section:features} we define the set of features used by the model. In Section \ref{section:dr_classification} we explain the machine learning classifier used in this work and evaluate its performance. In Section \ref{section:results} we show the results obtained when the model is applied to the ZTF light curves from the ZTF/4MOST sky. In Section \ref{section:discussion} we discuss our findings and compare our results with previous works. Finally, in Section \ref{section:sumary} we conclude and summarize the paper. 

The code used to process ZTF DR11 light curves, compute features, and cross-match ZTF DR11 objects with CatWISE and PS1 catalogs is publicly available\footnote{\url{https://github.com/alercebroker/ztf_dr}}, as well as the code used to train and evaluate this classifier, and to match external catalogs with our ZTF/4MOST classification catalog\footnote{\url{https://github.com/PaulaSanchezSaez/ZTF-DR11-classifier}}. Catalogs containing the LS for each band, the original master catalog used to cross-match with the ZTF objects, and the features and classifications for the ZTF/4MOST sky sources are available in electronic form at CDS and Zenodo.

\section{Data}\label{section:data}
\subsection{ZTF DR11}\label{section:ztf_dr11}

In this work, we used data from ZTF DR11, which includes observations taken from March 2018 to March 2022. We used the bulk download option described in Section 12.c of the ZTF DR11 documentation\footnote{\url{https://irsa.ipac.caltech.edu/data/ZTF/docs/releases/dr11/ztf_release_notes_dr11.pdf}} to recover the light curves of the full ZTF sky. Each ZTF field spans $\approx 7^{\circ}\times7^{\circ}$, and each file corresponds to a field/chip/quadrant/filter combination. The files include the PSF fit-based optical light curves in the $g$, $r$, and $i$ bands. The light curves contain the Heliocentric-based Modified Julian Date (hereafter MJD), which corresponds to the middle observing date of each exposure (with integration time $\geq30$ seconds), the PSF magnitude in a given band (calibrated for a source with color $g - r = 0$ in the AB photometric system), the magnitude error, a linear color coefficient (not used in this work as it requires previous knowledge of the simultaneous $g-r$ color of each source) and a quality score (\texttt{catflags}).

ZTF DR11 provides unique object IDs for unique combinations of RA, DEC, filter, field, CCD, and CCD-quadrant. Therefore, a single astronomical source may be associated with more than one light curve per band. This makes the position-matching of light curves from different bands or observed with different CCDs computationally expensive, especially when dealing with several millions of real astronomical sources. Hence, we decided to treat each light curve independently. This implies that some astronomical objects will have several object IDs associated with them.  

Since the aim of this work is to implement a variability-based classifier, we only considered light curves with 20 or more observations ($\text{N}_{\text{obs}}\geq20$). This minimum number of observations is more conservative than the one used by \citetalias{Sanchez-Saez21a} and is selected to ensure proper identification of intrinsic variations and proper computation of variability features from the DR light curves, which are noisier and prone to show spurious variations compared to the ZTF alert light curves. We also decided to exclude the $i$ band from the analysis, as it only covers two years of data, with fewer epochs, in the public ZTF DR11. There are 704,388,801 unique object IDs in the $g$-band and 1,251,889,824 in the $r$-band with $\text{N}_{\text{obs}}\geq20$. We call this set of sources with $\text{N}_{\text{obs}}\geq20$ in at least one band the ``long-lc-ZTF-DR11'' sample. Note that we avoided computing internal matches per band for this sample, as it is computationally expensive (although this was done for small samples, as described in Sections \ref{section:master_cat} and \ref{section:agns}). Therefore, from this work, a real astronomical object can have more than one set of features and classifications in a given band.

\subsection{Additional photometric data}\label{section:other_phot}

As mentioned in Section \ref{section:ztf_dr11}, sources with unique IDs are treated independently, thus we do not compute colors from the ZTF DR11 light curves. However, \citetalias{Sanchez-Saez21a} demonstrated the importance of including average optical colors, as well as near-infrared colors, as features for the classification of variable and transient objects. Therefore, in this work, we also included (when available) optical photometry in the $g$, $r$, and $i$ bands from PS1, and W1 and W2 photometry from CatWISE. To associate a ZTF DR11 object ID with a PS1 or CatWISE source, we used a radius of $1.5''$ (which is the default matching radius used by ZTF) and $2''$ (which is the standard matching radius adopted by ALeRCE for CatWISE data, considering the larger pixel scale of the CatWISE images compared to ZTF), respectively.

\section{Classification taxonomy and labeled set}\label{section:taxonomy_lb}

\subsection{Classification taxonomy}\label{section:taxonomy}

In this work, we use a classification taxonomy inspired by the one presented in \citetalias{Sanchez-Saez21a}; although we decided to change some of the classes presented there and add new ones. In particular, we are working with DR light curves, which include every observation performed on a source, therefore significantly differing from the alert light curves. Thus, it becomes necessary to modify the taxonomy presented in \citetalias{Sanchez-Saez21a} and for instance to include non-variable sources. In addition, the light curves of transient objects in the DRs look very different from the alert ones; thus, we decided to modify the taxonomy tree presented in \citetalias{Sanchez-Saez21a}, and use one that is more suitable for the DR light curves. For instance, the class CV/Nova is included in the transient subclass (in \citetalias{Sanchez-Saez21a} it was in the stochastic subclass), since in DR light curves they share more similarities with SNe light curves. In total, we consider 17 classes, including persistent variable objects, transients, and non-variable classes. As in \citetalias{Sanchez-Saez21a}, we divide the taxonomy in a hierarchical fashion; however, we include an additional hierarchy level in the taxonomy tree, as shown in Figure \ref{figure:taxonomy}. The division is done as follows:

\begin{itemize}
    \item [1.] Non-variable star: sources classified as stars that do not show variable behavior according to external catalogs (nonvar-star). 
    \item [2.] Non-variable galaxy: sources classified as passive or star-forming galaxies, or obscured (type 2) AGNs (nonvar-galaxy).
    \item [3.] Variable object: sources classified as persistent variable objects or transients (variable). This class is subdivided into the following:
    \begin{itemize}
        \item [(a)] Transient: sources that are not persistently variable. We include here the classes Type Ia supernova (SNIa); all the other SN types (SN-other), including Ibc supernova, type II supernova, or super luminous supernova; and cataclysmic variable or nova (CV/Nova).    
        \item [(b)] Stochastic: sources that show persistent stochastic variability. We include here AGN with $z \leq 0.5$ (lowz-AGN); AGN with $0.5 < z \leq 3$ (midz-AGN); AGN with $z > 3$ (highz-AGN); beamed jet-dominated AGN (Blazar); and YSO (class dominated by sources at evolutionary stages where the stochastic variations should dominate over the periodic ones).
        \item [(c)] Periodic: stars with periodic variable signal. We consider here long-period variable (LPV; including regular, semi-regular, and irregular variable stars); RR Lyrae (RRL); Cepheid (CEP); detached eclipsing binary (EA); semi-detached and contact variable (EB/EW); $\delta$ Scuti (DSCT); and other periodic variable stars (Periodic-other; this includes variable stars classified as miscellaneous, rotational, or RS Canum Venaticorum-type systems).
    
    \end{itemize}

\end{itemize}

\subsection{Labeled set construction}\label{section:dr_label_set}

We created a set of sources for training and testing (labeled set, LS) by cross-matching the sources of the long-lc-ZTF-DR11 sample with different catalogs of objects with known labels, obtained from multiple photometric and spectroscopic surveys. For this, we followed a priority strategy similar to that presented in \cite{Forster21}. However, since in this case we are dealing with several million objects, we decided to create first a master catalog of known objects, which was then cross-matched with the long-lc-ZTF-DR11 sample. In the following sections, we describe how this master catalog was created, and how we dealt with the cross-match of catalogs with a large number of sources. 

\subsubsection{Catalogs of non-variable stars}\label{section:nonvar_star_lb}

Catalogs of non-variable stars can be tricky to create, as a star can be constant in a given survey and variable in another one with better photometric precision. Thus, in order to generate a catalog of constant stars in ZTF, we decided to use surveys with photometric precision equal to or better than the ZTF precision. 

To create a master catalog of non-variable stars, we used two different surveys, namely the Sloan Digital Sky Survey (SDSS; \citealt{York00}) and the \textit{Gaia} early data release 3 (\textit{Gaia} eDR3; \citealt{GaiaCollaboration16,GaiaCollaboration21}). 
We first used the latest release of the SDSS Stripe 82 Standard Star Catalog\footnote{\url{https://faculty.washington.edu/ivezic/sdss/catalogs/stripe82.html}} (v4.2; \citealt{Ivezic07,Thanjavur21}), which contains 999,472 stars with at least 4 observations in each $gri$, and that are considered non-variable from their measured $\chi^2$ per filter. In order to reduce the contamination by variable stars in the sample, we only kept stars with 10 or more total observations ($N_{tot}$ in the original catalog) and with a mean magnitude per band brighter than 21 magnitudes, with a root-mean-square scatter (rms) lower than 0.16 (to reduce any possible contamination from variable stars), without classification of variable star in the SIMBAD database \citep{Wenger00}, and that are not classified as variable star in the LS used by \citetalias{Sanchez-Saez21a}. After all these filters, we ended up with a sample of 407,926 stars (SDSS\_calibstar catalog). 

\begin{figure}[tb]
    \centering
    \includegraphics[width=\linewidth]{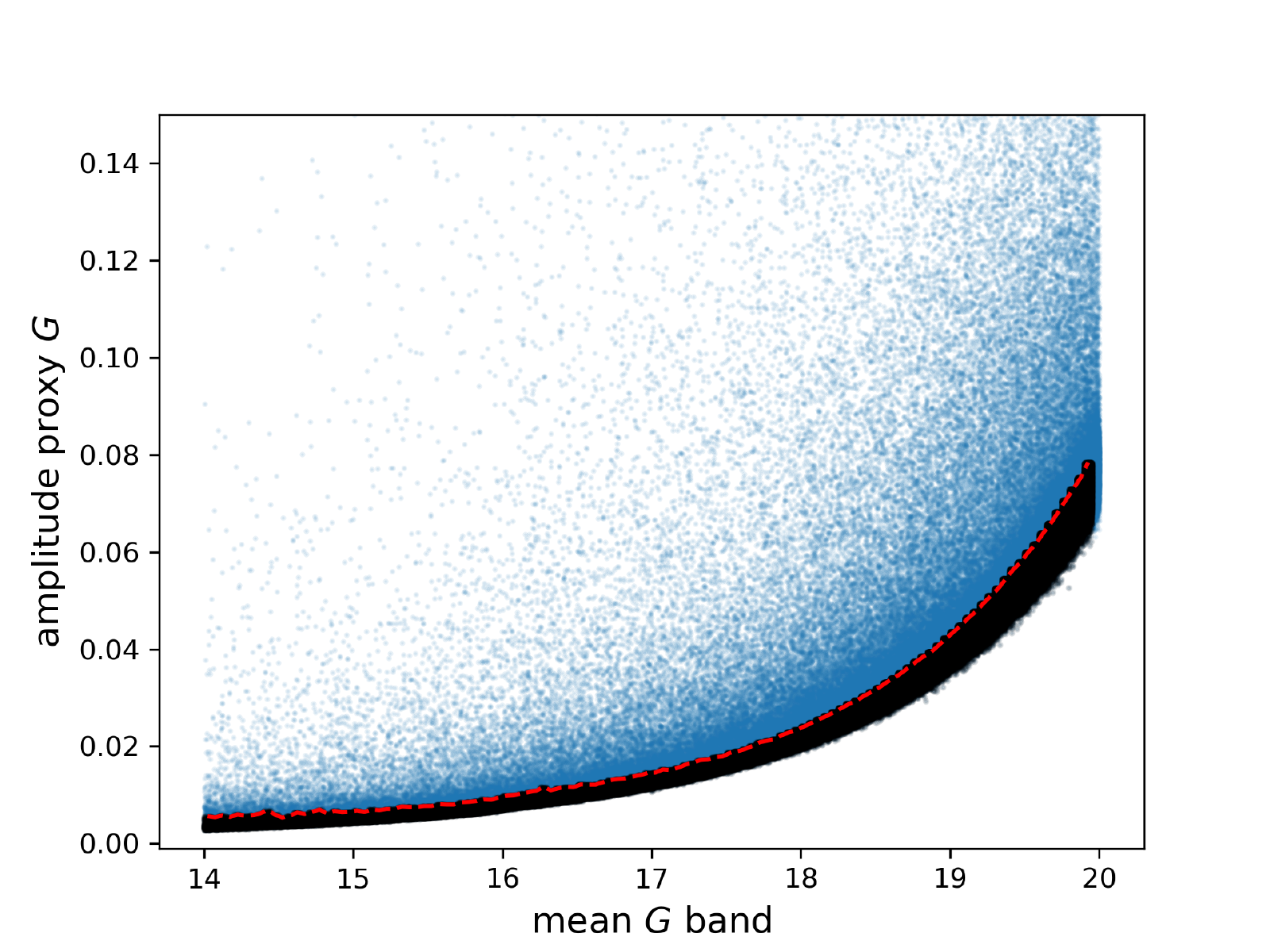}
    \caption{Amplitude proxy in the \textit{Gaia} $g$-band versus the mean $g$-band magnitude for a sample of potential non-variable stars. In blue, we show the sources that are removed from the sample, and in black the sources included in the final sample of non-variable stars. The red dashed line shows the mean $A_{\text{proxy}}$ per bin of magnitude in the $g$-band ($A_{mean}$).\label{figure:gaia_nv_star_G}}
\end{figure}

In addition, to include a sample of non-variable stars located at different positions in the sky, we used \textit{Gaia} eDR3, following the strategy presented in \cite{Mowlavi21}. We used the following advanced query (ADQL) in the \textit{Gaia} eDR3 database:
\begin{verbatim}
SELECT * 
FROM gaiaedr3.gaia_source
WHERE dec>Y AND dec<X 
AND abs(pmra)<1 AND abs(pmdec)<1 
AND phot_g_mean_mag>14 AND phot_g_mean_mag<20 
AND phot_g_n_obs>20
\end{verbatim} Here, \texttt{X} and \texttt{Y} correspond to the minimum and maximum DEC for the query.  We searched for sources in the ZTF sky using five different ranges of declination, querying 3 million objects per range, ending up with 15 million objects. To clean the sample and ensure that extended sources such as galaxies, unresolved AGNs, or sources with poor photometry are not included in the final set, we used the corrected {\em Gaia} blue (BP) and red (RP) flux excess factor $C^*$ (see Equation 6 and Table 2 in \citealt{Riello21}). \cite{Riello21} show in Figure 21 that galaxies and unresolved AGNs tend to present values of $C^*$ larger than 2; thus, we decided to only include sources with $-0.15 < C^* \leq 1$, a range that is expected to be dominated by stars. We also decided to remove from the sample all the sources with fewer than 60 observations in both BP and RP, to ensure that the final sample is as pure as possible. This filtering left us with 718,983 sources.

Then, following \cite{Mowlavi21}, we measured a proxy of the scatter in each \textit{Gaia} eDR3 band by using their equation (2): $ A_{\text{proxy}}= \sqrt{N} \frac{\epsilon(I)}{I}$, where $N$ corresponds to the number of observations in the band, $I$ to the weighted mean flux in the band, and $\epsilon(I)$ to the error in the mean flux.  We used $A_{\text{proxy}}$ to remove potential variable sources by measuring the mean $A_{\text{proxy}}$ per band and per magnitude bin of size 0.05~mag ($A_{mean}$), and then removing the sources from the sample with $A_{\text{proxy}}<A_{mean}$. Figure \ref{figure:gaia_nv_star_G} shows an example of this procedure for the $g$-band. The final sample of \textit{Gaia} eDR3 non-variable stars corresponds to the sources that satisfy this requirement in the three bands. After this, we ended up with a sample of 43,584 non-variable stars (Gaia\_nonvarstar catalog). None of these sources is located in the Stripe 82 area.

\subsubsection{Catalogs of non-variable galaxies}\label{section:nonvar_gal_lb}

Galaxies without AGN activity, as well as obscured or type 2 AGNs, are not expected to show persistent optical variability. Thus, to create a sample of non-variable galaxies, we used catalogs of known galaxies and known type 2 AGNs. 

Optical emission lines have been used for years to separate star-forming galaxies from objects with AGN activity. In particular, the BPT diagram (after Baldwin, Phillips, and Terlevich), consisting of indexes that relate the strengths of several mission lines  (H$_{\beta}$, [O III]$\lambda$5007, H$_{\alpha}$, and [N II]$\lambda$6583) as proxies for the state of ionization of a region \citep{Baldwin81}, has been widely used for this purpose. We used the Portsmouth emission line flux measurements \citep{Thomas13} to select a sample of 238,773 sources classified as star-forming galaxies from their location on the BPT diagram. Then, we removed from the sample all the sources with public alerts reported by ZTF (including data obtained until November 2021), as some of these galaxies could be hosts of SNe events, or have a weak AGN in their nuclei. After this, we ended up with 221,805 star-forming galaxies (SDSS\_galaxy catalog).

Furthermore, we use the Million Quasars Catalog (MILLIQUAS Catalog v7.4c; \citealt{Flesch19}) to select a sample of type 2 AGNs. We selected 42,639 sources classified as type 2 narrow-line core-dominated (K) or as type 2 Seyferts/host-dominated (N) in MILLIQUAS v7.4c. 

\subsubsection{Catalogs of variable and transient sources}\label{section:var_classes_lb}

To create the sample of variable and transient sources, we used the same catalogs described in \cite{Forster21} and \citetalias{Sanchez-Saez21a}: the ASAS-SN catalog of variable stars (ASASSN; \citealt{Jayasinghe18,Jayasinghe19,Jayasinghe19b,Jayasinghe20}), the Catalina Surveys Variable Star Catalogs (CRTS; \citealt{Drake14,Drake17}), the LINEAR catalog of periodic light curves (LINEAR; \citealt{Palaversa13}), the {\em {\em \textit{Gaia}}} Data Release 2 sample of LPVs and other variable stars (LPV\_GAIA and GAIADR2VS, respectively; \citealt{Mowlavi18,Rimoldini19}), the Transient Name Server database (TNS, only SNe classes were included),\footnote{\url{https://www.wis-tns.org/}} the Roma-BZCAT Multi-Frequency Catalog of Blazars (ROMABZCAT; \citealt{Massaro15}), the MILLIQUAS catalog (version 7.4c), the New Catalog of Type 1 AGNs (Oh2015; \citealt{Oh15}), and the SIMBAD database (SIMBAD\_variables; \citealt{Wenger00}). Some additional CV labels were obtained from different catalogs (including \citealt{Ritter03}), compiled by \cite{Abril20} (called JAbril).

In addition, we included two new catalogs: the new catalog of 78 hydrogen-poor superluminous SNe observed by ZTF (SLSN\_ZTF; \citealt{Chen23}), and a set of YSO catalogs curated by the ALeRCE collaboration (YSO\_ALeRCE).  The YSO compilation set consisted of several primary sources of targets that were then matched against the ZTF data release. These primary sources include catalogs that are not necessarily based on variability features and cover large portions of the sky, such as \cite{SPICY, SACY1, SACY2, SACY3} or \cite{HerbigsML}. But also others fully motivated by variability, such as the AAVSO database, or more concrete studies focused on specific star forming regions. Among the later we covered a large range of ages, environments and wavelengths at which variability was confirmed, namely: NGC 1333 \citep{NGC1333}, GGD12-15 \citep{GGD12-15}, the cluster IRAS 20050+2720 \citep{IRAS20050}, Lynds 1688 \citep{Lynds1688}, the Cep OB3b OB association \citep{CepOB3b}, the Pelican Nebula \citep{PelicanNebula}, the Cygnus OB2 association \citep{CygnusOB2}, Praesepe \citep{Praesepe}, Upper Scorpius and $\rho$ Ophiuchus \citep{USCOAnsdell, USCORebull}, and Taurus \citep{TaurusRodriguez, TaurusRebull}.

\subsubsection{Master catalog generation and cross-match with ZTF DR11 sources}\label{section:master_cat}

As mentioned in Section \ref{section:data}, ZTF DR11 contains hundreds of millions of light curves per band, making the generation of a label set a computationally expensive task. Thus, to generate an LS, instead of cross-matching each catalog with the ZTF sample and then combining the results, as done in \cite{Forster21}, we first created a master catalog that was then cross-matched against the long-lc-ZTF-DR11 sample. To avoid having repeated objects with different labels in the master catalog, we used a priority order, similar to the one presented in \cite{Forster21}, which includes the new catalogs considered in this work. These priorities were defined considering the methods used to generate each catalog and the level of disagreement in the classifications provided by them.

The master catalog was generated by concatenating the different catalogs presented in Sections \ref{section:nonvar_star_lb}, \ref{section:nonvar_gal_lb}, and \ref{section:var_classes_lb}, in the following priority order (from highest to lowest): JAbril, SLSN\_ZTF, RomaBZCAT, Oh2015, MILLIQUAS, TNS, YSO\_ALeRCE, SIMBAD\_variables, CRTSnorth, CRTSsouth, LINEAR, LPV\_GAIA, GAIADR2VS, ASASSN, SDSS\_calibstar, GAIA\_nonvarstar, SDSS\_galaxy. There are sources that can appear in more than one catalog; therefore, we used the internal match available in TOPCAT \citep{topcat} to remove all but the first element of a group of sources within a 3$''$ radius (and keeping the order of priorities). This master catalog contains 1,907,096 sources in total (1,903,799 when only including the classes considered in this work), and it can be used as an LS in future works. The catalog can be downloaded from CDS or Zenodo. 

\begin{figure*}[tb]
    \centering
    \includegraphics[width=0.8\linewidth]{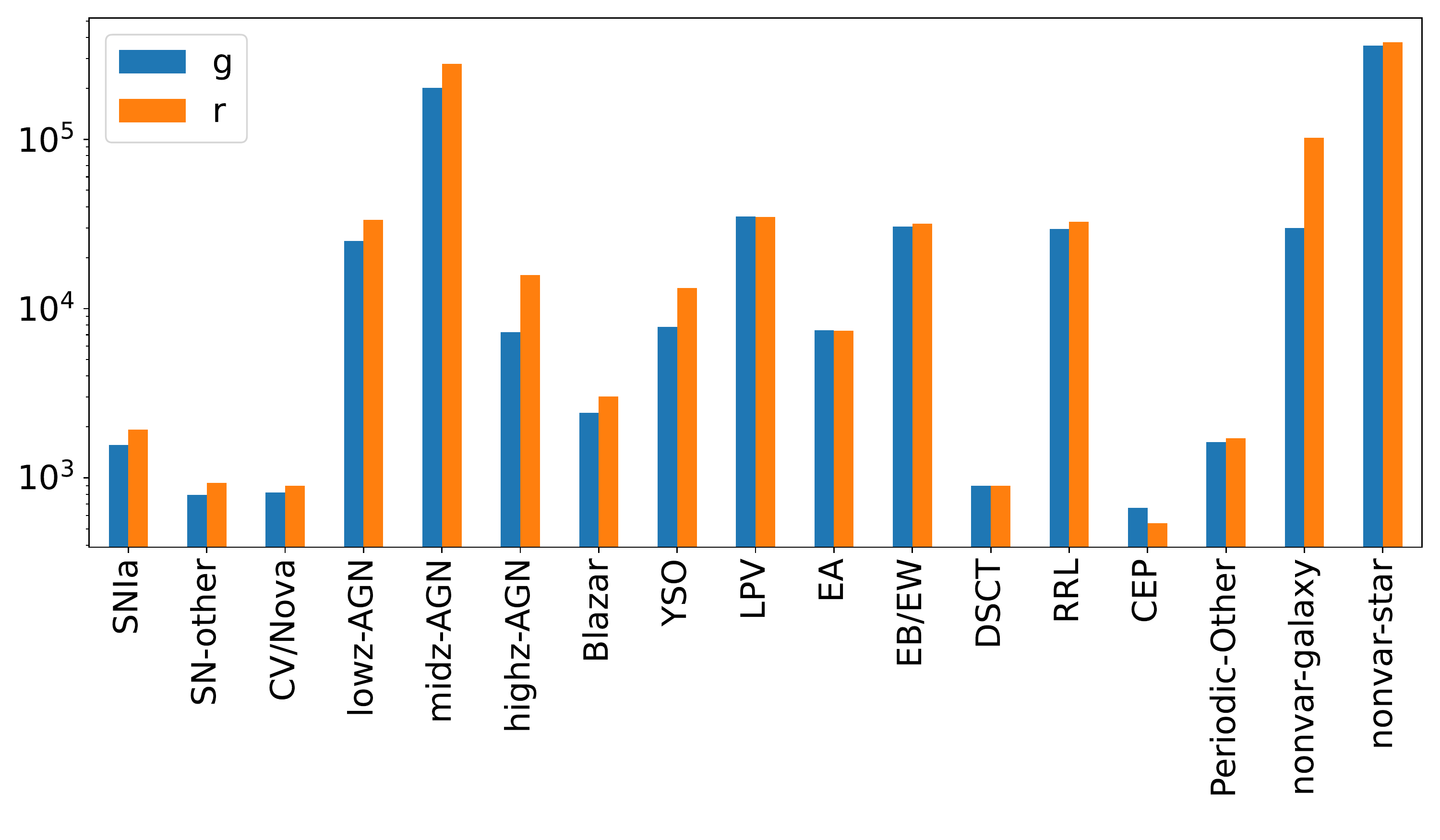}
    \caption{Number of sources per class in the LS of the $g$ and $r$ bands (note log scaling in the y axis).\label{figure:labeled_set}}
\end{figure*}
    
To generate the LS for ZTF DR11, we cross-matched the master catalog with the long-lc-ZTF-DR11 sample using a radius of 1.5$''$. There are more than 700 million sources and more than 1.2 billion sources in the long-lc-ZTF-DR11 sample with $g$-band light curves and $r$-band light curves, respectively; therefore, to cross-match the catalogs, we used the ASTROIDE extension \citep{Brahem18} for Apache Spark \citep{spark}, which is designed to deal with large-volume astronomical data sets and includes a cross-match tool. However, instead of using ASTROIDE with Spark, we used a Python wrapper of ASTROIDE developed by the ALeRCE broker, which is publicly available.\footnote{ \url{https://github.com/alercebroker/minimal_astroide}} Only sources with more than 20 observations, after removing epochs with \texttt{catflags}$>0$, are considered for the cross-match. Once the variability features were computed (see Section \ref{section:var_feats}, we also removed from the long-lc-ZTF-DR11 sample all light curves with an average magnitude greater than 20.5 ($\bar{m}> 20.5$). After cross-matching, in order to avoid having repeated light curves in a given band for a given target, we kept the longest light curve associated with each source in each bandpass. This internal match was done using TOPCAT, and was possible due to the small size of the LS. From this, we ended up with 741,263 and 936,145 sources in the LS of the $g$-band and the $r$-band, respectively. Figure \ref{figure:labeled_set} shows the number of unique astronomical targets per class in each band. Table \ref{table:catalogs_ls} presents the number of sources from each of the catalogs presented in Sections \ref{section:nonvar_star_lb}, \ref{section:nonvar_gal_lb}, and \ref{section:var_classes_lb}, included in the master catalog, and in the labeled set of the $g$-band and the $r$-band. Catalogs containing the LS for each band can be downloaded from CDS or Zenodo.

\begin{table}[htpb]
  \begin{center}
    \caption{Number of objects (\#) per catalog.}
    \label{table:catalogs_ls}
    \begin{tabular}{cccc} 
   
\hline
\hline

Catalog & \# master & \# LS  & \# LS \\
 &catalog &$g$-band & $r$-band \\

\hline

CVsJavierAbril &	1140 &	494 &	548 \\
SLSN\_ZTF &	78 &	14 &	22 \\
RomaBZCAT &	3443 &	1803 &	2173 \\
Oh2015 &	5533 &	4046 &	4610 \\
MILLIQUAS &	868952 &	247561 &	349640 \\
TNS &	13215 &	2386 &	2890 \\
YSO\_ALeRCE &	14800 &	6037 &	9989 \\
Simbad\_variables &	12887 &	1769 &	3234 \\
CRTSnorth &	42947 &	32451	 &34801 \\
CRTSsouth &	34066 &	3316 &	3420 \\
LINEAR &	3044 &	2116	 &2438 \\
LPV\_GAIA	 &75560	 &20448	 &20659 \\
GAIADR2VS	 &105294	 &31718	 &33535 \\
ASASSN	 &56006 &	16002 &	14945 \\
SDSS\_calibstar &	407874 &	323488 &	338572 \\
GAIA\_nonvarstar &	43583	 & 34777 &	36997 \\
SDSS\_galaxy	 &215377	 &12837 &	77672 \\

\hline
\hline

  \end{tabular}
  \end{center}
\end{table}

\section{Features}\label{section:features}

The main purpose of this work is to classify objects in the ZTF/4MOST sky for ChANGES; therefore, we restricted the computation of features to the ZTF fields located in this region. ZTF divides the sky into two grids. In this work, we focus on the primary grid fields (637 in total), selecting the fields centered in the sky region with $-28^{\circ} \leq \text{DEC} \leq 8.5^{\circ}$ (from field 245 to field 498). We also excluded fields in the ZTF/4MOST area with Galactic  longitude $0^{\circ} \leq gal\_l\leq 50^{\circ}$, and Galactic latitude in the range $-9^{\circ} \leq gal\_b \leq 9^{\circ}$, in order to avoid the extremely dense Milky Way bulge, where there is a high probability of having blended sources, and the high-extinction regions of the Milky Way disk. The features were computed for all the light curves from the long-lc-ZTF-DR11 sample located in this sky region. In contrast, we computed the features of the sources in the LS, regardless of their location in the ZTF sky. For each individual light curve, a total of 61 variability features, nine colors, and one morphology feature are included per source, giving a total of 71 features per target.

We computed features for 86,576,577 and 140,409,824 individual light curves in the $g$ and $r$ bands, respectively, in the ZTF/4MOST area. Due to the large size of these samples, we did not perform internal cross-matches to remove repeated objects, thus a given astronomical object can have more than one set of features associated per band.

\subsection{Variability Features}\label{section:var_feats}

In this work, we used several variability features taken from \citetalias{Sanchez-Saez21a}. In particular, we used the ``Detection Features'' presented in Table 2 of \citetalias{Sanchez-Saez21a}, excluding those that are computed using both the $g$ and $r$ bands at the same time since we treat these two bands independently. The features are computed for all the $g$ and $r$-band light curves, as described in Section \ref{section:features}. The following 61 features are computed for each light curve: \texttt{MHPS\_ratio}, \texttt{MHPS\_low}, \texttt{MHPS\_high}, \texttt{SPM\_A}, \texttt{SPM\_t0}, \texttt{SPM\_gamma}, \texttt{SPM\_beta}, \texttt{SPM\_tau\_rise}, \texttt{SPM\_tau\_fall}, \texttt{SPM\_chi}, \texttt{Amplitude}, \texttt{AndersonDarling}, \texttt{Autocor\_length}, \texttt{Beyond1Std}, \texttt{Con}, \texttt{Eta\_e}, \texttt{Gskew}, \texttt{MaxSlope}, \texttt{Meanvariance}, \texttt{MedianAbsDev}, \texttt{MedianBRP}, \texttt{PairSlopeTrend}, \texttt{PercentAmplitude}, \texttt{Q31}, \texttt{Rcs}, \texttt{Skew}, \texttt{SmallKurtosis}, \texttt{Std}, \texttt{StetsonK}, \texttt{Pvar}, \texttt{ExcessVar}, \texttt{SF\_ML\_amplitude}, \texttt{SF\_ML\_gamma}, \texttt{IAR\_phi}, \texttt{LinearTrend}, \texttt{GP\_DRW\_sigma}, \texttt{GP\_DRW\_tau}, \texttt{Period}, \texttt{PPE}, \texttt{Power\_rate\_1/4}, \texttt{Power\_rate\_1/3}, \texttt{Power\_rate\_1/2}, \texttt{Power\_rate\_2}, \texttt{Power\_rate\_3}, \texttt{Power\_rate\_4}, \texttt{Psi\_CS}, \texttt{Psi\_eta}, \texttt{Harmonics\_mag\_n} (with n = 1..7), \texttt{Harmonics\_phase\_n} (with n = 2..7), and \texttt{Harmonics\_mse}. For a detailed description of these features, see \citetalias{Sanchez-Saez21a} and the references therein. 

All the features were computed as described in \citetalias{Sanchez-Saez21a}, except for the period -- which is computed here using the single-band Multi-Harmonic Analysis of Variance (MHAOV) periodogram available in the \texttt{P4J} Python package \citep{Huijse18}\footnote{\url{https://github.com/phuijse/P4J}} -- and the features obtained from the supernova parametric model (SPM) --- which were originally based on the model presented in \cite{Villar19}. We use a version of the SPM model that can properly deal with DR light curves, which can include a non-variable component associated with the host galaxy. This SPM model adds a constant value ($C_{flux}$) to the modeled flux: 
\begin{eqnarray}
   F \: = C_{flux} + \:    \cfrac{A \left(1 - \beta' \frac{t - t_0}{t_1 - t_0}\right)}{1 + \exp{\left(-\frac{t - t_0}{\tau_{\rm rise}}\right)}} \,  \left[1 - \sigma \, \left( \frac{t-t_1}{3}\right) \right] 
   \\\nonumber
          + \:  \cfrac{A (1 - \beta') \exp{\left( - \frac{t - t_1}{ \tau_{\rm fall}} \right)}}{1 + \exp{\left(-\frac{t - t_0}{ \tau_{\rm rise}}\right)}} \, \left[\sigma \left(\frac{t - t_1}{3}\right)\right].
\end{eqnarray} This $C_{flux}$ parameter, however, is not included in the classifier as a feature, to avoid any bias in the LS magnitude distribution. 

The large number of light curves for which features have to be calculated (several tens of millions), and the diversity of target densities in different regions of the ZTF/4MOST sky, make the computation of features a challenging task. For this aim, we used the super-computing infrastructure of the National Laboratory for High Performance Computing (NLHPC) in Chile, which allowed us to compute features using 300 independent cores, with the ``general" partitions,\footnote{\url{https://www.nlhpc.cl/infraestructura/}} each with an allocated memory of 4.3 Gigabytes.

\subsection{Color and morphology Features}\label{section:color_feats}

As mentioned in Section \ref{section:other_phot}, colors can improve the classification of variable and transient objects. Therefore, we used PS1 and CatWISE photometry to compute the following colors: $g-r$, $r-i$, $g-\text{W1}$, $g-\text{W2}$, $r-\text{W1}$, $r-\text{W2}$, $i-\text{W1}$, $i-\text{W2}$, and $\text{W1}-\text{W2}$. Note that these colors are not corrected for Galactic extinction. We also used a morphological star/galaxy score (\texttt{ps\_score}) obtained from \cite{Tachibana18}, which used PS1 data to separate stars from galaxies.

\section{Classification algorithm}\label{section:dr_classification}

\subsection{The Hierarchical Balanced Random Forest}\label{section:bhrf}

Taking into account the hierarchical structure of the taxonomy presented in Section \ref{section:taxonomy}, we decided to construct a hierarchical classifier, where each ZTF band is treated independently. There are various approaches to performing a hierarchical classification (e.g., \citealt{Silla11,HierarchicalClassificationBook}); in this work, we used a \textit{local classifier per parent node} approach, where each node of the hierarchical classifier corresponds to a multi-class classifier. The model follows the hierarchy presented in Figure \ref{figure:taxonomy}. In the first level of the model, we classify each source as nonvar-star, nonvar-galaxy, or variable (node\_init). All the sources classified as variable go to the second level, where we classify the sources as transient, stochastic, or periodic (node\_variable). Finally, in the third level, we further classify the transient sources into SN Ia, SN-other, or CV/Nova (node\_transient); the stochastic sources into lowz-AGN, midz-AGN, highz-AGN, Blazar, or YSO (node\_stochastic); and the periodic sources into LPV, RRL, CEP, EA, EB/EW, DSCT, or Periodic-other (node\_periodic). 

Considering the high class imbalance of our LS, as shown in Figure \ref{figure:labeled_set}, we decided to use a Balanced Random Forest (BRF\footnote{For further details about the BRF classifier, see \citetalias{Sanchez-Saez21a}}; \citealt{Chen04}) in each node of the hierarchical classifier. Therefore, we call this model the Hierarchical Balanced Random Forest (HBRF). In particular, we used the BRF implementation available in the \texttt{Imbalanced-learn} Python package \citep{imblearn}. We trained independent BRF classifiers for each node; thus, the HBRF of each band is composed of five BRF classifiers.

We used 80\% of the LS as training set, and 20\% as testing set. In order to maintain the original percentage of samples per class, we used the \texttt{train\_test\_split} method available in \texttt{scikit-learn} \citep{Pedregosa12}, in a stratified fashion. For instance, for the $r$-band, in the training set there are 1,545 SNIa, 223,680 midz-AGN, and 27,772 LPV, and in the testing set there are 386 SNIa, 55,920 midz-AGN, and 6,943 LPV. 

We optimize the hyperparameters of each BRF classifier independently. We used the K-Fold Randomized Cross-Validation procedure available in \texttt{scikit-learn}, with $k=5$ folds (from the training set) and using the ``F1-macro'' as target score. We searched for the optimal number of trees in the forest (number of estimators: [100, 200, 300, 400, 500]), the fraction of features to consider in each split (max features: [0.2, 0.4, \texttt{auto}, \texttt{sqrt}]), and the maximum depth of each tree (max depth: [10, 30, 50, 70, 90, \texttt{None}]). Table \ref{table:hyper-parameters} shows the optimized hyper-parameters used by each node of the HBRF classifier. The rest of the parameters were not optimized; we used \textit{entropy} as the split criterion and \textit{balanced\_subsample} for the class\_weight parameter; for the rest, we used the default values.

\begin{table}[htpb]
  \begin{center}
    \caption{Hyper-parameters of the HBRF.}
    \label{table:hyper-parameters}
    \begin{tabular}{ccccc} 
   
   \hline

Band & Node & Number of & Max & Max \\
 &  & estimators & features & depth \\

\hline

\hline
 
        & node\_init       & 500 & 0.4 & None  \\
        & node\_variable   & 400 & 0.4 & 30 \\
 $g$ & node\_transient  & 300 & 0.4 & 10 \\
        & node\_stochastic & 300 & 0.4 & 30 \\
        & node\_periodic   & 300 & 0.4 & None \\

\hline

        & node\_init       & 300 & 0.4 & 70  \\
        & node\_variable   & 500 & 0.4 & 70 \\
    $r$ & node\_transient  & 200 & auto & 10 \\
        & node\_stochastic & 200 & 0.4 & 70 \\
        & node\_periodic   & 400 & 0.4 & 50 \\

\hline

  \end{tabular}
  \end{center}
\end{table}

From this model, we classify objects among 17 classes, which correspond to the final leaves of each node, and we provide the probabilities obtained by the corresponding node ($P$). For instance, for SN Ia, we report the probability according to the HBRF of the node\_transient; for YSOs we report the probability of the node\_stochastic; for LPVs, we report the probability of the node\_periodic; and for nonvar-stars, we report the probability of the node\_init.  For the sources classified as variable in the node\_init, we also report the probabilities and classifications obtained from this node ($P_{init}$), and the probabilities and classifications of the node\_variable.

\begin{table}[htpb!]
  \begin{center}
    \caption{HBRF classifier metrics per band and class, and macro-averaged scores.}
    \label{table:scores}
    \begin{tabular}{ccccc} 
   
   \hline

Band & Class & Precision & Recall & F1-score \\

\hline

\hline
 
   &     SNIa   &    0.72   &   0.38   &   0.50    \\
   &   SN-other   &  0.34  &    0.35   &   0.35    \\
   &    CV/Nova    &   0.24   &   0.86   &   0.37 \\
   &   lowz-AGN   &    0.47   &   0.53  &    0.50 \\
   &   midz-AGN   &    0.99   &  0.79  &    0.88  \\
   &   highz-AGN   &    0.22   &   0.89   &   0.36  \\
   &   Blazar  &   0.12   &   0.56   &   0.20  \\
   &   YSO   & 0.88   &   0.43     & 0.58 \\
 $g$   & LPV  &     0.98    &  0.98    &  0.98      \\
   &  EA  &     0.76   &   0.90   &   0.82  \\
   &  EB/EW  &    0.95  &    0.82 &     0.88   \\
   &  DSCT    &    0.39   &   0.83    &  0.53     \\
   &   RRL    &   0.98   &   0.90  &    0.94   \\
   &  CEP     & 0.30   &   0.86   &   0.45 \\
   & Periodic-other  &     0.25   &   0.65   &   0.36   \\
   &  nonvar-galaxy   &    0.71   &   0.99   &   0.83 \\
   &   nonvar-star    &   0.99   &  1.00   & 0.99 \\
   
   \cline{2-5}
   
   & macro-av &  0.61    &  0.75  &    0.62 \\

\hline

  &       SNIa    &    0.62  &    0.35  &    0.44  \\
  &    SN-other   &     0.29   &   0.32 &     0.30   \\
  &     CV/Nova   &     0.19 &     0.80   &   0.31    \\
  &    lowz-AGN   &    0.48   &   0.51   &   0.49    \\
  &    midz-AGN   &   0.99    &  0.80    &  0.88 \\
  &   highz-AGN   &   0.32   &   0.89    &  0.47       \\
  &      Blazar   &   0.10   &   0.50   &   0.16 \\
  &         YSO   &   0.91   &   0.58    &  0.71    \\
 $r$ &     LPV   &   0.98   &   0.99    &  0.98  \\
  &        EA   &    0.73   &   0.89   &   0.80   \\
  &       EB/EW   &  0.95    &  0.81   &   0.88  \\
  &        DSCT   &   0.30   &   0.84  &    0.44  \\
  &         RRL   &   0.97  &    0.88   &   0.92  \\
  &         CEP   &   0.18    &  0.83   &   0.30  \\
& Periodic-other   &   0.27  &    0.68   &   0.39    \\
 & nonvar-galaxy   &    0.86  &    0.99 &     0.92   \\
  & nonvar-star   &    0.99   &   1.00  &    0.99   \\

   \cline{2-5}

   & macro-av &  0.60  &    0.74  &    0.61  \\

\hline

  \end{tabular}
 \end{center}
\end{table}

\subsection{Classification Performance}\label{section:performance}

\begin{figure*}[htbp!]
\begin{center}
\begin{tabular}{cc}
\includegraphics[scale=0.35]{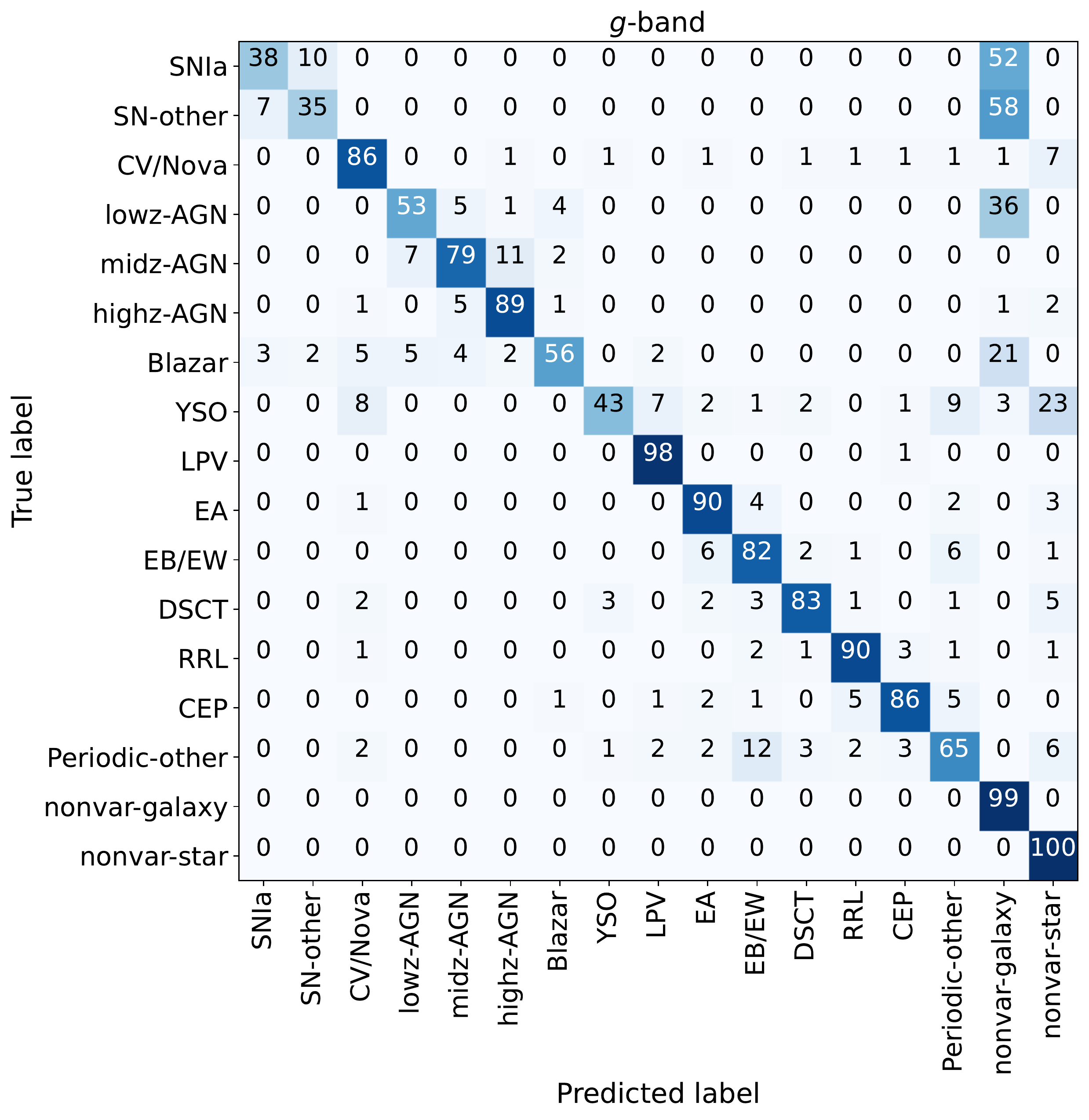} &
  \includegraphics[scale=0.35]{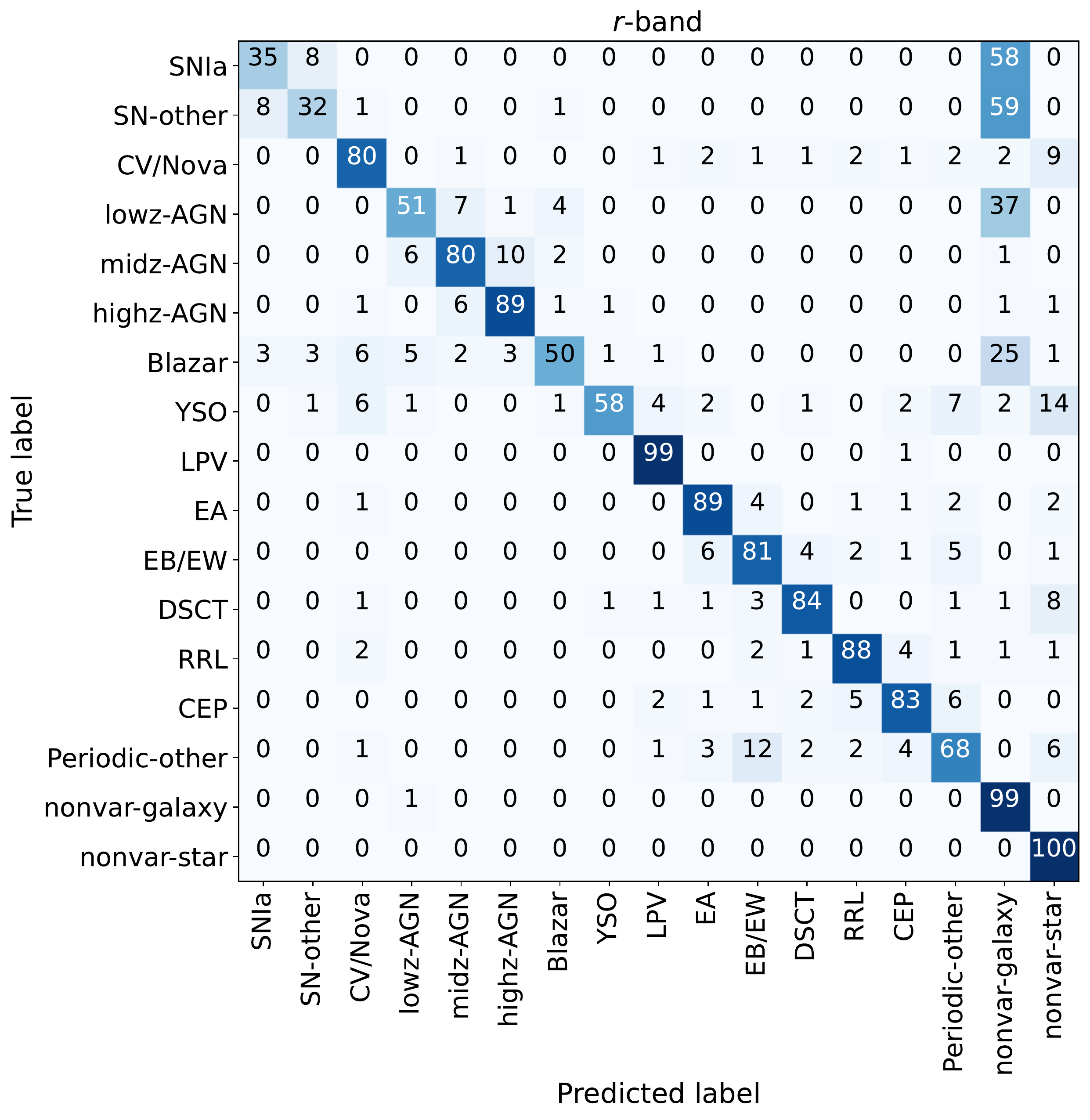} \\
\end{tabular}
\caption{Confusion matrices of the final 17 classes obtained by using the HBRF in the testing set. The matrix for the $g$-band is shown in the left panel, and for the $r$-band in the right panel. To normalize the confusion matrix results as percentages, we divided each row by the total number of objects per class with known labels, and we rounded these percentages to integer values.  \label{figure:conf_max_third}}
\end{center}
\end{figure*}

To evaluate the performance of the HBRF, we applied the model to the testing set, which corresponds to the 20\% of the LS not used during training. In particular, we used the \texttt{classification\_report} method available in \texttt{scikit-learn}, to recover the precision, recall, and F1-score for each of the 17 classes independently, and for the full testing set (macro-averaged scores). These scores are defined as: 

\begin{equation}\label{eq:sing_precision} 
\text{Precision}_{i} = \frac{TP_i}{TP_i+FP_i},
\end{equation} 

\begin{equation}\label{eq:sing_recall}
\text{Recall}_i =  \frac{TP_i}{TP_i+FN_i},
\end{equation} 

\begin{equation}\label{eq:sing_f1}
\text{F1-score}_i = 2 \times \frac{\text{Precision}_i \times \text{Recall}_i}{\text{Precision}_i + \text{Recall}_i},
\end{equation}
where $i$ corresponds to a particular class, and $TP_i$, $FP_i$, $FN_i$ are the corresponding numbers of true positives, false positives, and false negatives, respectively.

From these scores per class, we compute the macro-averaged scores as: 

\begin{equation}\label{eq:precision} 
\text{Precision}_{\text{macro}} = \frac{1}{n_{cl}} \sum_{i=1}^{n_{cl}} \text{Precision}_{i},
\end{equation} 

\begin{equation}\label{eq:recall}
\text{Recall}_{\text{macro}} =  \frac{1}{n_{cl}} \sum_{i=1}^{n_{cl}} \text{Recall}_{i},
\end{equation} 

\begin{equation}\label{eq:f1}
\text{F1-score}_{\text{macro}} = \frac{1}{n_{cl}} \sum_{i=1}^{n_{cl}} \text{F1-score}_{i},
\end{equation} 
where $n_{cl}$ is the total number of classes (17 in this work). Table \ref{table:scores} shows the scores obtained per class and per band for the final classification (final leaves of each node). From the table, we can see that the classification of nonvar-star has high precision, recall, and F1-score, in both ZTF bands. The same is observed for nonvar-galaxies. The lowest scores are for SN-other, Blazar, CEP, and Periodic-other.

The low scores observed in some classes can be explained by the confusion with classes that show similar variability behaviors. To better illustrate this, we show in Figure \ref{figure:conf_max_third} the confusion matrices for each band. We can see in both bands that SNe classes tend to be miss-classified as nonvar-galaxy; the same is observed for lowz-AGN and Blazar. On the other hand, YSO are confused with nonvar-stars and Periodic-other. Periodic-other are confused with binary classes. These confusions between persistent variable and transient classes are expected and have been observed before in other works that use ZTF light curves (e.g., \citealt{Sanchez-Saez21a,vanRoestel21}). They can be explained by the similarities between classes and the lack of precise variability features when measured from the sparse and irregularly sampled DR light curves. For the case of YSOs the confusion with periodic classes can be explained by the diversity of variability behaviors of YSOs, which can present stochastic and periodic signals \citep{Lakeland22}. The confusion with non-variable classes can be explained by the fact that we are using DR light curves, which can hinder variations typically seen in the light curves of extended sources (like lowz-AGNs), or in low-amplitude point sources located in crowded fields (like YSOs). Moreover, the confusion between SNe classes and the nonvar-galaxy class is expected, since a transient event does not necessarily occur in the nucleus of the host, and thus, the DR light curves do not necessarily include the transient signal.

Regardless of these classification confusions, the fraction of non-variables classified as variables is close to zero. Among nonvar-galaxies, 1\% of the testing set are classified as AGNs, but this confusion is expected since the class nonvar-galaxy includes type 2 AGNs, which can, in fact, correspond to misidentified type 1 AGNs \citep{LopezNavas23}, or that can correspond to former type 2 AGNs that have transitioned to type 1 (CSAGNs; e.g., \citealt{LopezNavas22}). We also expect a small fraction of nonvar-galaxies to be classified as SNe when they are the latter's hosts. 

Moreover, when grouping the four AGN classes into one single class (lowz-AGN, midz-AGN, highz-AGN, and Blazar), its precision, recall, and F1-score are 1.00, 0.95, and 0.97, respectively, in both $g$ and $r$ bands. Therefore the low scores measured for some of the AGN classes, can be explained by confusion within these classes. This confusion is expected, since these four classes represent intrinsically the same astrophysical class (AGN), but observed under different conditions and with different properties (like the redshift or the presence or not of a relativistic jet). 

These results demonstrate that we can use ZTF DR light curves to recover clean samples of variables and transients objects, and particularly of AGNs, which is encouraging, considering that we are using PSF fit-based light curves, which are very sensitive to seeing and sky condition variations when dealing with extended sources.

\begin{table*}[htpb]
  \begin{center}
    \caption{Feature ranking (top 15) for each node of the HBRF and for each band, ordered by relevance.}
    \label{table:feat_rank}
    \begin{tabular}{cccccc} 
   
   \hline

Band & node\_init & node\_variable & node\_transient & node\_stochastic & node\_periodic\\

\hline

\hline
 
   & \texttt{ps\_score}      & $\text{W}1-\text{W}2$   &  \texttt{ps\_score}       & $\text{W}1-\text{W}2$  & \texttt{Period}  \\
   & $i-\text{W}2$           & \texttt{SPM\_chi}       &    \texttt{SPM\_chi}      &   $g-r$                & \texttt{Skew}  \\
   & $\text{W}1-\text{W}2$   &   \texttt{Psi\_eta}     & \texttt{MHPS\_high}       &    $r-\text{W}1$       & \texttt{IAR\_phi} \\
   &  \texttt{PPE}           &  \texttt{PPE}           &  \texttt{SPM\_A}          & $g-\text{W}1$          &  \texttt{Gskew} \\
   & $i-\text{W}1$           &   \texttt{ps\_score}    &  $g-\text{W}1$            &    \texttt{ps\_score}  & \texttt{MedianAbsDev}  \\
   & \texttt{Psi\_eta}       &   $i-\text{W}2$         &    $g-r$                  & $i-\text{W}2$          &  \texttt{Q31}  \\
$g$   & $r-\text{W}2$        & \texttt{SPM\_A}         & \texttt{SPM\_tau\_fall}   &   $r-i$                & \texttt{GP\_DRW\_tau}  \\
   & \texttt{IAR\_phi}       &    $g-\text{W}1$        & $r-i$                     &  \texttt{SPM\_chi}     &  $g-\text{W}1$   \\
   & \texttt{GP\_DRW\_tau}   &  $g-r$                  &  \texttt{Skew}            &    $r-\text{W}2$       &    \texttt{MedianBRP}  \\
   &  \texttt{ExcessVar}     &   $r-i$                 & \texttt{Gskew}            &   $i-\text{W}1$        & \texttt{Beyond1Std}  \\
   &\texttt{SPM\_chi}        & \texttt{Skew}           & \texttt{PercentAmplitude} & \texttt{ExcessVar}     & \texttt{MHPS\_ratio} \\
   &\texttt{Autocor\_length} & $g-\text{W}2$           &  \texttt{PPE}             & $g-\text{W}2$          & $r-i$    \\
   &\texttt{GP\_DRW\_sigma}  & $r-\text{W}2$           & $g-\text{W}2$             & \texttt{MHPS\_low}     & $g-\text{W}2$ \\ 
   &  $g-r$                  & \texttt{MHPS\_low}      & $i-\text{W}1$             &\texttt{IAR\_phi}       & $i-\text{W}2$ \\
   & $r-\text{W}1$           & \texttt{Harmonics\_mse} & \texttt{GP\_DRW\_tau}     & \texttt{Amplitude}     &  \texttt{SPM\_chi} \\

\hline

 & \texttt{ps\_score}        & $\text{W}1-\text{W}2$    & \texttt{ps\_score}        & $r-i$                 & \texttt{Period}  \\
 & $i-\text{W}2$             & \texttt{SPM\_chi}        & $i-\text{W}1$             & $\text{W}1-\text{W}2$ &  \texttt{Skew} \\
 & $\text{W}1-\text{W}2$     &  \texttt{Psi\_eta}       & \texttt{MHPS\_high}       &  $g-r$                & \texttt{Gskew} \\
 & $i-\text{W}1$             &  \texttt{ps\_score}      & \texttt{SPM\_A}           & $r-\text{W}1$         &  $g-\text{W}1$\\
 & \texttt{GP\_DRW\_sigma}   & \texttt{SPM\_A}          & \texttt{SPM\_chi}         & \texttt{ps\_score}    & \texttt{GP\_DRW\_tau} \\
 & \texttt{Psi\_eta}         & \texttt{Eta\_e}          & $g-\text{W}1$             & \texttt{SPM\_chi}     & \texttt{MHPS\_ratio} \\
 $r$ &  \texttt{PPE}         & $i-\text{W}2$            & $r-i$                     & $r-\text{W}2$         & $g-\text{W}2$\\
 & $r-\text{W}2$             &\texttt{Harmonics\_mse}   & \texttt{Harmonics\_mse}   & $i-\text{W}2$         & $i-\text{W}1$  \\
 &  \texttt{Meanvariance}    & $g-\text{W}2$            & $g-r$                     &  $i-\text{W}1$        & \texttt{IAR\_phi} \\
 & \texttt{Q31}              & $g-\text{W}1$            &  $g-\text{W}2$            &  $g-\text{W}1$        & $i-\text{W}2$  \\ 
 &  $g-r$                    & $r-i$                    & \texttt{SPM\_tau\_fall}   &  \texttt{IAR\_phi}    & \texttt{Beyond1Std}  \\
 & \texttt{GP\_DRW\_tau}     & $r-\text{W}2$            & \texttt{PercentAmplitude} & \texttt{ExcessVar}    & $r-\text{W}1$\\
 &\texttt{Harmonics\_mag\_1} & $g-r$                    & $r-\text{W}2$             &  \texttt{Eta\_e}      & \texttt{MedianBRP}   \\
 & \texttt{Eta\_e}           & \texttt{Amplitude}       & $r-\text{W}1$             & \texttt{Amplitude}    &\texttt{SPM\_chi}  \\
 &  \texttt{SPM\_chi}        & \texttt{PPE}             & $i-\text{W}2$             & $g-\text{W}2$         & \texttt{MedianAbsDev}\\

\hline

  \end{tabular}
 \end{center}
\end{table*}

\subsection{Feature ranking}\label{section:feat_rank}

The HBRF is composed of five different BRF classifiers. Each of these classifiers will have different feature rankings, since the classes they consider are different. In Table \ref{table:feat_rank} we show the top 15 features per node and per band, ordered by relevance. From the table, we can see that the PS1 and CatWISE colors are relevant in all the classifiers. The PS1 morphology score (\texttt{ps\_score}) is relevant for all classifiers except the node\_periodic; this is expected, as this is the only classifier that considers purely point sources. The most relevant variability features remain distinct for each classifier; among the most important ones, we highlight the following: features related to the amplitude of the variability (\texttt{ExcessVar}, \texttt{GP\_DRW\_sigma}, \texttt{Meanvariance}, \texttt{Amplitude}, among others); features related to the timescale of the variations or the level of autocorrelation (\texttt{GP\_DRW\_tau}, \texttt{Autocor\_length}, and \texttt{IAR\_phi}); the Mexican hat power spectrum (MHPS) features; and the SPM features. For the case of the node\_periodic, the \texttt{Period} is the most relevant feature, which is expected from previous works (e.g., \citetalias{Sanchez-Saez21a}).

The differences observed in the ranking of features are produced by the complexity of the model and the diversity of classes considered. Therefore, we decided to keep all the features described in Section \ref{section:features} for the final model used in this work.  

\subsection{Calibration of the BRF probabilities}\label{section:prob_cal}

Random Forest (RF) models tend to provide uncalibrated probabilities \citep{Niculescu-Mizil05}. For binary classification, techniques like Platt scaling \citep{Platt99} or the isotonic method \citep{Zadrozny01,Zadrozny02} can be used to calibrate the output probabilities. However, for multiclass, and/or hierarchical, and highly imbalanced classification, the use of these techniques is not straightforward.

To understand how well-calibrated are the output probabilities of each node of the HBRF classifier, we computed the Reliability Diagrams \citep{Guo17} and Expected Calibration Error (ECE; \citealt{Naeini15}) for each node of the classifier independently, using the testing set for each band. To construct the Reliability Diagram, we used 10 bins of probabilities (each one with a width of 0.1) and measured the positive fraction (PF; the number of correctly classified samples versus the total of samples; equivalent to the micro-averaged precision) per bin of probability and the average predicted probability of the bin (confidence). The Reliability Diagram corresponds to the comparison of the PF versus the confidence of each bin. When this comparison is close to the identity function, we can say that the probabilities are well-calibrated. Then we measured the gap, which corresponds to the absolute value of the difference between the PF and the confidence in each bin ($\text{gap}=|\text{PF}-\text{confidence}|$), and calculated the ECE as $$\text{ECE}=\sum_{m=1}^M \frac{N_{bin}}{N} \text{gap}.$$ Where $N$ corresponds to the total of samples, $N_{bin}$ to the total of samples in the probability bin, and $M$ to the total number of probability bins. When the ECE score is close to zero, we can say that the model is well-calibrated.

\begin{figure*}[htbp]
\begin{center}
\begin{tabular}{ccc}
   \includegraphics[width=0.3\linewidth]{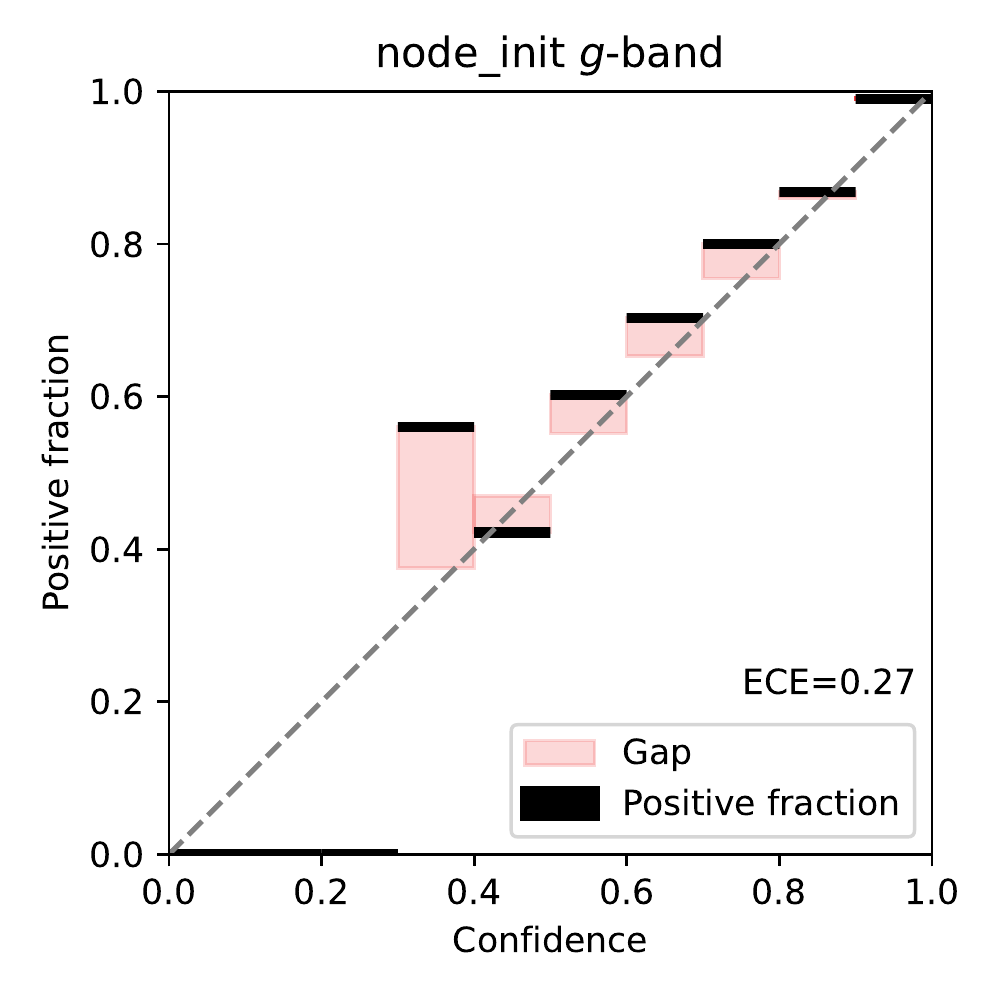} &

  \includegraphics[width=0.3\linewidth]{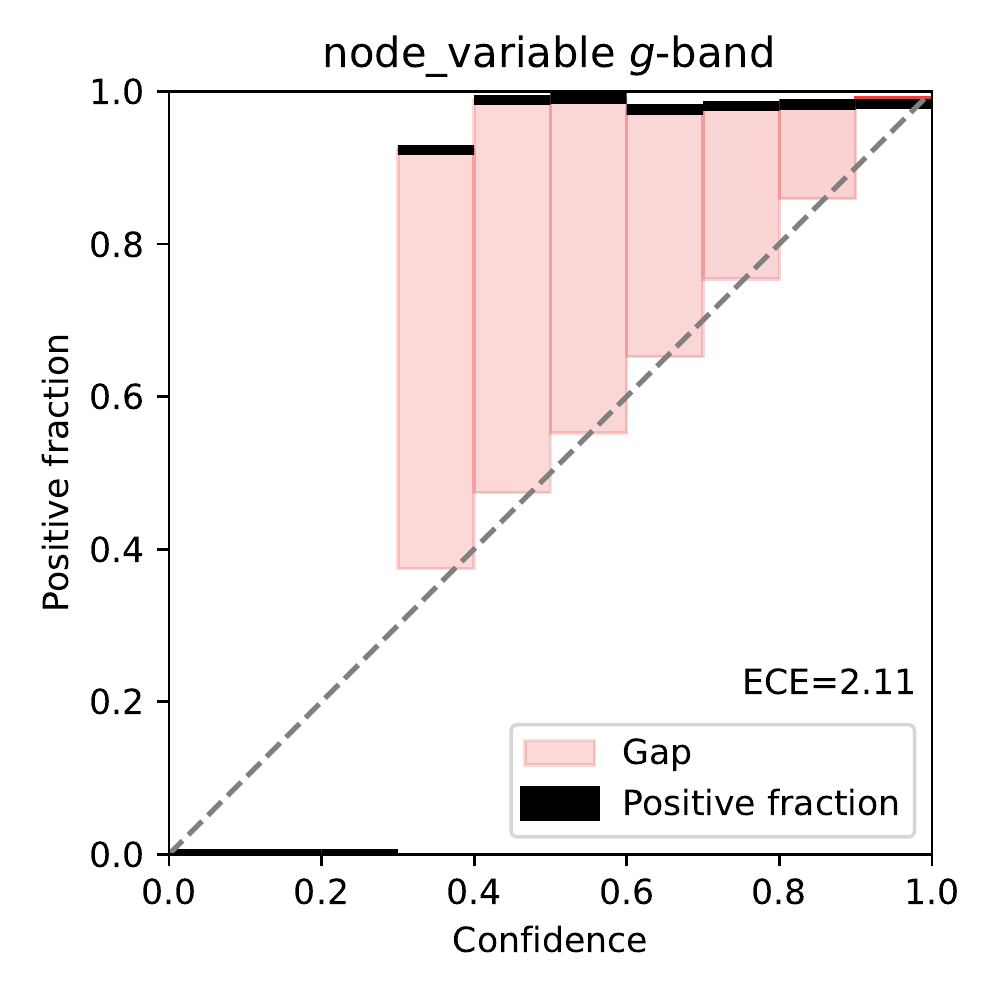} &

  \includegraphics[width=0.3\linewidth]{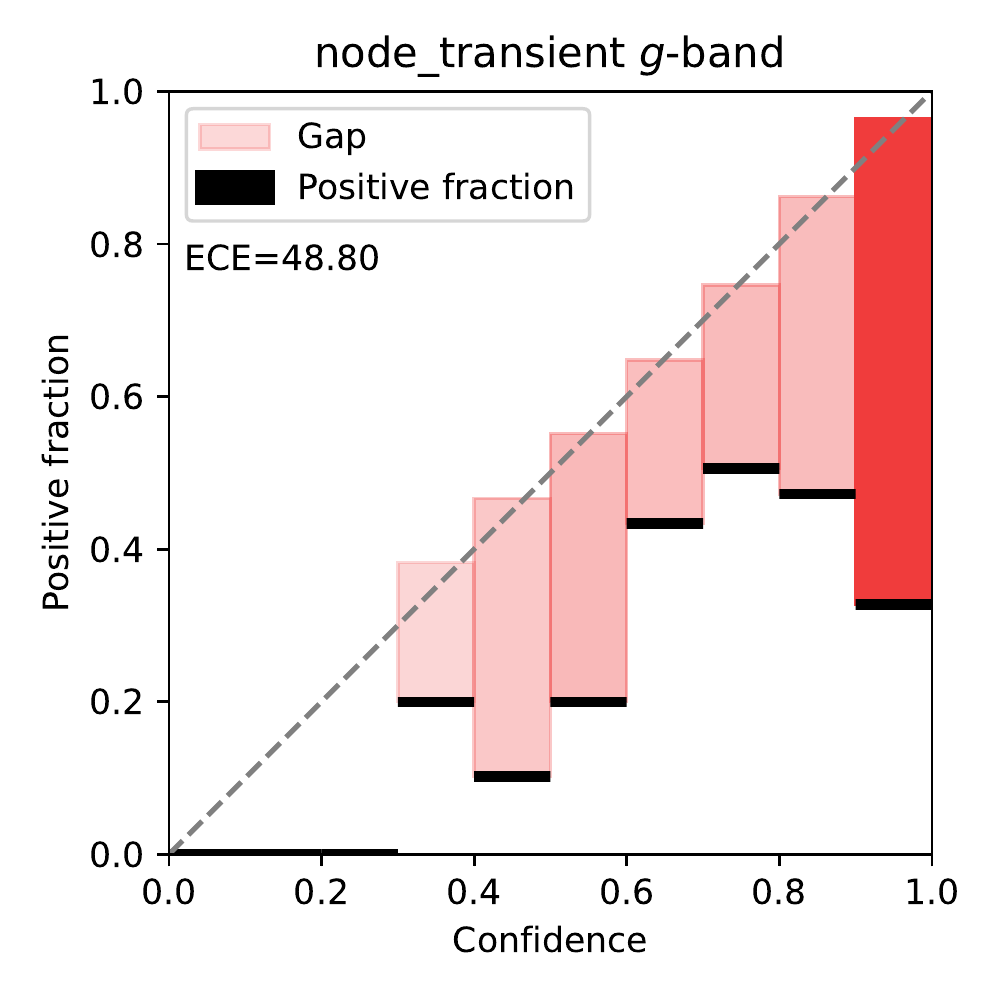} \\
\end{tabular}

\begin{tabular}{cc}
  \includegraphics[width=0.3\linewidth]{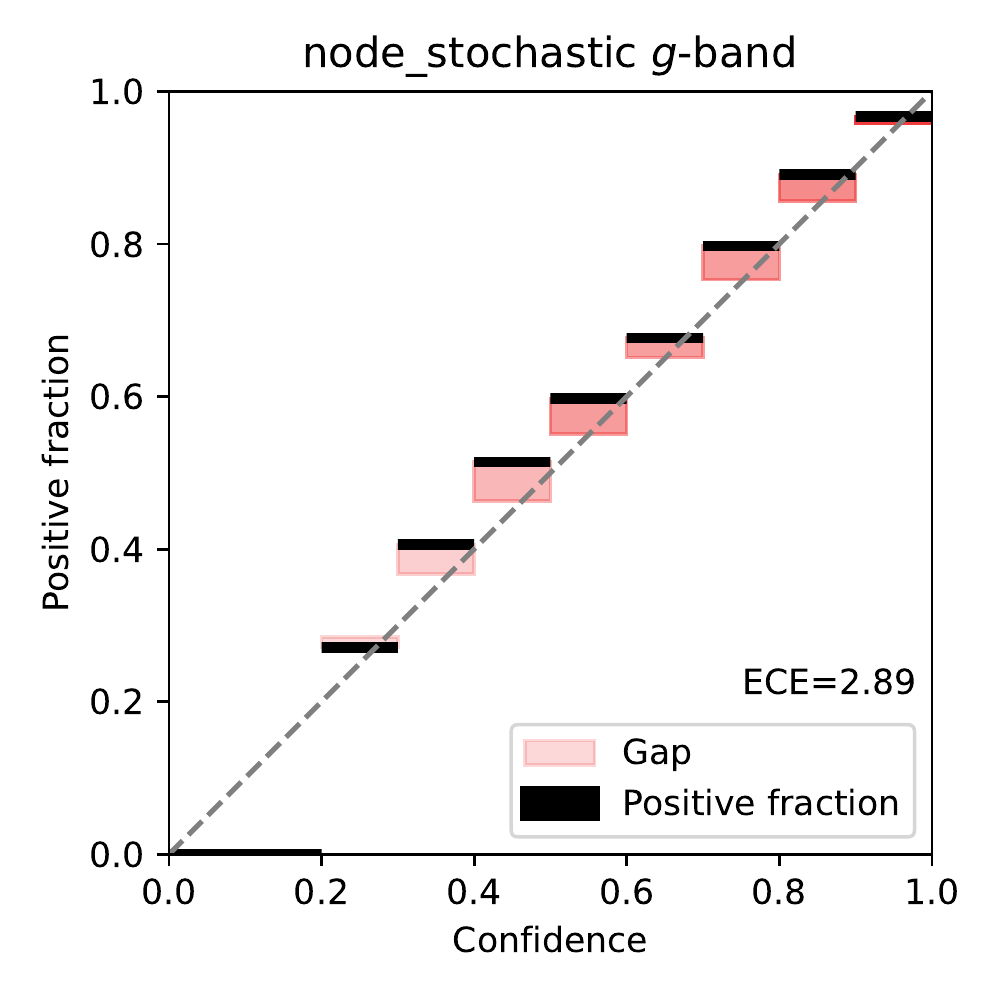} &

  \includegraphics[width=0.3\linewidth]{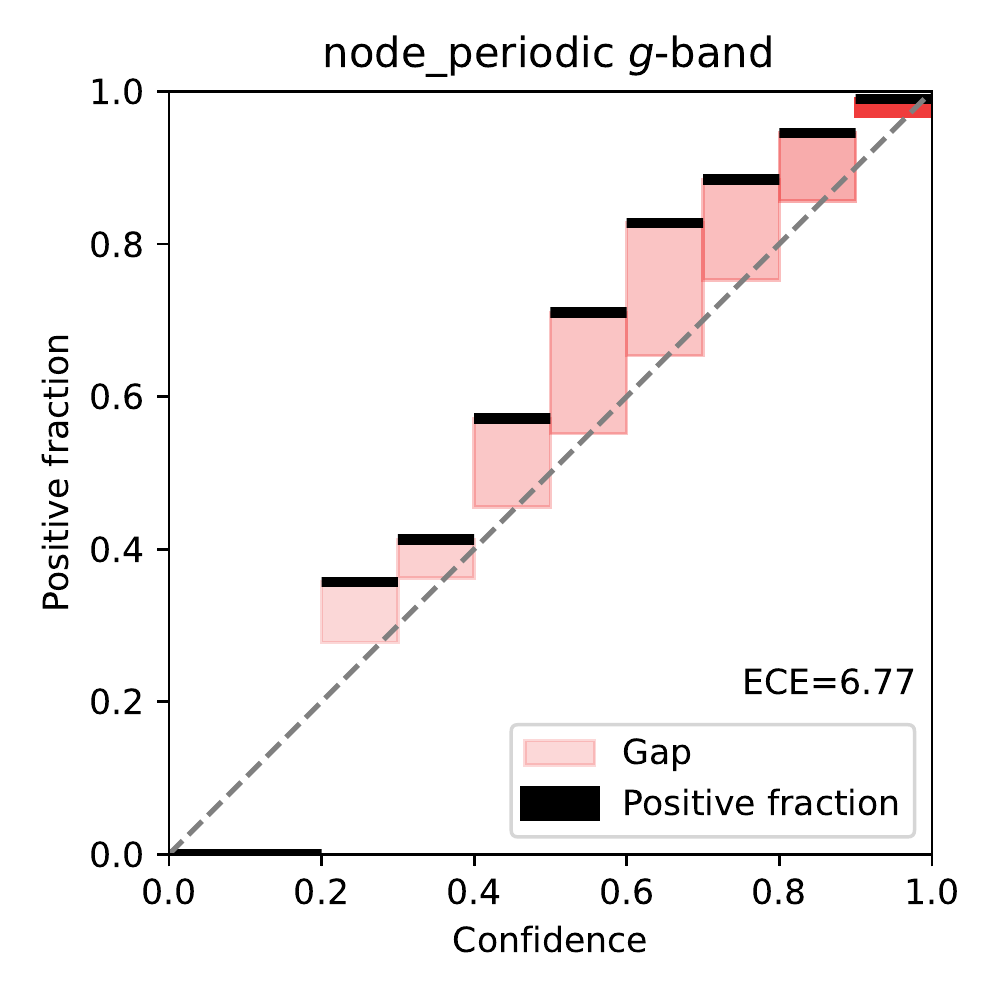} \\

\end{tabular}
\caption{Reliability Diagrams for each node of the $g$-band HBRF classifier. We plot the positive fraction (shown in black) and the gap (in red) versus the confidence. The transparency of the gap represents how much each bin contributes to the ECE score (less transparent for bins with more samples). The identity function is shown as a reference. The ECE score is shown next to the legend for each node.
\label{figure:reliability_diagram_gband}}
\end{center}
\end{figure*}

\begin{figure*}[htbp]
\begin{center}
\begin{tabular}{ccc}
   \includegraphics[width=0.3\linewidth]{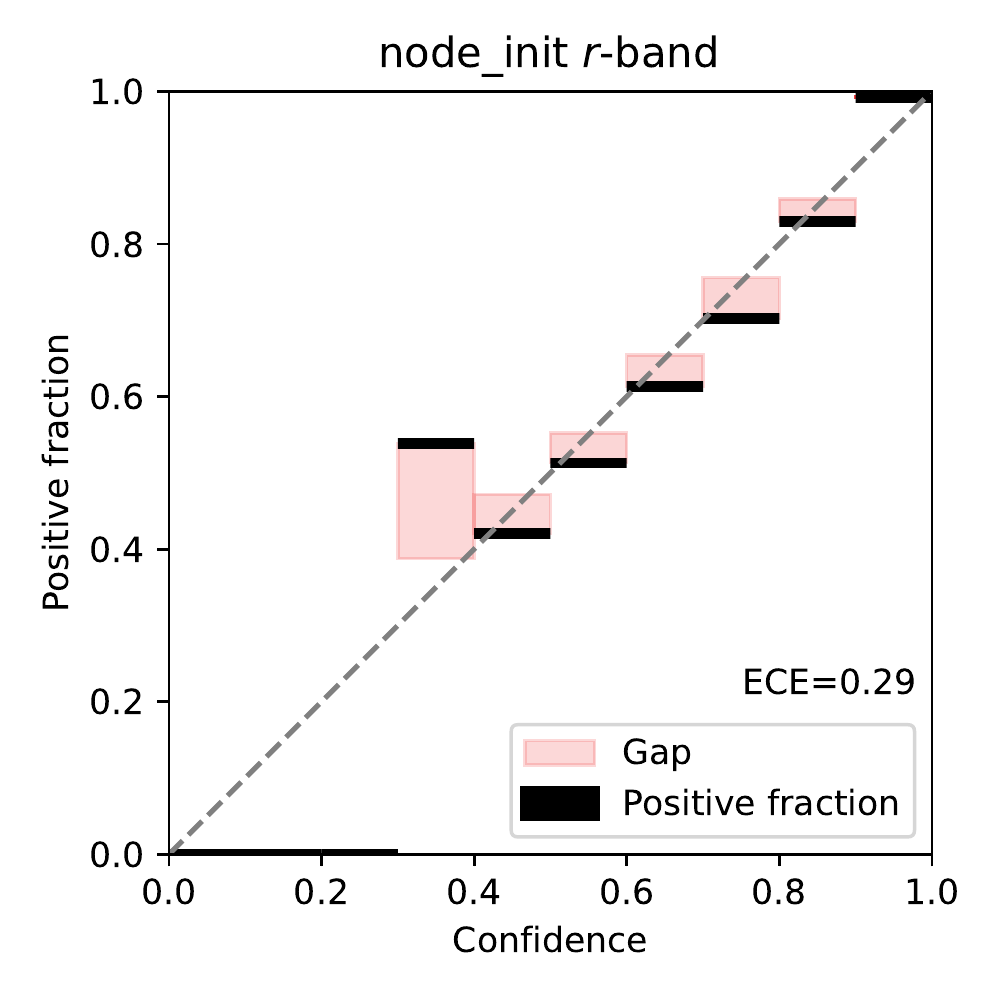} &

  \includegraphics[width=0.3\linewidth]{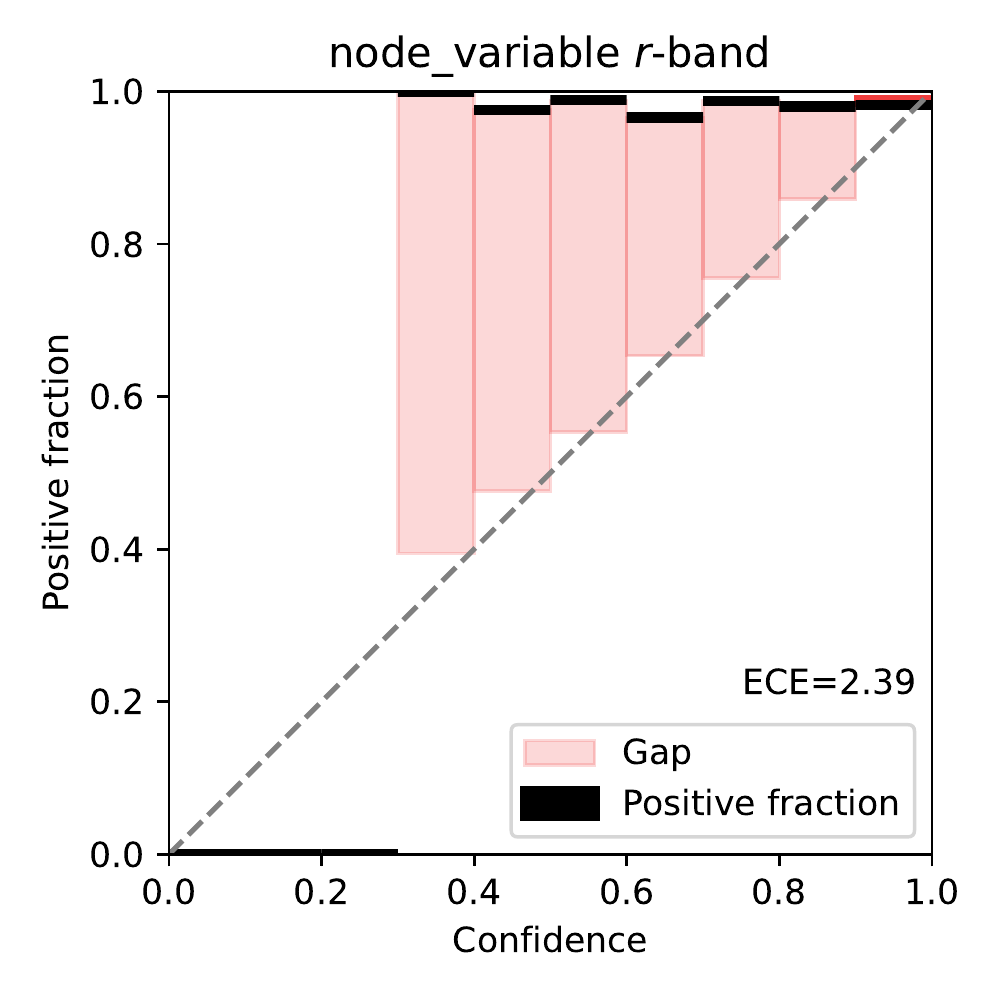} &

  \includegraphics[width=0.3\linewidth]{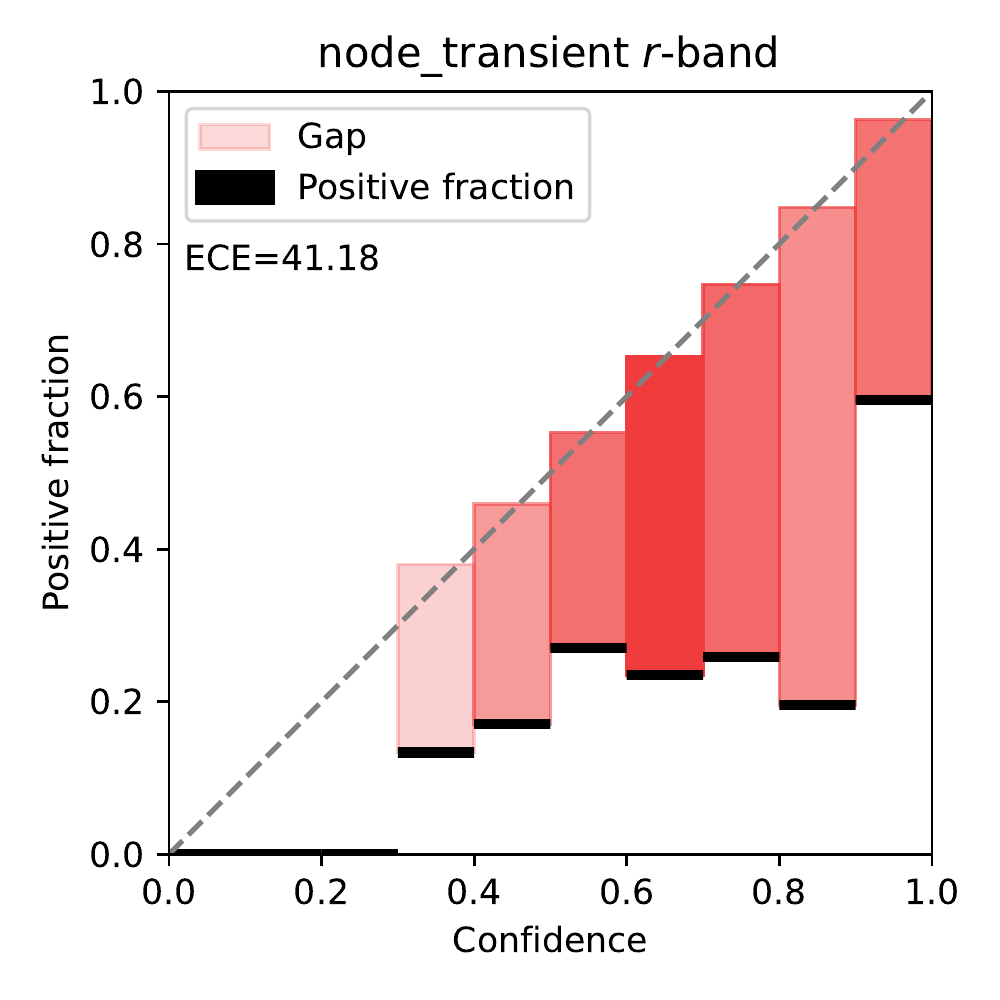} \\
\end{tabular}

\begin{tabular}{cc}
  \includegraphics[width=0.3\linewidth]{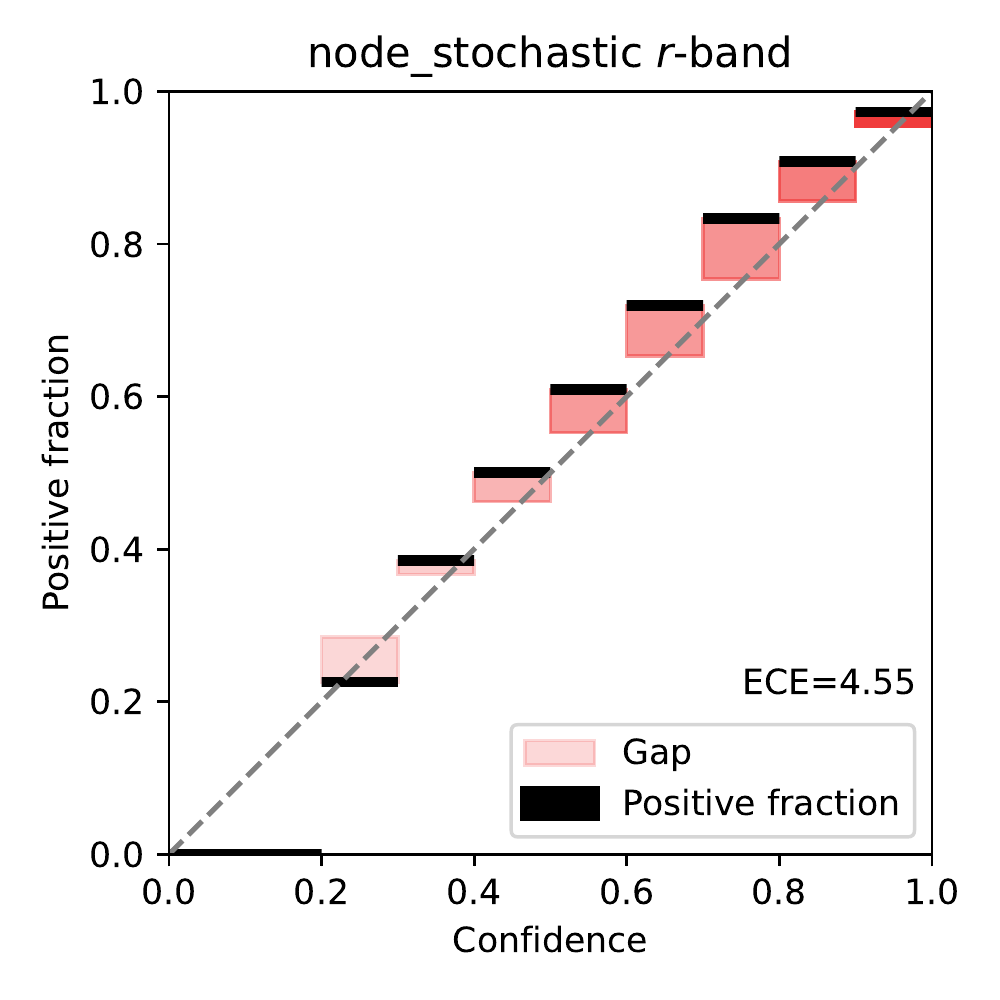} &

  \includegraphics[width=0.3\linewidth]{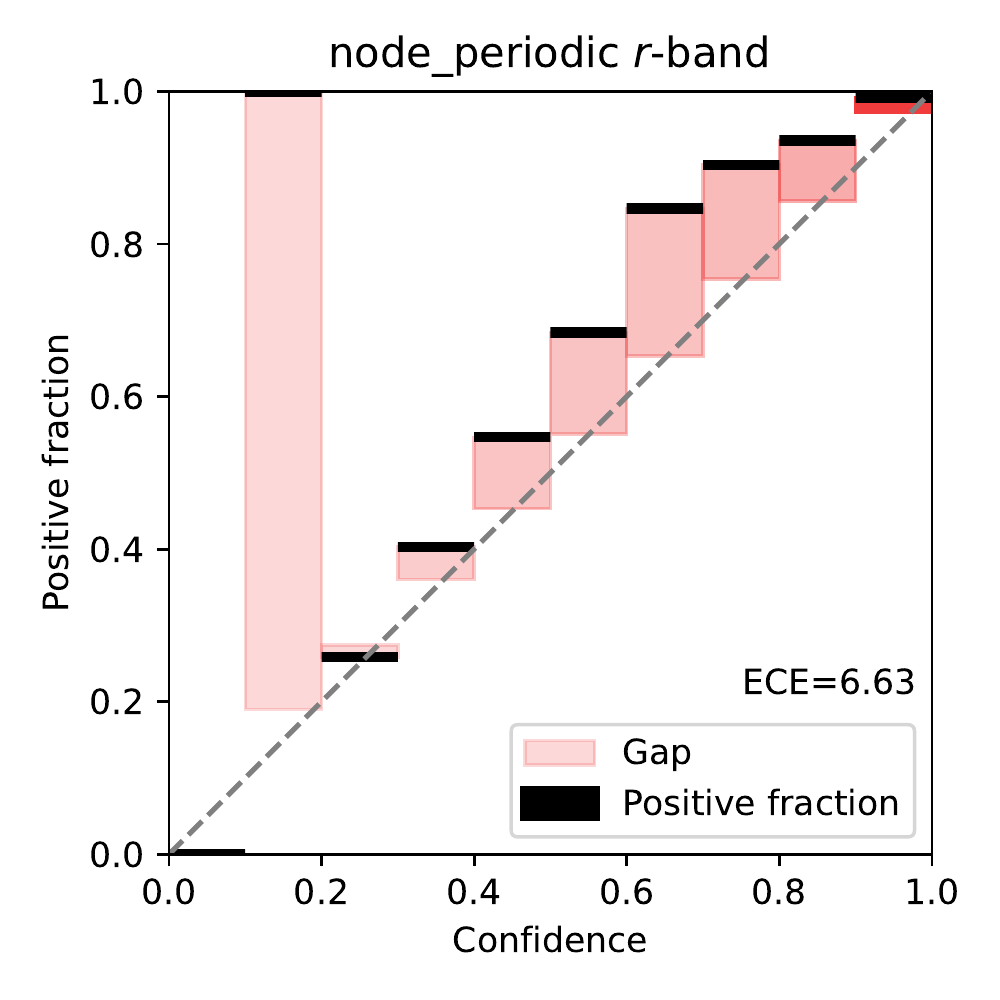} \\

\end{tabular}
\caption{As in Figure \ref{figure:reliability_diagram_gband}, but for the $r$-band HBRF classifier. 
\label{figure:reliability_diagram_rband}}
\end{center}
\end{figure*}

Figures\footnote{The code used to generate these plots corresponds to a modified version of the code presented in \url{https://github.com/hollance/reliability-diagrams}} \ref{figure:reliability_diagram_gband} and \ref{figure:reliability_diagram_rband} present the Reliability Diagrams and the ECE scores for each node of the HBRF, for the $g$-band and the $r$-band testing sets, respectively. For each bin, we plot the PF (in black) and the gap (in red) versus the confidence. From these figures, we can see that the node\_init, node\_stochastic, and node\_periodic of both bands are well-calibrated (except for one bin of the node\_periodic in the $r$-band, but this bin contains only a few samples). For the case of the node\_transient, the PF is smaller than the average probability, which implies that the probabilities overestimate the reliability of the classifications. These results are expected, though, as the quality of the ZTF DR light curves is not adequate for the study of transients. For the node\_variable, the PF is larger than the average probability in most of the probability bins, however, only a small fraction of the predictions have a probability smaller than 0.9, and thus the ECE score for the node\_variable is smaller than the score obtained for the node\_transient, node\_stochastic, and node\_periodic. From this we can conclude that when the predicted probabilities of the node\_variable are lower than 0.9, there is an underestimated reliability of the classifications. Since this is a hierarchical model, this result favors the quality of the classifications, because we can expect that the sources that go to each of the following nodes (node\_transient, node\_stochastic, and node\_periodic) are properly classified. We suggest to the users of our classifications that they take these results into account when filtering the candidates from each node by probability. 

\section{Results for the ZTF/4MOST sky}\label{section:results}

\begin{figure*}[htbp]
\begin{center}
\begin{tabular}{c}
   \includegraphics[width=1\linewidth]{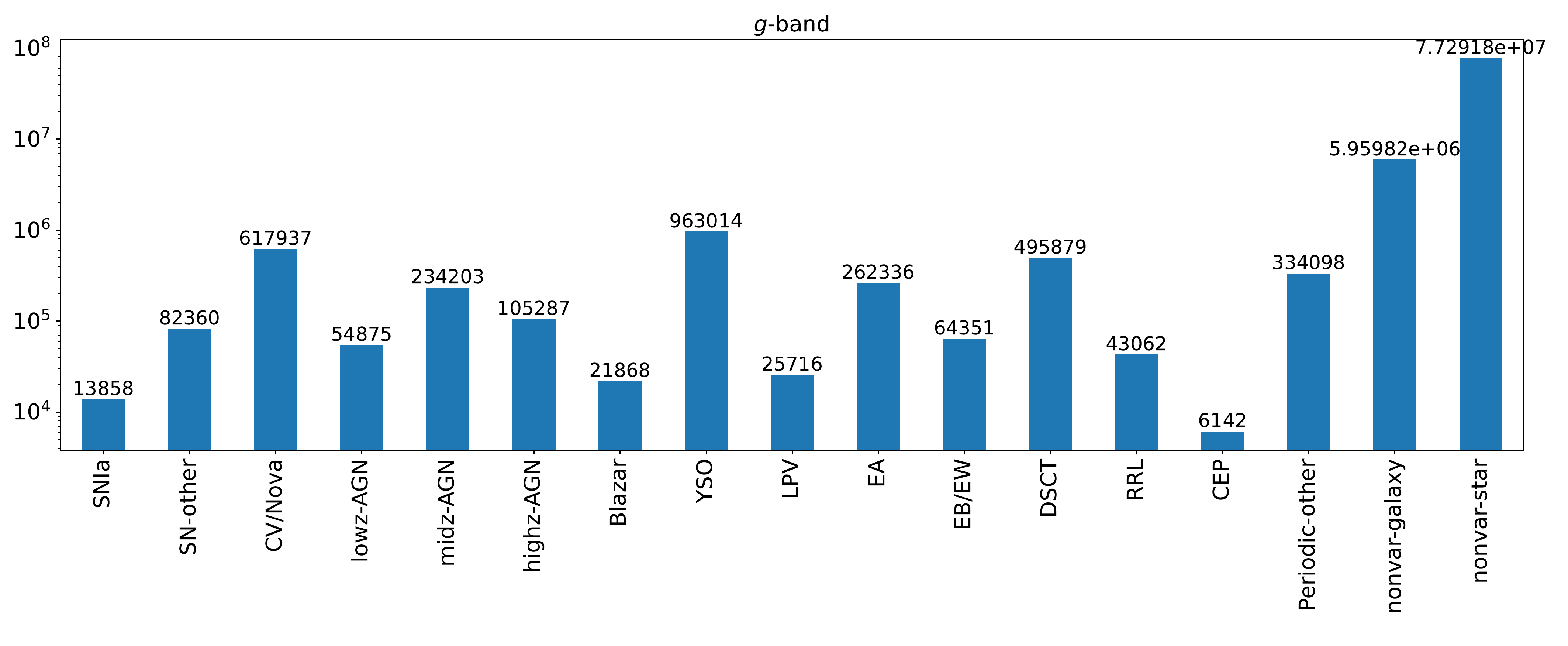} \\

  \includegraphics[width=1\linewidth]{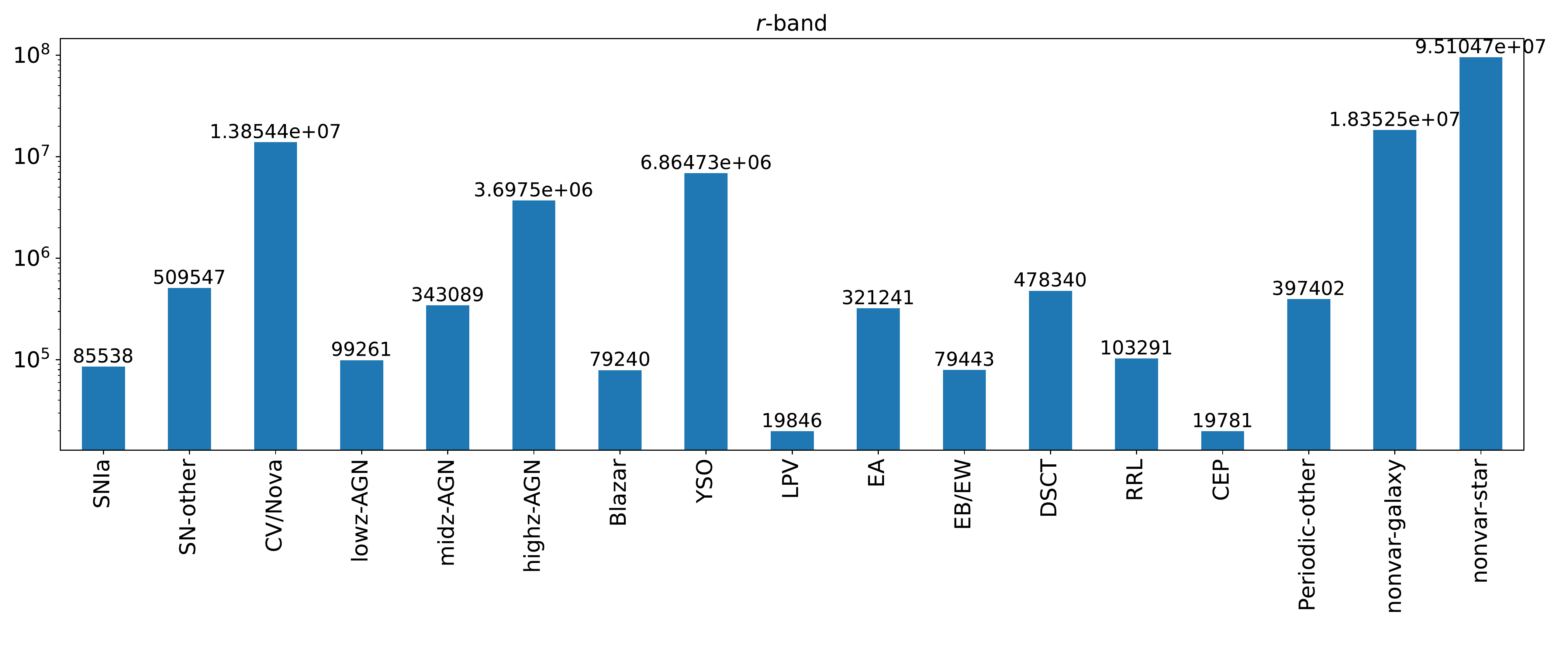} \\

\end{tabular}
\caption{Number of candidates per class for all the sources in the ZTF/4MOST sky with more than 20 epochs. The results for the $g$-band are shown in the top panel, and for the $r$-band in the bottom panel. The number of sources per band is specified on top of each bar.
\label{figure:unlabeled_classifications}}
\end{center}
\end{figure*}

\begin{figure*}[htbp]
\begin{center}
\begin{tabular}{c}
   \includegraphics[width=1\linewidth]{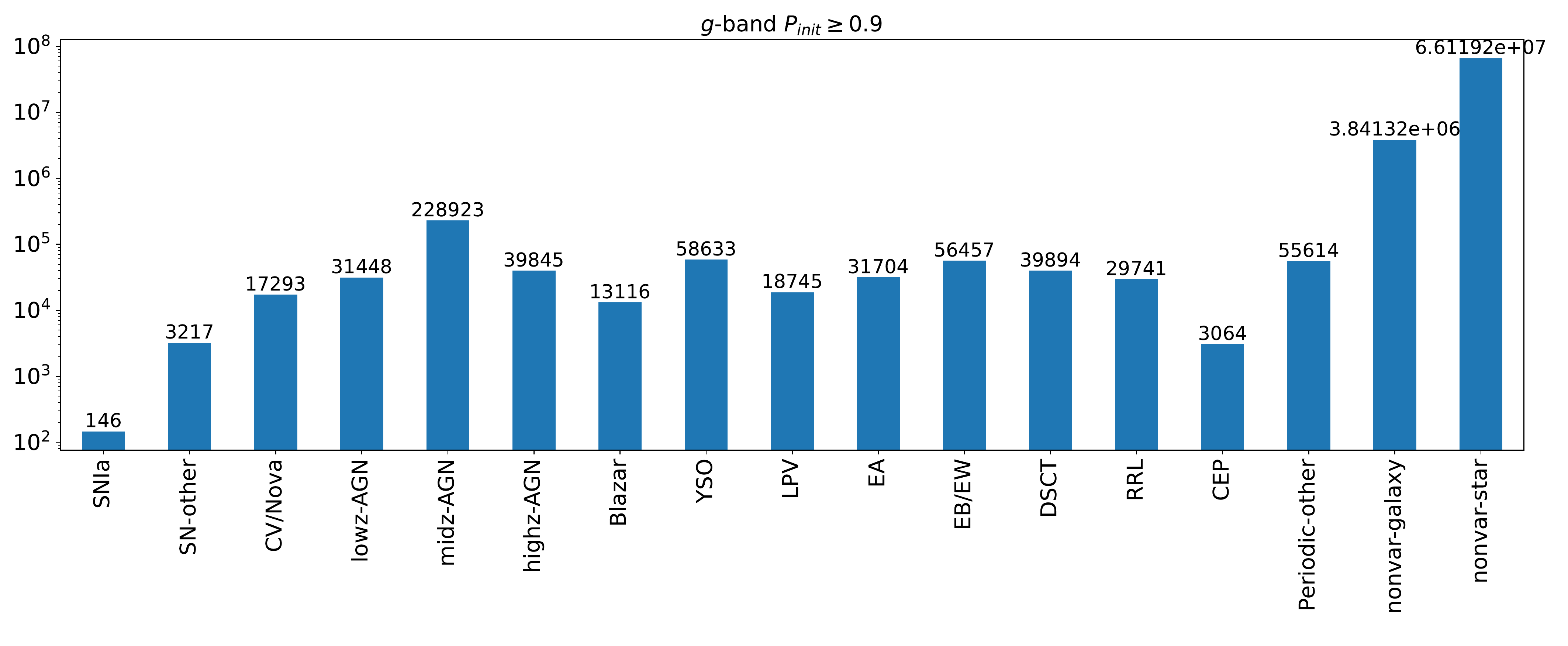} \\

  \includegraphics[width=1\linewidth]{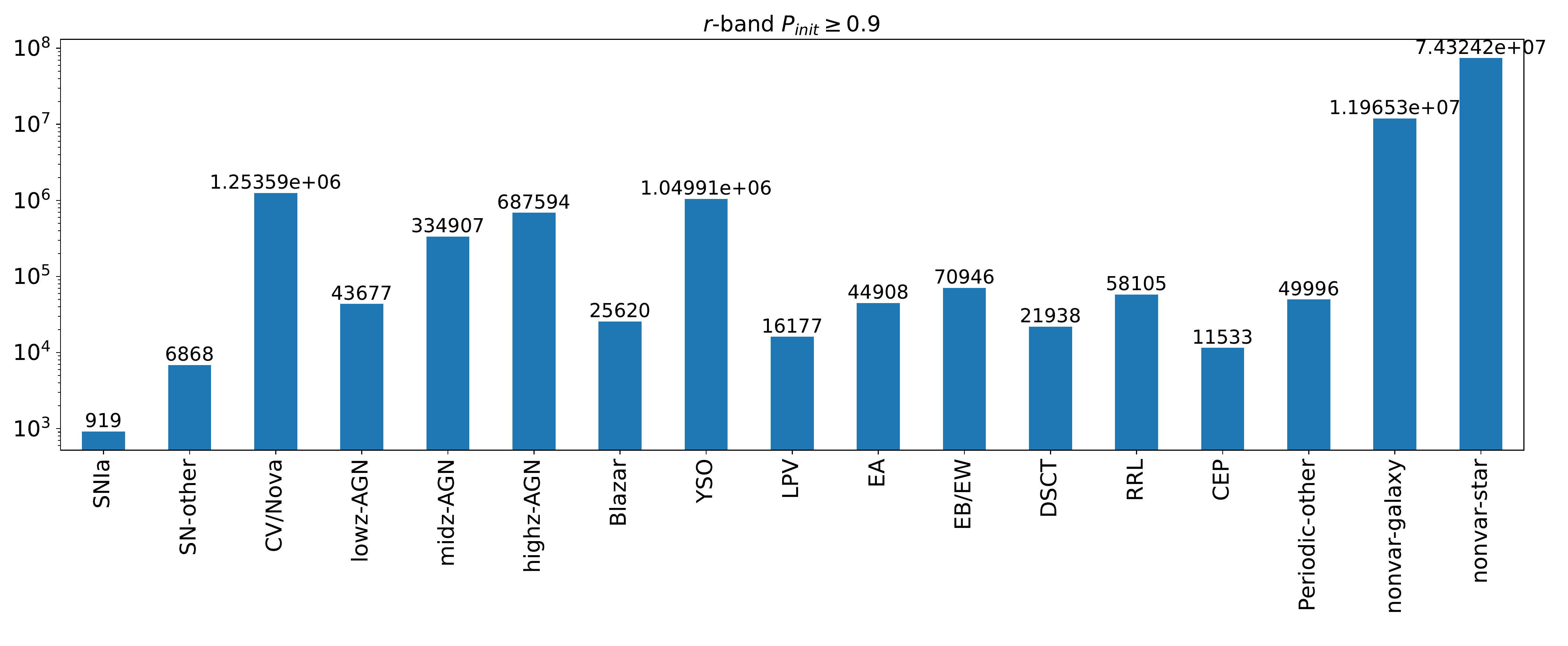} \\

\end{tabular}
\caption{Number of candidates per class for all the sources in the ZTF/4MOST sky with more than 20 epochs, and with a classification probability in the node\_init greater or equal than 0.9. The results for the $g$-band are shown in the top panel, and for the $r$-band in the bottom panel. The number of sources per band is specified on top of each bar.
\label{figure:unlabeled_classifications_hp}}
\end{center}
\end{figure*}

We applied the $g$-band and $r$-band HBRF models to the sources in the ZTF/4MOST sky described in Section \ref{section:features}, which includes 86,576,577 and 140,409,824 individual light curves in the $g$-band and the $r$-band, respectively. Figure \ref{figure:unlabeled_classifications} shows the number of candidates per class for these sources. The fraction of sources classified as transient or persistently variable is 3.84\% in the $g$-band, and 19.20\% in the $r$-band. This is a surprising difference and is not reflected at all in the results obtained for the testing set (which corresponds to 20\% of the LS), presented in Section \ref{section:performance}. Figure \ref{figure:unlabeled_classifications_hp} shows the number of sources per class, when only the sources with a classification probability in the node\_init equal to or greater than 0.9 are considered ($P_{init} \geq 0.9$). The fraction of sources classified as transient or persistently variable drops to 0.73\% and 2.62\% in the $g$ and $r$ bands, respectively. In this case, the differences between the results of the $g$ and $r$ bands have narrowed considerably, but still remain much higher when using the $r$-band light curves. From Figures \ref{figure:unlabeled_classifications} and \ref{figure:unlabeled_classifications_hp}, we can notice that the main difference in the results, when going from the $g$ to the $r$ band, occurs in the cases of the SN-other, CV/Nova, highz-AGN, and YSO classes.

\begin{figure*}[htbp]
\begin{center}
\begin{tabular}{cc}
   \includegraphics[scale=0.46]{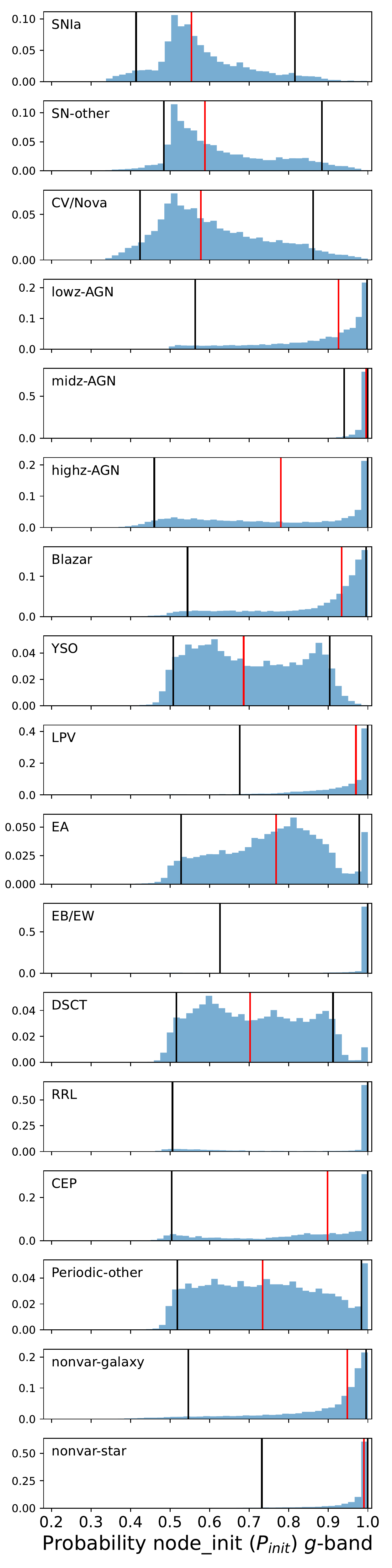} &

  \includegraphics[scale=0.46]{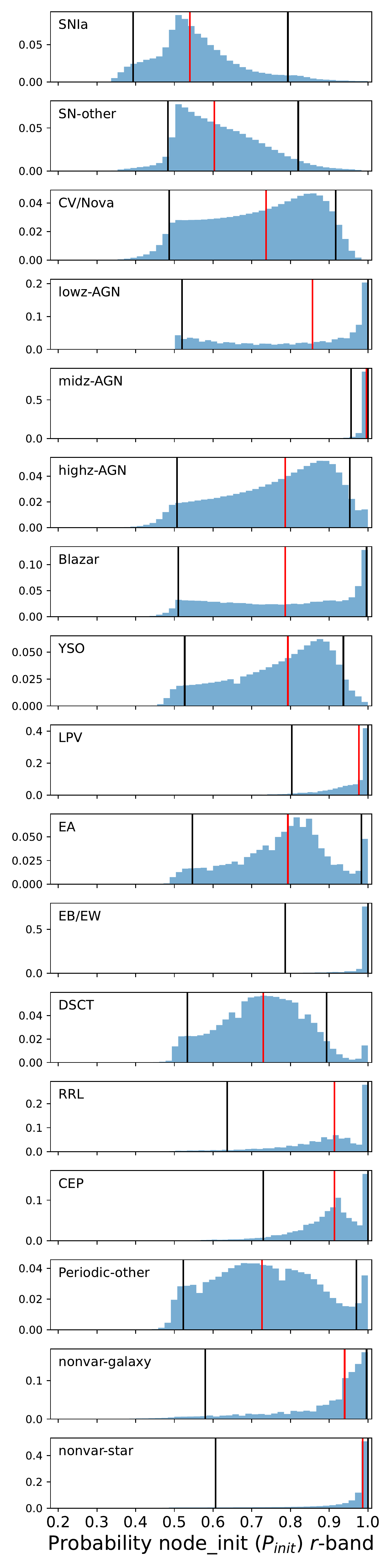} \\

\end{tabular}
\caption{Normalized probability distributions of the node\_init ($P_{init}$) for the 17 classes considered in this work. The results for the $g$-band are shown in the left panel, and for the $r$-band in the right panel. The red lines show the median probability for each class. The black lines show the 5th and 95th percentiles of the probabilities.
\label{figure:prob_hists_init}}
\end{center}
\end{figure*}

\begin{figure*}[htbp]
\begin{center}
\begin{tabular}{cc}
   \includegraphics[scale=0.455]{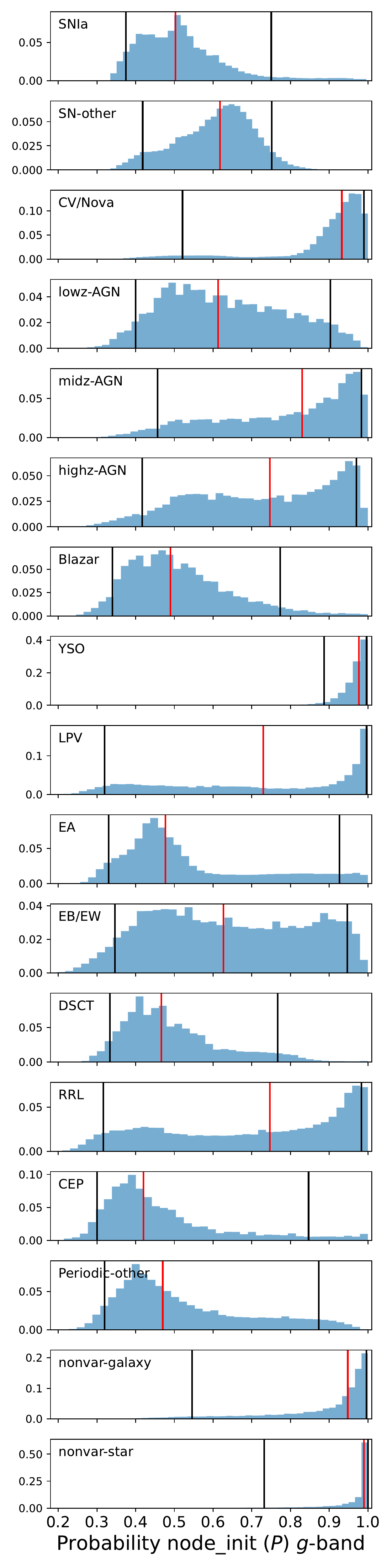} &

  \includegraphics[scale=0.455]{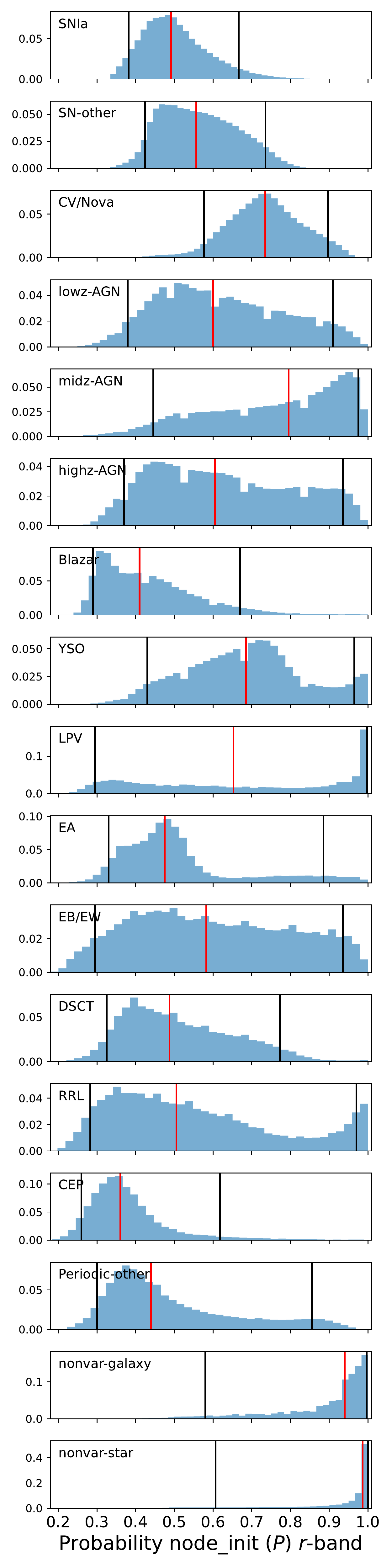} \\

\end{tabular}
\caption{As in Figure \ref{figure:prob_hists_init}, but for the final leaves (final prediction) of the model ($P$).
\label{figure:prob_hists_final}}
\end{center}
\end{figure*}

In Figures \ref{figure:prob_hists_init} and \ref{figure:prob_hists_final}, we show the probability distributions obtained for all the sources in the ZTF/4MOST sky, in the $g$ and $r$ bands. Figure \ref{figure:prob_hists_init} shows the probability obtained by the node\_init, $P_{init}$, that separates the sources into nonvar-star, nonvar-galaxy, and variable, while Figure \ref{figure:prob_hists_final} shows the final prediction of the model, $P$, which corresponds to the final leaves of each node (i.e., the probability of the node\_init for nonvar-star and nonvar-galaxy, and the probability of the node\_transient, node\_stochastic or node\_periodic, for sources classified as variable in the node\_init). From these figures, it can be noticed that in the node\_init the classes with the lowest probabilities are the transients, and the ones with the highest probabilities are the non-variable and AGN classes. We can also notice that the distribution of $P_{init}$ for most classes in both bands presents a bimodal distribution, with several sources having $P_{init}\sim 1$, and others having lower values. The final predicted probabilities $P$ vary strongly between classes, as was also observed in \citetalias{Sanchez-Saez21a}, and is explained by the confusion between some similar classes. The classes with the highest $P$ in both bands are CV/Nova, midz-AGN, LPV, and the non-variable classes.

\begin{figure*}[]
\begin{center}
   \includegraphics[width=0.95\linewidth]{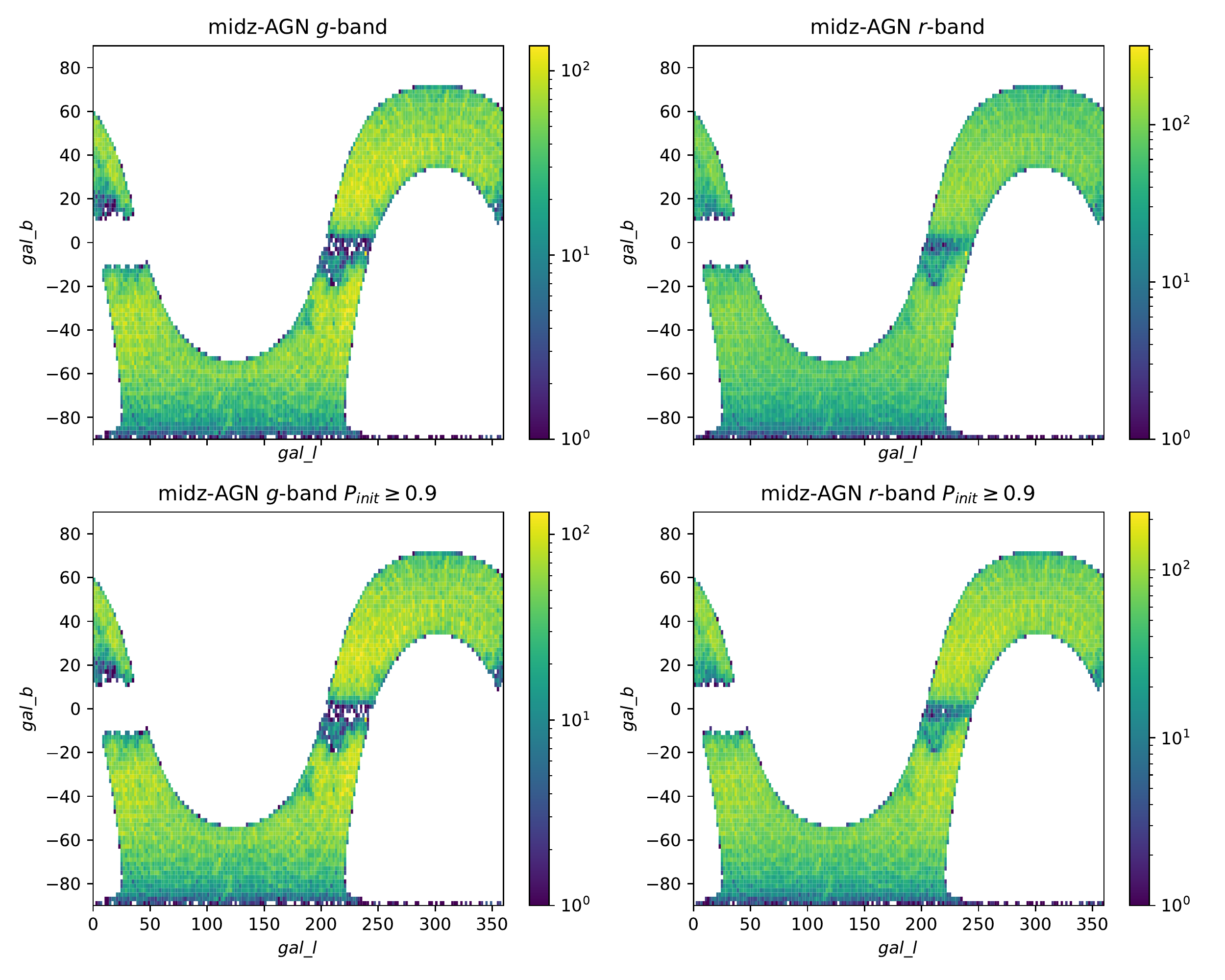} 

\caption{Number of targets (see color bars) per 4 deg$^2$ for the midz-AGN class, in Galactic  coordinates. The upper panels show the densities for all the selected candidates in the $g$-band (upper left) and the $r$-band (upper right). The bottom panels show the densities for candidates with large probability in the node\_init ($P_{init}\geq0.9$) for the $g$-band (bottom left) and the $r$-band (bottom right). Note that the range of the number of targets covered by each color bar can differ in each panel.
\label{figure:density_midz-AGN}}
\end{center}
\end{figure*}

\begin{figure*}[]
\begin{center}
   \includegraphics[width=0.95\linewidth]{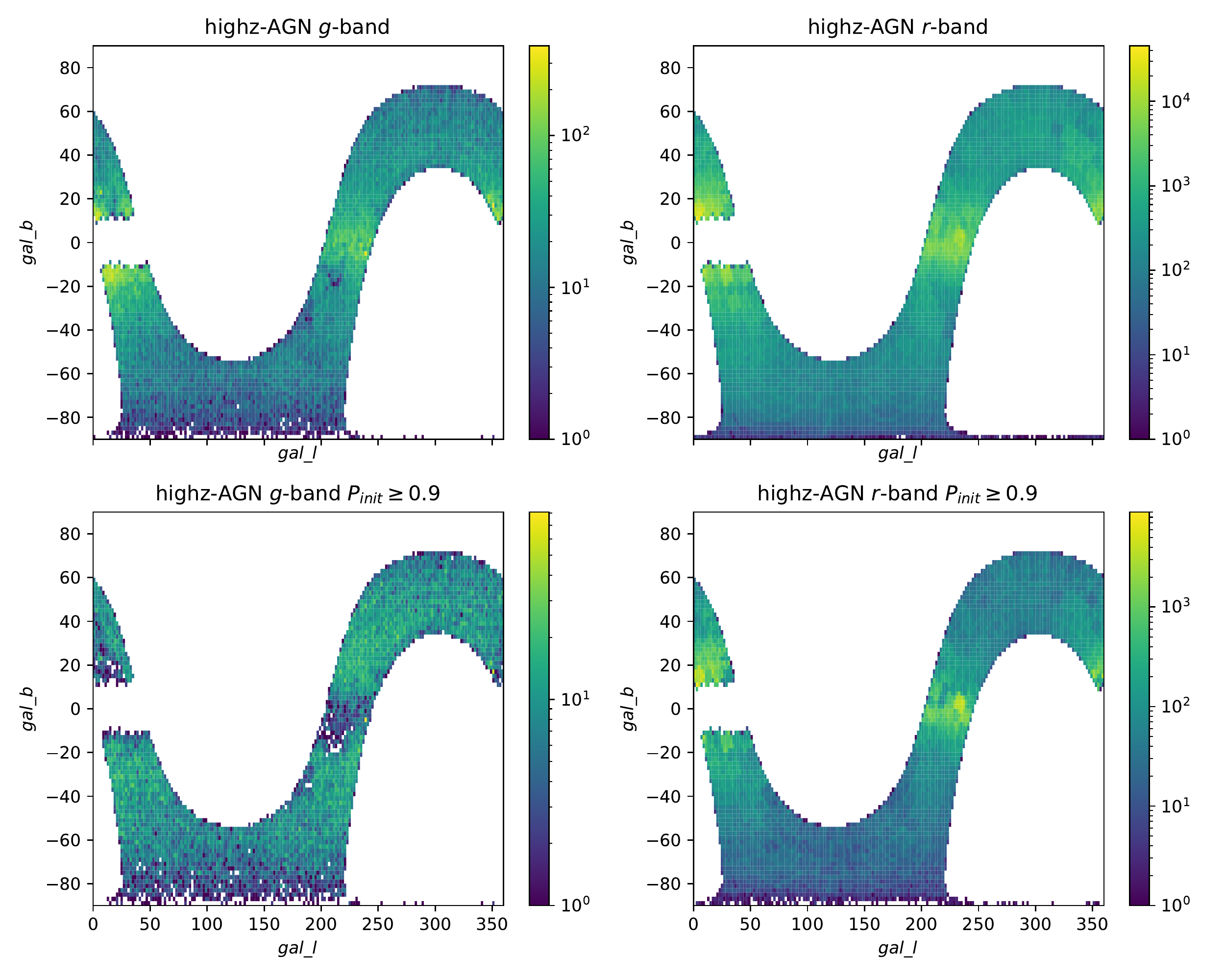} 

\caption{As in Figure \ref{figure:density_midz-AGN} but for the highz-AGN class.
\label{figure:density_highz-AGN}}
\end{center}
\end{figure*}

The coordinates of the objects are not included as features in the HBRF model, and thus we use them to inspect the quality of the classifier by exploring the sky densities of each class in the Galactic  coordinate space. Figures \ref{figure:density_midz-AGN} and \ref{figure:density_highz-AGN} show the sky density of midz-AGN and highz-AGN candidates, respectively, in Galactic  coordinates. The density maps of the remaining 15 classes are shown in Appendix \ref{app:sky_densities}. From these maps, we can notice that for some extragalactic classes, like highz-AGN (and SNIa, SN-other, or Blazar), there is a high overdensity of candidates in the Galactic plane ($gal\_b\sim0$), while for others, like midz-AGN (and lowz-AGN) the densities drop at lower Galactic latitudes. The overdensities of extragalactic classes in the Galactic plane are unexpected, as we should see a decrease in the number of sources belonging to these classes in this region due to Galactic extinction. On the other hand, the enormously larger numbers of stars in the Galactic plane provide a natural source of false positives in this region. In the case of the $g$-band, these overdensities disappear when only sources with a high probability in the node\_init ($P_{init}\geq0.9$) are selected. For classes that are expected to be more prominent in the Galactic plane, like YSO, CEP, and LPV, we see a much lower density at larger Galactic latitudes. For all classes, we obtain more reliable results in the $g$-band, when $P_{init}\geq0.9$. We, therefore, recommend giving priority to candidates selected from the $g$-band light curves, when using ZTF DR light curves to identify variable and transient objects.

\section{Discussion}\label{section:discussion}

\subsection{Differences in the number of variables and transient candidates selected using the $g$ and $r$ ZTF bands}\label{section:num_band_comp}

For some classes, like SNIa, SN-other, CV/Nova, highz-AGN, and YSO, there is a clear difference in the number of persistent variables and transients selected using the $g$ and $r$ ZTF bands. Moreover, as previously mentioned, some of the extragalactic classes (SNIa, SN-other, highz-AGN, and Blazar) show large overdensities in the Galactic plane.

Our hypothesis is that these results can be either explained by the very intense observing seasons carried out in the $r$-band by the ZTF project, as can be seen in the ZTF DR11 documentation\footnote{\url{https://irsa.ipac.caltech.edu/data/ZTF/docs/releases/dr11/src/TspanVsDT\_fid2.png}}, or by source crowding.  

In order to disentangle what is producing these differences in the classification of the $g$-band and $r$-band light curves, we first visually inspected the science images of the ZTF $r$-band in the zones with a high density of extragalactic candidates (around $gal\_l\sim220$ and $gal\_b\sim0$), but we did not find evidence of large numbers of source blends, or other photometric issues (like ghosts). To perform a more quantitative analysis, we selected the ZTF field 360, which is centered at $gal\_l\sim225$ and $gal\_b\sim2.6$, and measured the number of sources with unique ZTF ID in the $r$-band that have a neighboring source within a radius of 3$''$. Of the 3,356,426 sources in this field in the $r$-band, only 61,314 (2\%) have a neighboring source within this radius. This corresponds to a low number of sources with close neighbors, and thus source crowding cannot explain the high density of extragalactic candidates in the galactic plane when using the $r$-band. 

\begin{figure*}[htbp]
\begin{center}
\begin{tabular}{c}
   \includegraphics[scale=0.62]{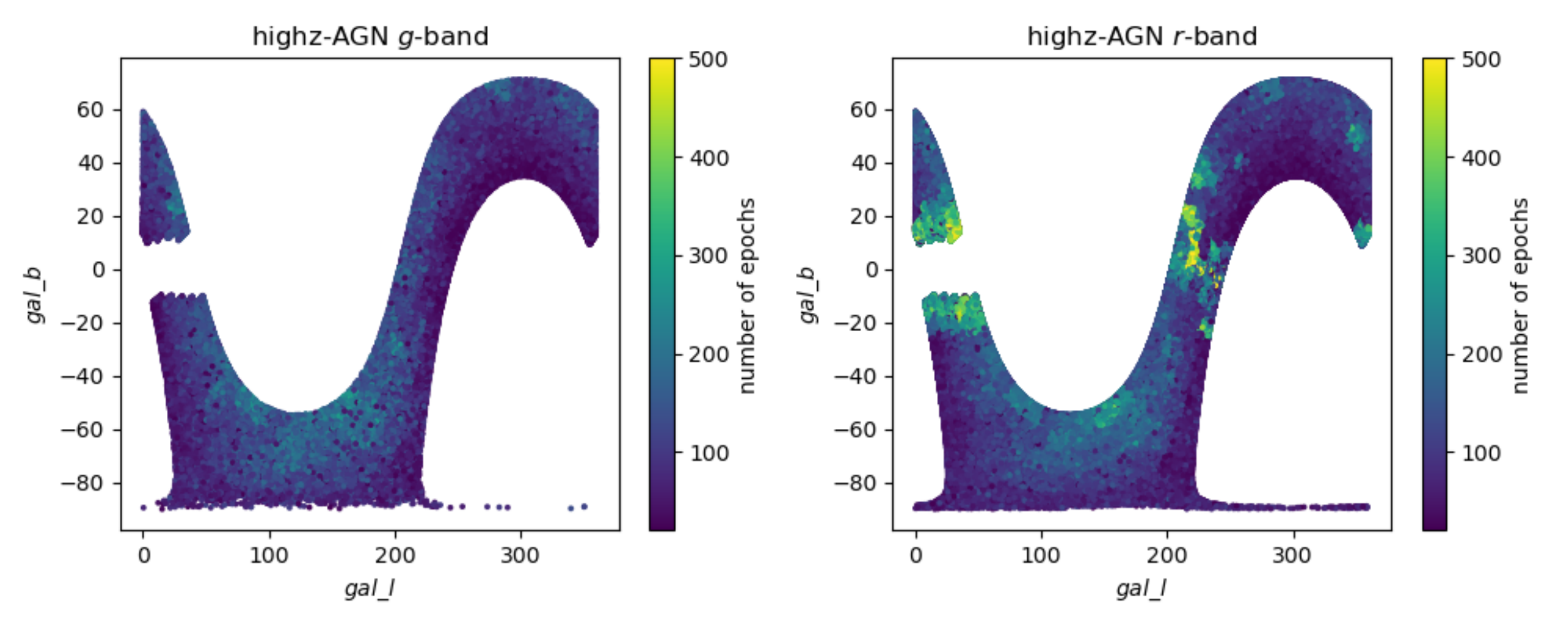} \\

  \includegraphics[scale=0.62]{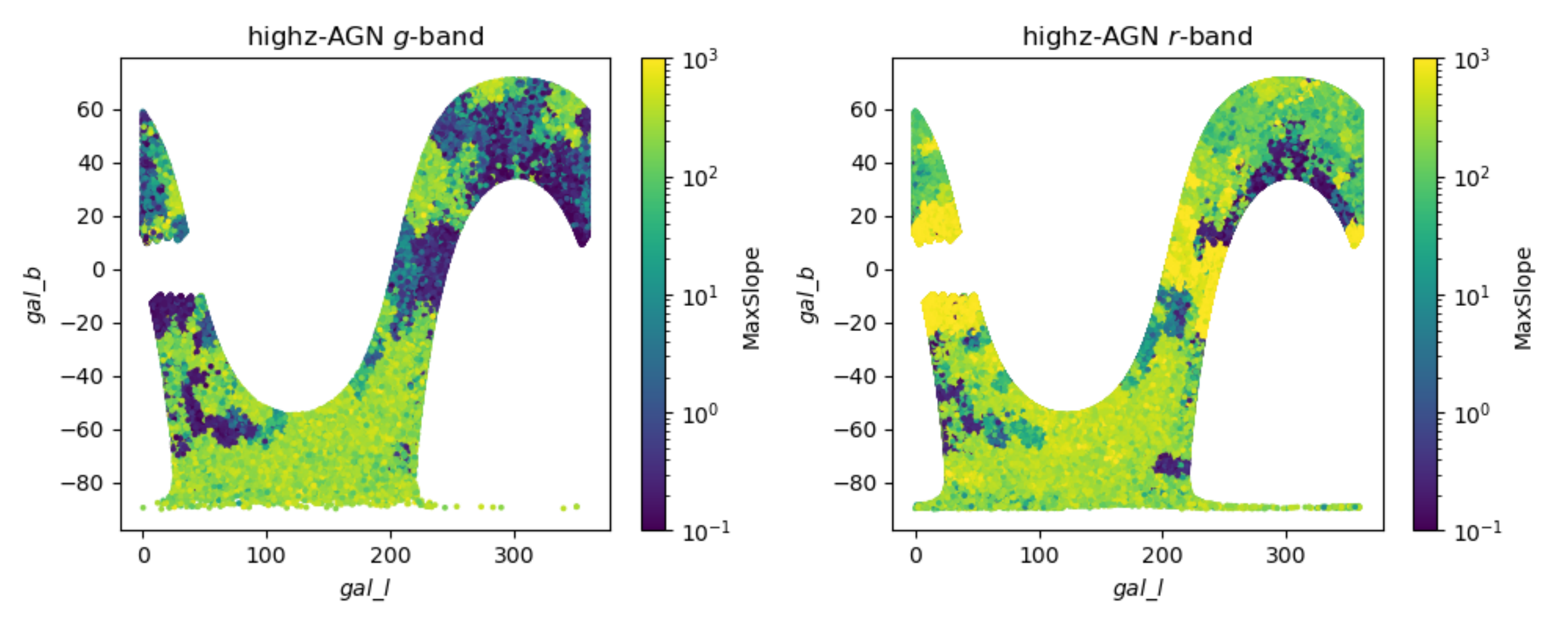} \\

\end{tabular}
\caption{Distribution of light curve properties as a function of the Galactic coordinates for highz-AGN candidates. The top panels show the number of epochs, and the bottom panels the \texttt{MaxSlope}, for the $g$ (left) and $r$ (right) bands.
\label{figure:highz-agn-feats}}
\end{center}
\end{figure*}

\begin{figure*}[htbp]
\begin{center}
\begin{tabular}{cc}
   \includegraphics[scale=0.48]{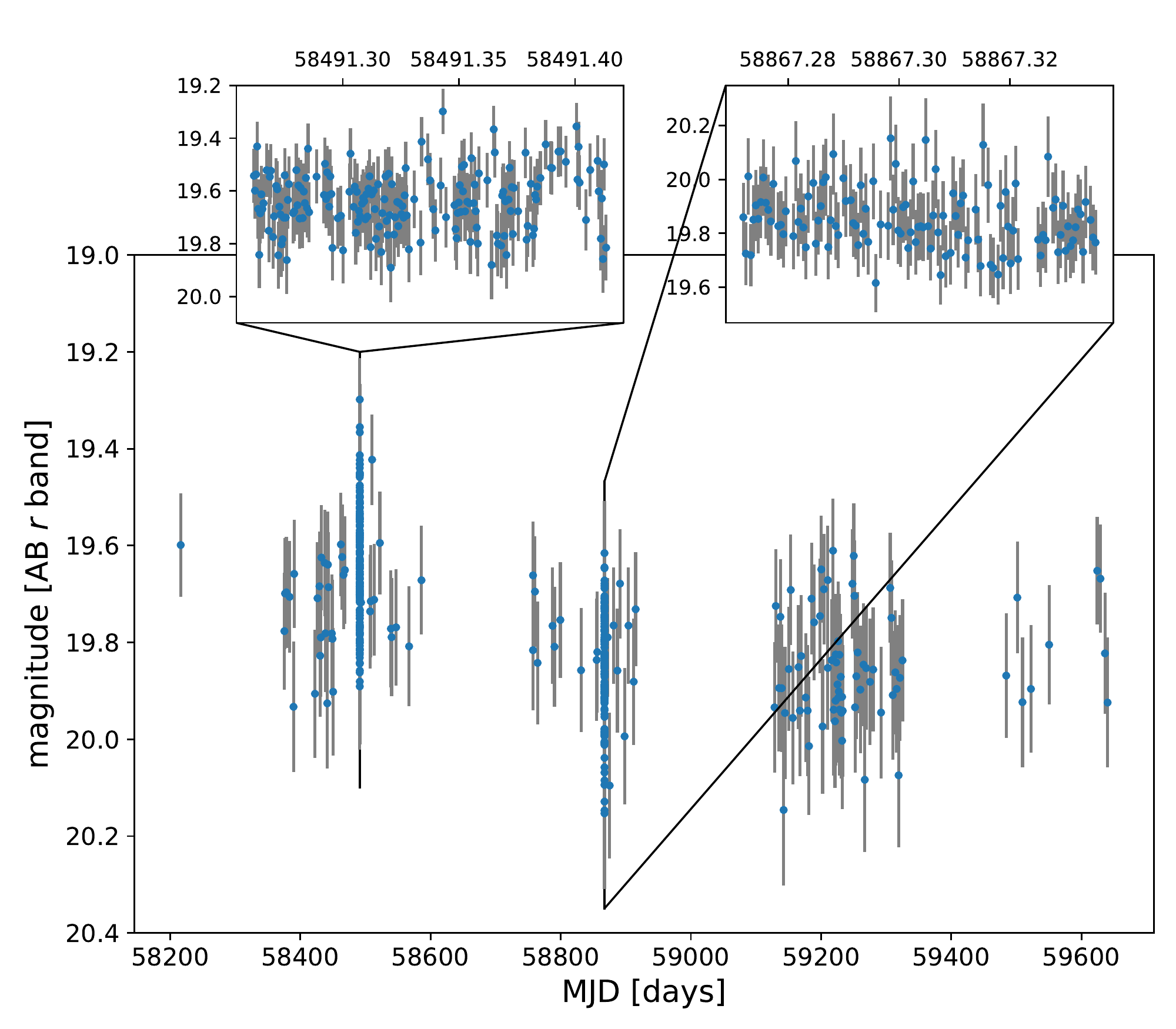} &

  \includegraphics[scale=0.62]{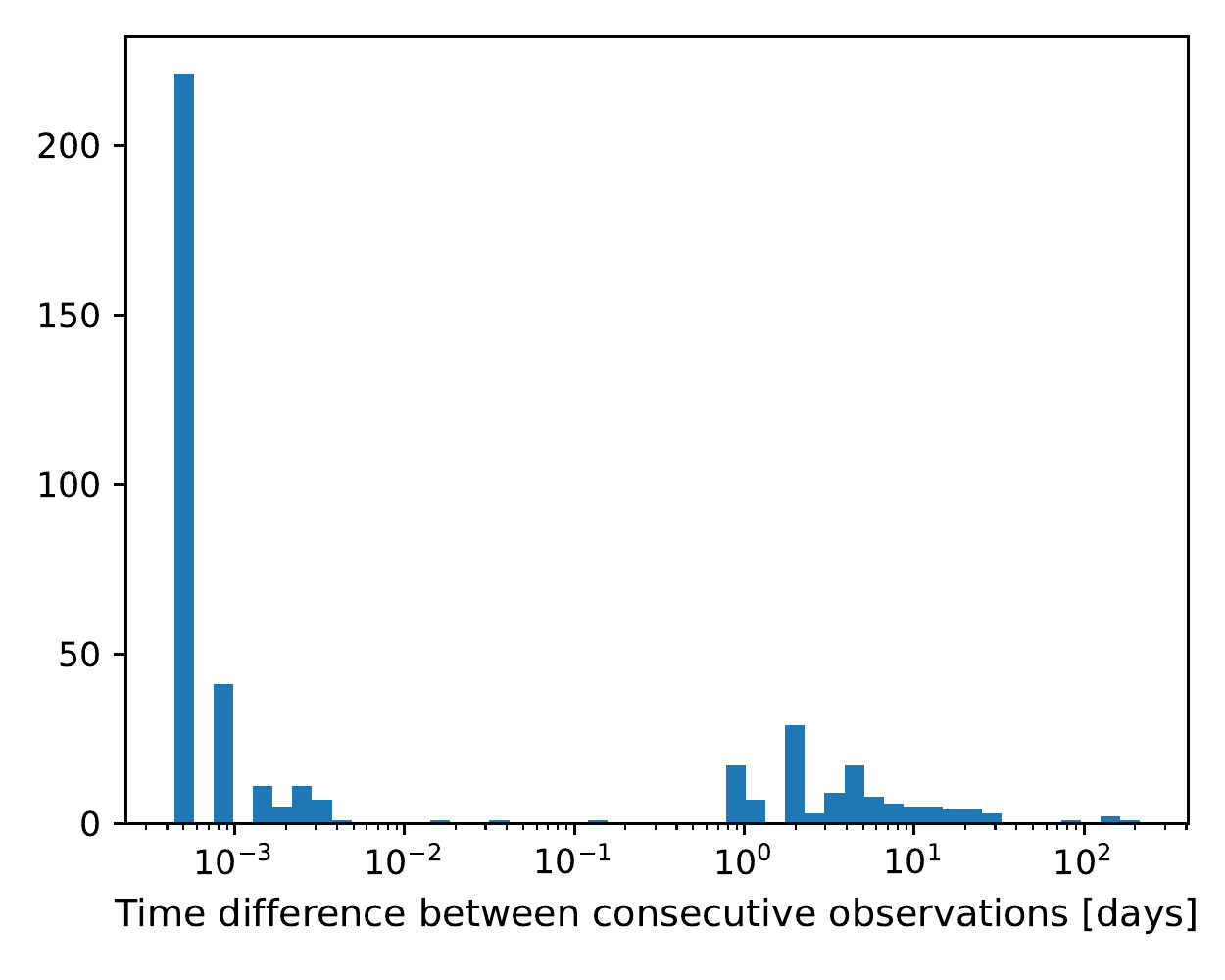} \\

\end{tabular}
\caption{Highz-AGN candidate with ZTF DR11 ID 360206100033262. The left panel shows its $r$-band light curve. Zoom-in at MJD$=58491$ and MJD$=58867$ are shown. The right panel shows the histogram of the time difference between consecutive observations for the $r$-band light curve.
\label{figure:highz-agn-example}}
\end{center}
\end{figure*}

We then inspected whether the regions showing the largest number of extragalactic candidates are consistent with regions of very rapid cadence in the ZTF light curves. Figure \ref{figure:highz-agn-feats} shows the number of epochs and the \texttt{MaxSlope} (maximum absolute magnitude slope between two consecutive observations) versus the Galactic  coordinates for highz-AGN candidates. We can see that the sky regions with a higher number of candidates coincide with regions with a larger number of observations and extreme values of \texttt{MaxSlope}. When \texttt{MaxSlope} is large, it could mean that there are two data points very close in time, but with different magnitude values. From the figure, we can see that around the Galactic plane, there is a high concentration of sources with more than 400 epochs and that \texttt{MaxSlope} has extreme values in this region. A similar behavior is observed in the $r$-band for the other extragalactic classes (SNIa, SN-other, lowz-AGN, midz-AGN, and Blazar). These large values of \texttt{MaxSlope} can be produced by the very rapid cadence seasons in the $r$-band, where in some cases the separation between two consecutive observations can even be less than a minute. This can be seen in Figure \ref{figure:highz-agn-example}, which shows the light curve in the $r$-band of a highz-AGN candidate with ZTF ID 360206100033262, located in the field 360, and the histogram of the time difference between consecutive observations for its light curve. This light curve has a total of 445 observations. During the day with MJD$=58491$, a total of 189 observations were obtained (42\% of the total observations), and the day with MJD$=58867$, there were 124 observations (28\%). In the histogram of the time difference between consecutive observations, we can notice that most of the epochs are separated by a few minutes. 

In order to test the effect of the rapid cadence in some ZTF fields, we recomputed the features for the field 360 in the $r$-band, but keeping only one observation per night (the first observation), and we use the model trained for the $r$-band (without any modifications) to classify the new set of features. There were 2,107,241 light curves with more than 20 epochs (after filtering), and of these, 12,961 were classified as one of the AGN classes (lowz-, midz-, highz-AGN, or Blazar), with 1024 (8\%) of them having $P_{init}\geq0.9$. In the same field, before the modification of the features, there were 54,981 sources classified as AGN (with 21\% having $P_{init}\geq0.9$), and with the modification of the light curves, only 5\% of them are classified as AGN.

From these results, we recommend to give priority to the classifications obtained in the $g$-band. When using the $r$-band classifications, we suggest to filter the catalogs by the classification probability in the node\_init ($P_{init}\geq0.9$), and/or to avoid the use of our model in regions with rapid cadence observations (which normally have more than 400 epochs in the $r$-band light curves, according to the ZTF DR11 documentation). We also suggest taking into account the probability distribution shown in Figures \ref{figure:prob_hists_init} and \ref{figure:prob_hists_final}, and the reliability diagrams presented in Figures \ref{figure:reliability_diagram_gband} and \ref{figure:reliability_diagram_rband} when filtering our classifications by probability.

\subsection{AGN candidates}\label{section:agns}

As mentioned above, the main motivation for the development of this classifier was to identify AGN and Blazar candidates at different redshifts that will be followed up by the ChANGES survey. There are 416,233 ($g$-band) and 4,219,094 ($r$-band) AGN and Blazar candidates in total. When only sources with $P_{init}\geq0.9$ are considered, we have 313,332 candidates in the $g$-band and 1,091,798 in the $r$-band. This difference in the number of candidates can be explained by the larger number of light curves available in the $r$-band, but also by the issues described in Section \ref{section:num_band_comp}; there is a large overdensity in the Galactic plane of highz-AGN candidates detected in the $r$-band.  

Since a source can have more than one light curve per band in the ZTF DRs, we did an internal match of the candidates in each band, using a radius of 1.54$''$, and kept only one light curve per source per band. After this, we ended up with 384,242 candidates in the $g$-band, with 287,156 having $P_{init}\geq0.9$, and 4,048,299 candidates in the $r$-band, with 1,020,327 having $P_{init}\geq0.9$. We cross-matched these two samples using a radius of 1.5$''$, and found that there are 356,631 candidates that are classified as AGN or Blazar in both bands. This corresponds to 92.8\% of the $g$-band candidates and only 8.8\% of the $r$-band candidates. 

To have an idea of the quality of the AGN selection, we cross-matched the set of AGN candidates in both bands, and the AGN sample from the LS, with the \textit{Gaia} DR3 catalog of proper motions \citep{GaiaCollaboration22}. From this, for each sample we measured the signal-to-noise of the proper motion (PMsig), by dividing the proper motion (PM) of each source by its error in the following way: $\text{PMsig}=\text{pm}/\sqrt{\text{e\_pmRA}^2+\text{e\_pmDE}^2}$, where e\_pmRA and e\_pmDE are the PM errors in RA and DEC, respectively. We would expect to measure low PMsig values for AGNs, since they are distant sources, but a fraction of them will present large PMsig values when they look like extended sources in the \textit{Gaia} DR3 images. We noted that 84\% of the AGNs in the LS have $\text{PMsig}\leq3$, the same fraction is observed for the candidates selected in the $g$-band, while only 39\% of the candidates selected from the $r$-band have $\text{PMsig}\leq3$. Visual inspection indicates that the vast majority of the significant PM sources appear to have a stellar origin, although a slim minority seem to be legitimate AGN located in the centers of resolved, host-dominated nearby galaxies. 

\begin{figure*}[htbp!]
\begin{center}
\begin{tabular}{ccc}
\includegraphics[scale=0.38]{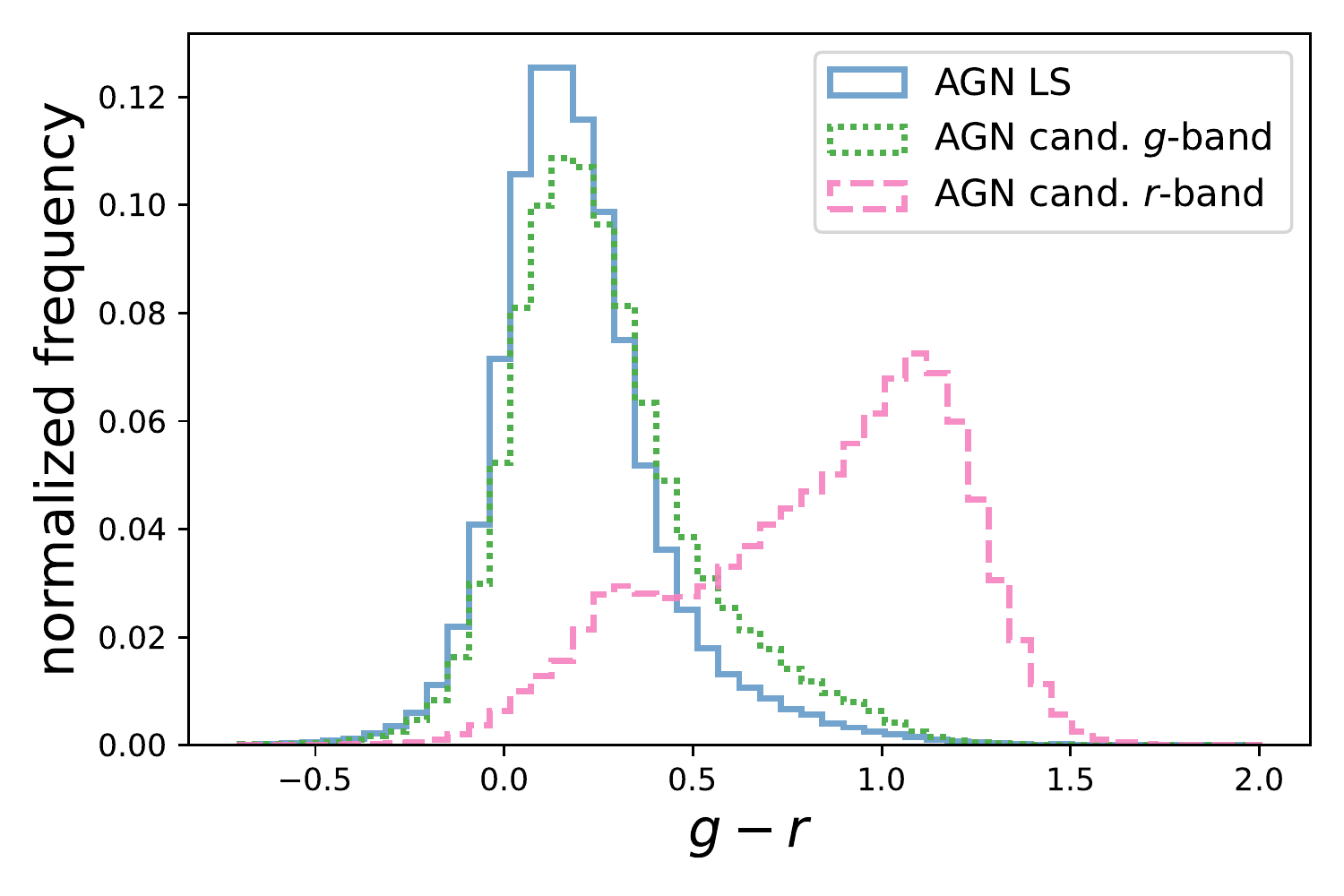} &
\includegraphics[scale=0.38]{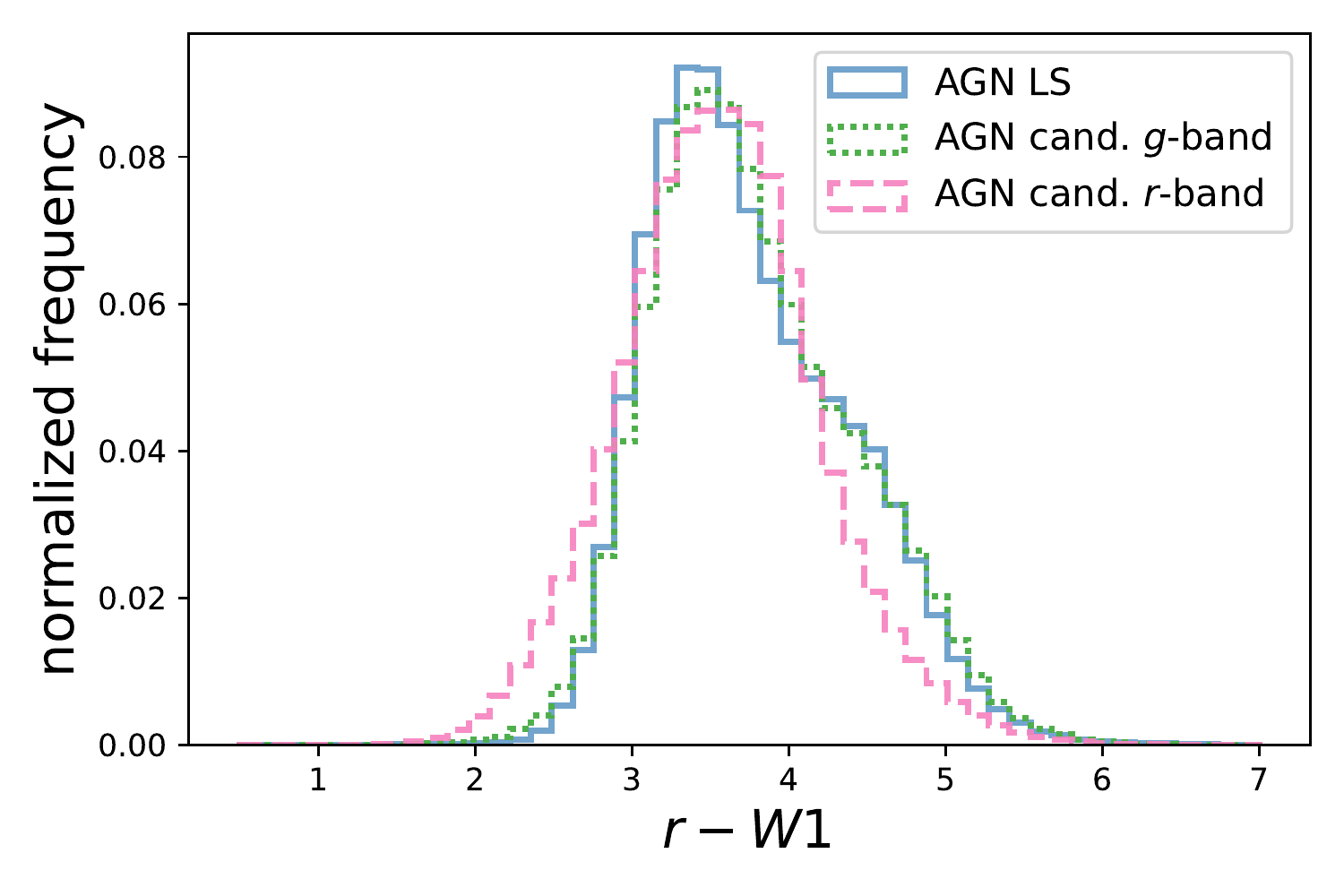} &
  \includegraphics[scale=0.38]{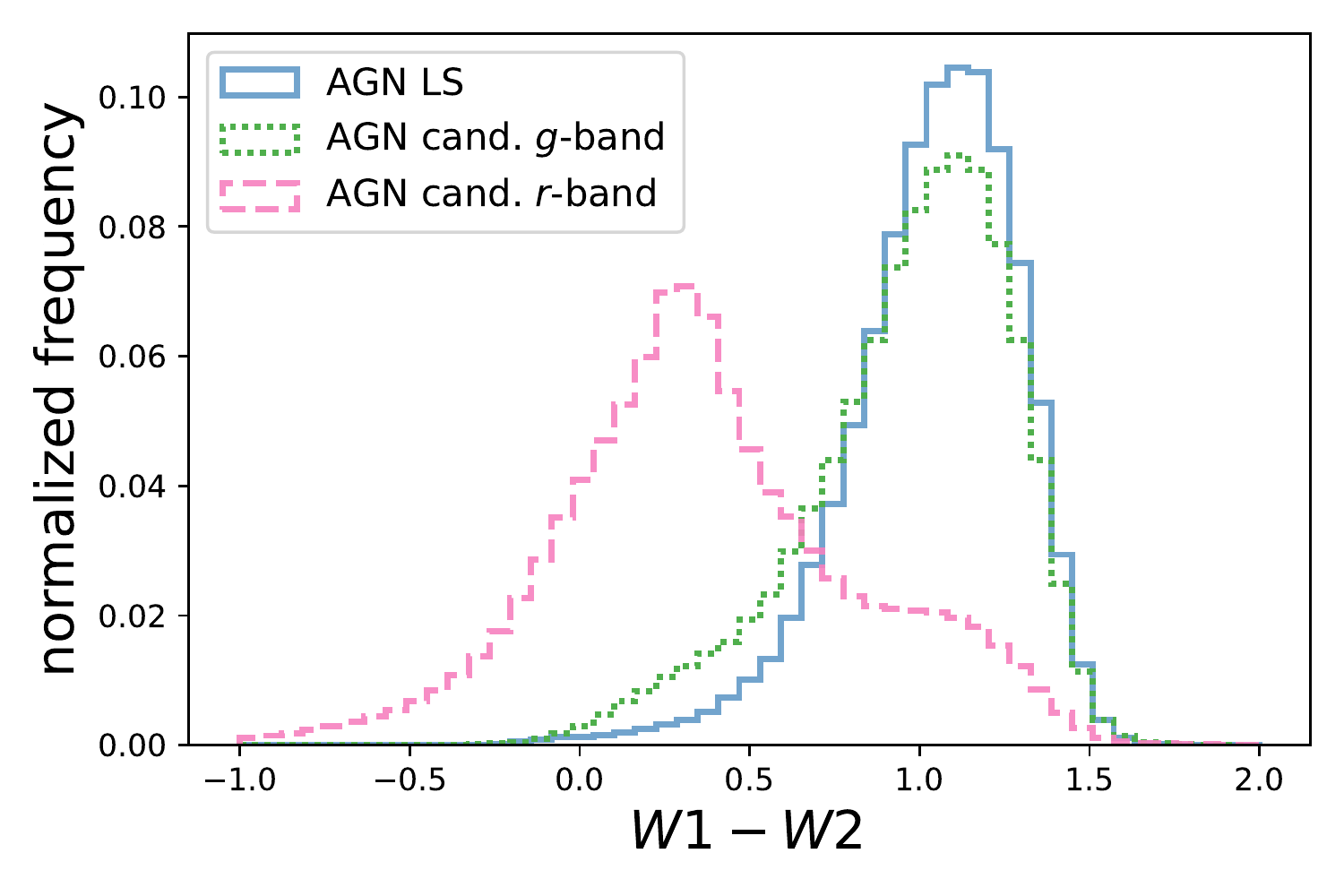} \\
\includegraphics[scale=0.38]{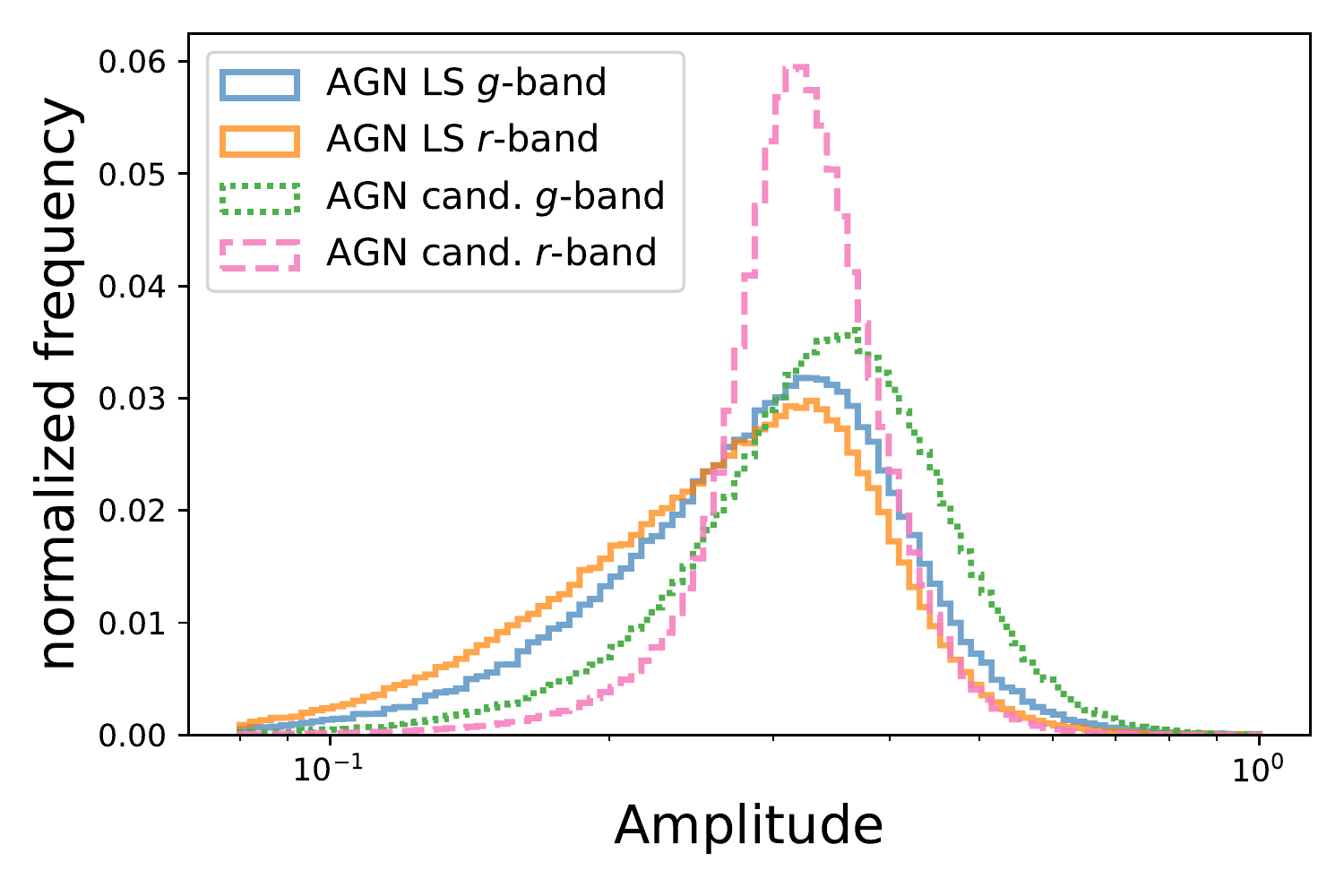} &
\includegraphics[scale=0.38]{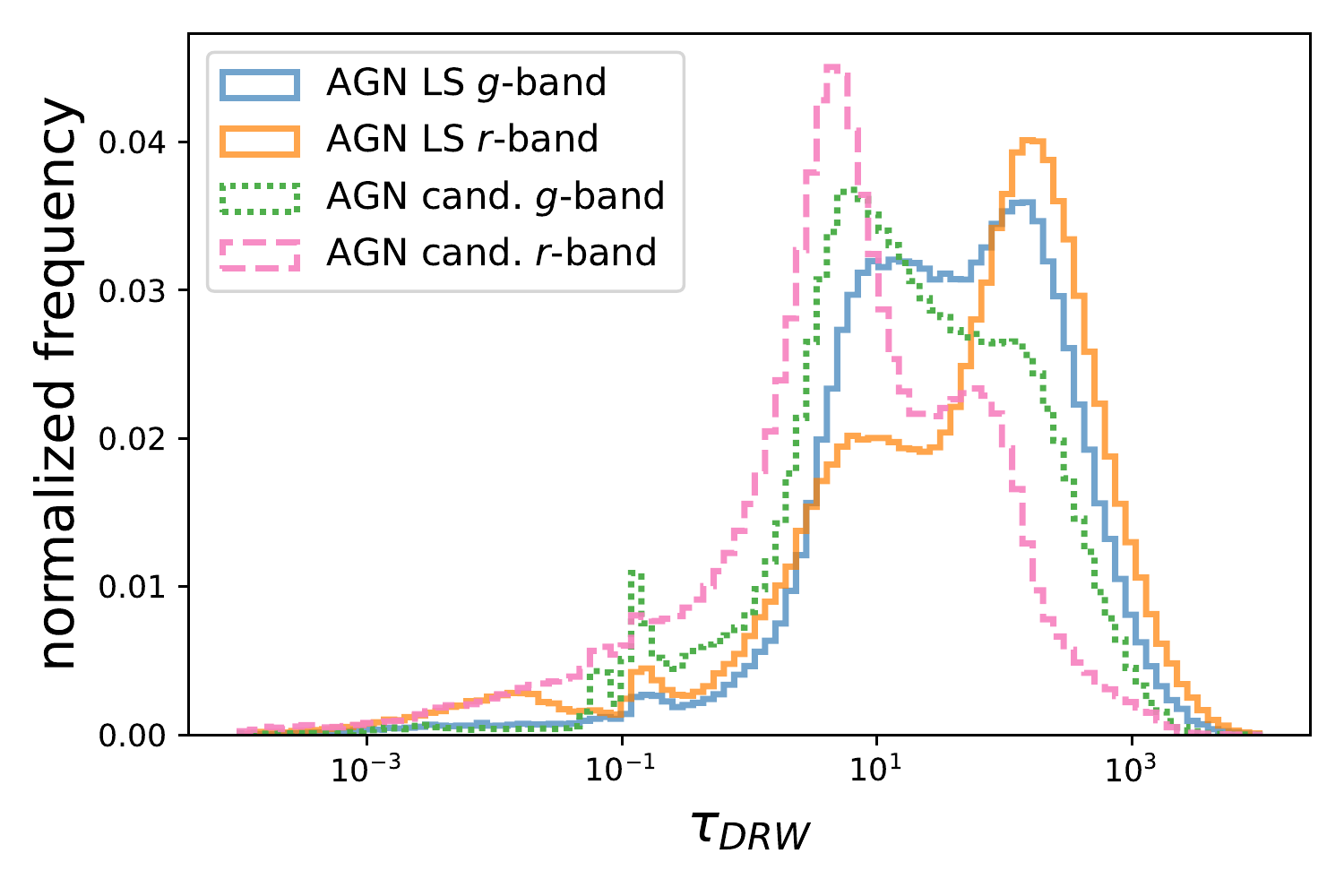} &
  \includegraphics[scale=0.38]{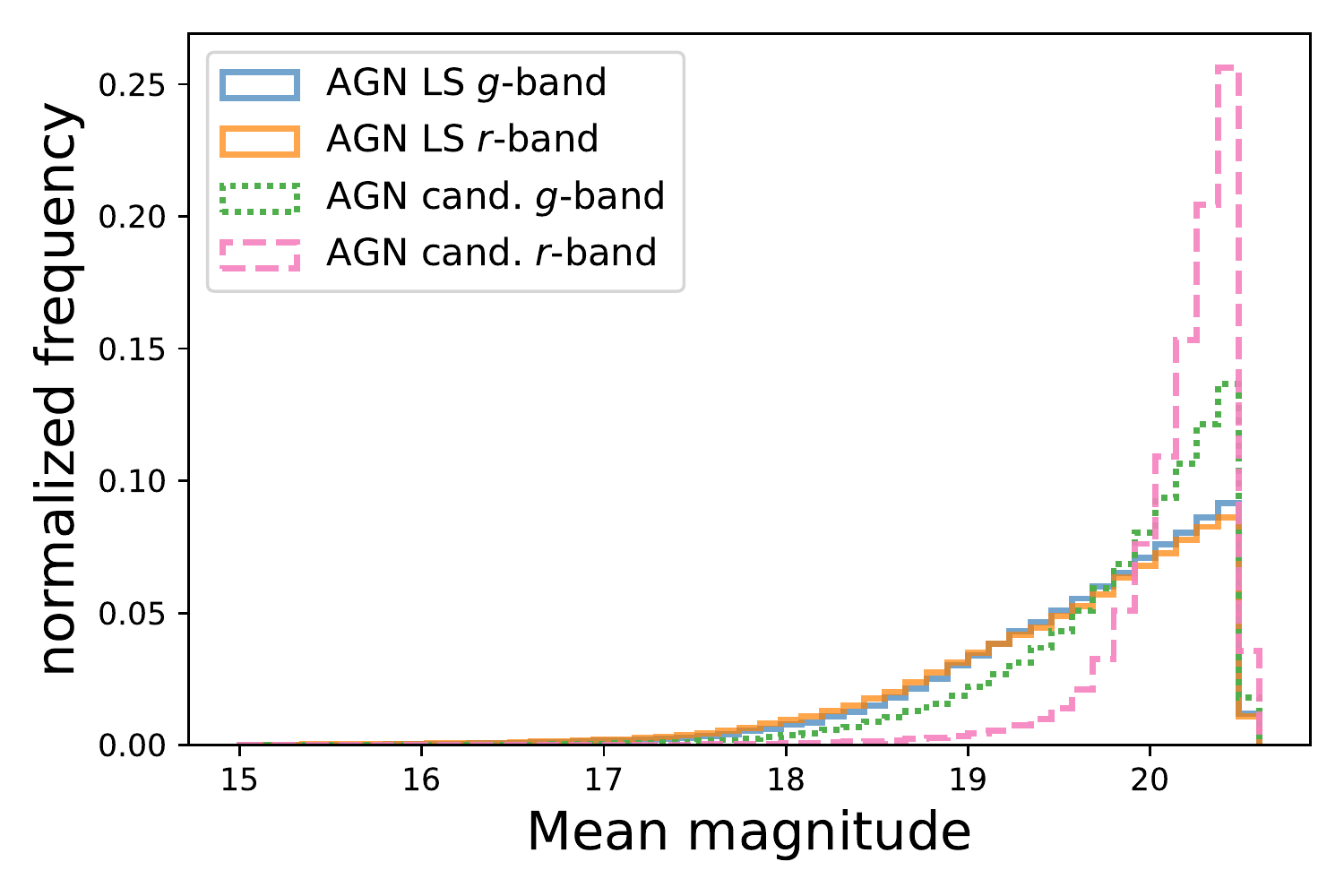} \\

\end{tabular}
\caption{Feature distributions of AGN candidates and AGN from the LS. \textit{Top panels}: $g-r$ (left), $r-W1$ (center), and $W1-W2$ (right) color distributions of AGN from the LS (blue), AGN candidates selected from the $g$-band (green), and AGN candidates selected from the $r$-band (pink). \textit{Bottom panels}: \texttt{Amplitude} (left), \texttt{GP\_DRW\_tau} (timescale of the DRW model, center), and mean magnitude (right), for AGN from the LS in the $g$-band (blue), AGN from the LS in the $r$-band (yellow), AGN candidates selected from the $g$-band (green), and AGN candidates selected from the $r$-band (pink).
\label{figure:comp_AGN_color}}
\end{center}
\end{figure*}

Moreover, we compared the optical and MIR color distributions of these three samples. These are shown in the top panels of Figure \ref{figure:comp_AGN_color}. The left panel shows the $g-r$ distribution for AGNs from the LS, and AGN candidates in both bands, the central panel shows the $r-W1$ color, and the right panel shows their $W1-W2$ distributions. We can note that the AGN candidates selected from the $g$-band have very similar colors to those from the LS, while the candidates selected from the $r$-band have different color distributions. We also compared the feature distributions in each band. Some of these are shown in the lower panels of figure \ref{figure:comp_AGN_color}. The bottom panel shows the feature \texttt{Amplitude} (left; half of the difference between the median magnitudes obtained with the 5\% brighter and 5\% fainter measurements), \texttt{GP\_DRW\_tau} (center; relaxation time, or timescale, from DRW modeling), and the mean magnitude (right), in the $g$-band, for AGNs in the LS (blue) and AGN candidates (green), and in the $r$-band, for AGNs in the LS (yellow) and AGN candidates (pink). From these figures we can see that in general the $g$-band candidates have feature distributions closer to the ones of the AGNs in the LS, compared to the $r$-band, although they are not exactly the same. This is expected, as the LS is dominated by blue and bright AGNs.

From these results, we can conclude that the AGN sample selected with the $g$-band is consistent with the AGN sample from the LS, and we can suspect that the number of false positives selected from the $r$-band are non-negligible. Therefore, for the selection of candidates to be observed by ChANGES, we decided to give priority to the candidates selected using the $g$-band, and for objects without $g$-band classification, the classification from the $r$-band was used, but including only objects with high probability in the node\_init ($P_{init}\geq0.9$).

\subsection{Variable star candidates}\label{section:vs}

\begin{figure*}[htbp!]
\begin{center}
\begin{tabular}{cc}
\includegraphics[scale=0.35]{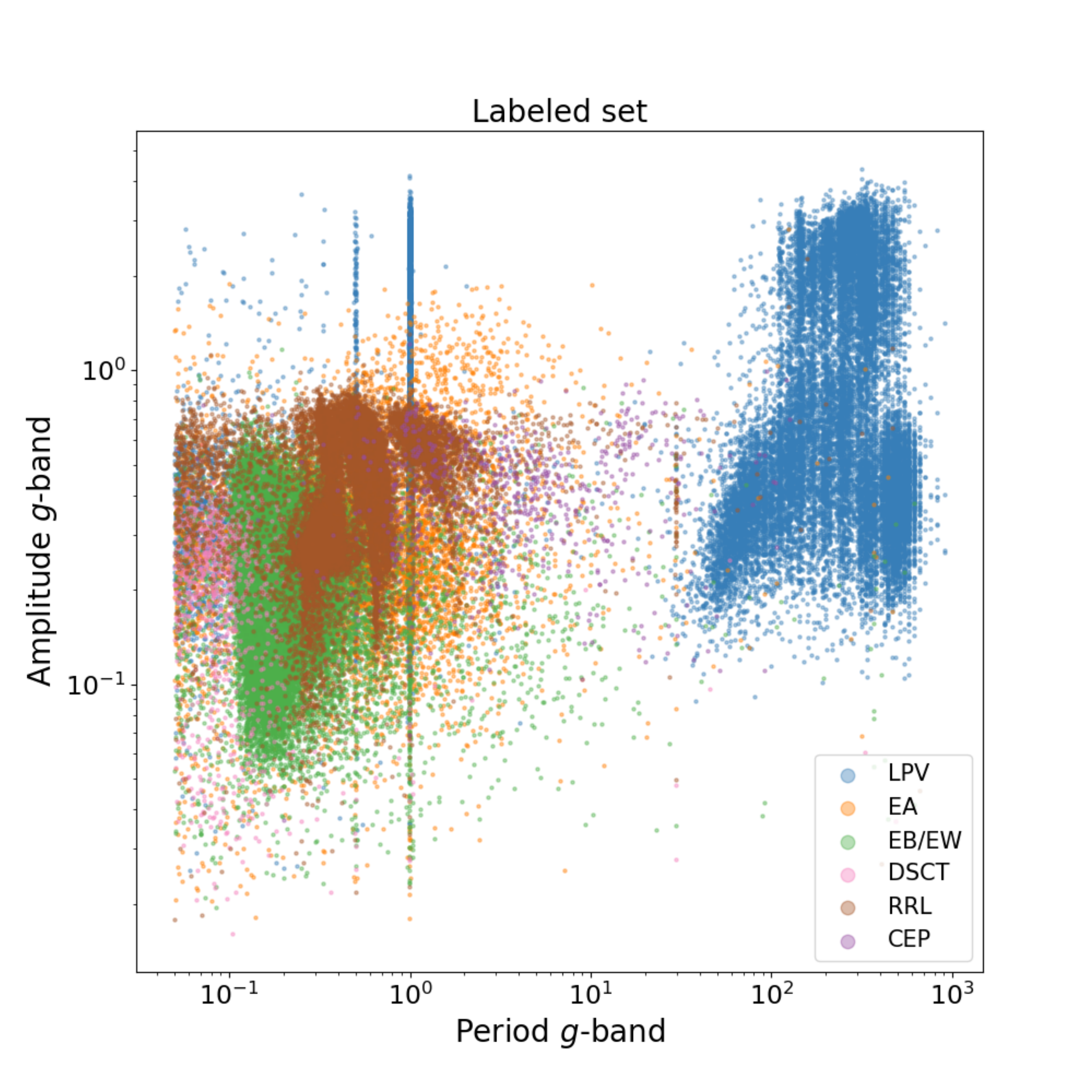} &
  \includegraphics[scale=0.35]{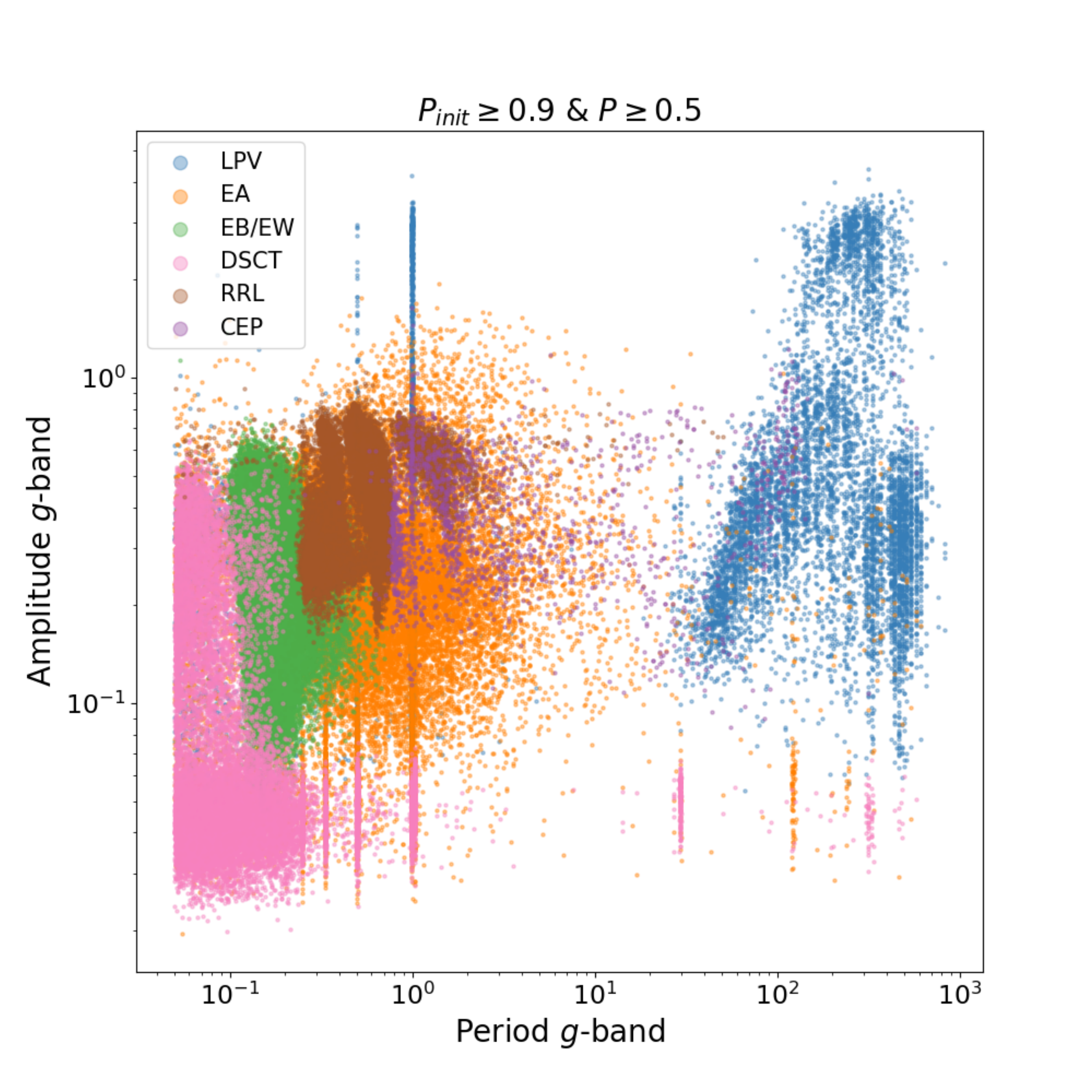} \\

\end{tabular}
\caption{\texttt{Period} versus \texttt{Amplitude} in the $g$-band for different periodic variable star classes. The left panel shows sources from the LS, and the right panel shows periodic variable star candidates, filtered by the probability of the node\_init $P_{init}\geq0.9$ and by the final probability $P\geq0.5$.\label{figure:period_amplitude}}
\end{center}
\end{figure*}

The variable star classes considered in this work can be roughly distinguished in a period versus amplitude diagram. In Figure \ref{figure:period_amplitude} we show the feature \texttt{Period} versus the feature \texttt{Amplitude} 
for the sources in the LS (left panel), and for candidates in the ZTF/4MOST sky filtered by their probability in the node\_init and their final predicted probability ($P_{init}\geq0.9$ and $P\geq0.5$; right panel). We can see in the figure that the periods of some classes have distributions that are different from expectations (e.g., \citealt{CS15}). This is produced by aliasing, when the periodogram finds a multiple of the true period, or spurious periods related to the light curve cadence (most prominently, 0.5 day, 1 day, and a month). For instance, some LPVs show periods of one day, which can affect their proper classification. Fortunately, the model uses other features that are related to the timescale of the variations, which can compensate for these miscalculated periods, such as \texttt{IAR\_phi} (level of autocorrelation using a discrete-time representation of a damped random walk or DRW model) and \texttt{GP\_DRW\_tau} (relaxation time from a DRW model), which are among the most relevant features in the node\_periodic (see Table \ref{table:feat_rank}). We can also see that there is a large fraction of DSCT candidates with very low amplitudes and short periods: these sources drop out when only sources with $P\geq0.9$ are selected. In general, the distribution of the high probability candidates (right panel in Figure \ref{figure:period_amplitude}) is consistent with previous findings (e.g., \citealt{CS15,Chen20}). Therefore, in order to obtain reliable candidates from this model, we suggest applying probability cuts, taking into account the probability distribution per class shown in Figures \ref{figure:prob_hists_init} and \ref{figure:prob_hists_final}, and the Reliability Diagrams presented in Figures \ref{figure:reliability_diagram_gband} and \ref{figure:reliability_diagram_rband}.

\subsection{Comparison with other works}\label{section:others_comparison}

The ALeRCE broker alert light curve classifier (\citetalias{Sanchez-Saez21a}) introduced a taxonomy similar to the one used in this work, but their model does not include non-variable classes, as they were dealing with alerts and not with DR light curves. We find that our precision, recall, and F1-score are quite similar to those obtained by them. Excluding the non-variable classes, the confusion matrices presented in Figure \ref{figure:conf_max_third} are comparable to the one presented in Figure 7 of \citetalias{Sanchez-Saez21a}. However, for some classes, like SNIa, lowz-AGN, or YSO, the recall is lower when using DR light curves; this result is expected, since the noisy DR light curves hinder the detection of the low-amplitude variations from AGNs and YSOs, while the alert light curves are dominated by the most variable sources belonging to these classes. Moreover, the DR light curves of transients have more noise compared to their alert light curves, since the photometry of the former is centered in the host galaxy, and the contribution from the host is removed in the alert light curves. 

\begin{figure*}[htbp!]
\begin{center}
\begin{tabular}{cc}
\includegraphics[scale=0.28]{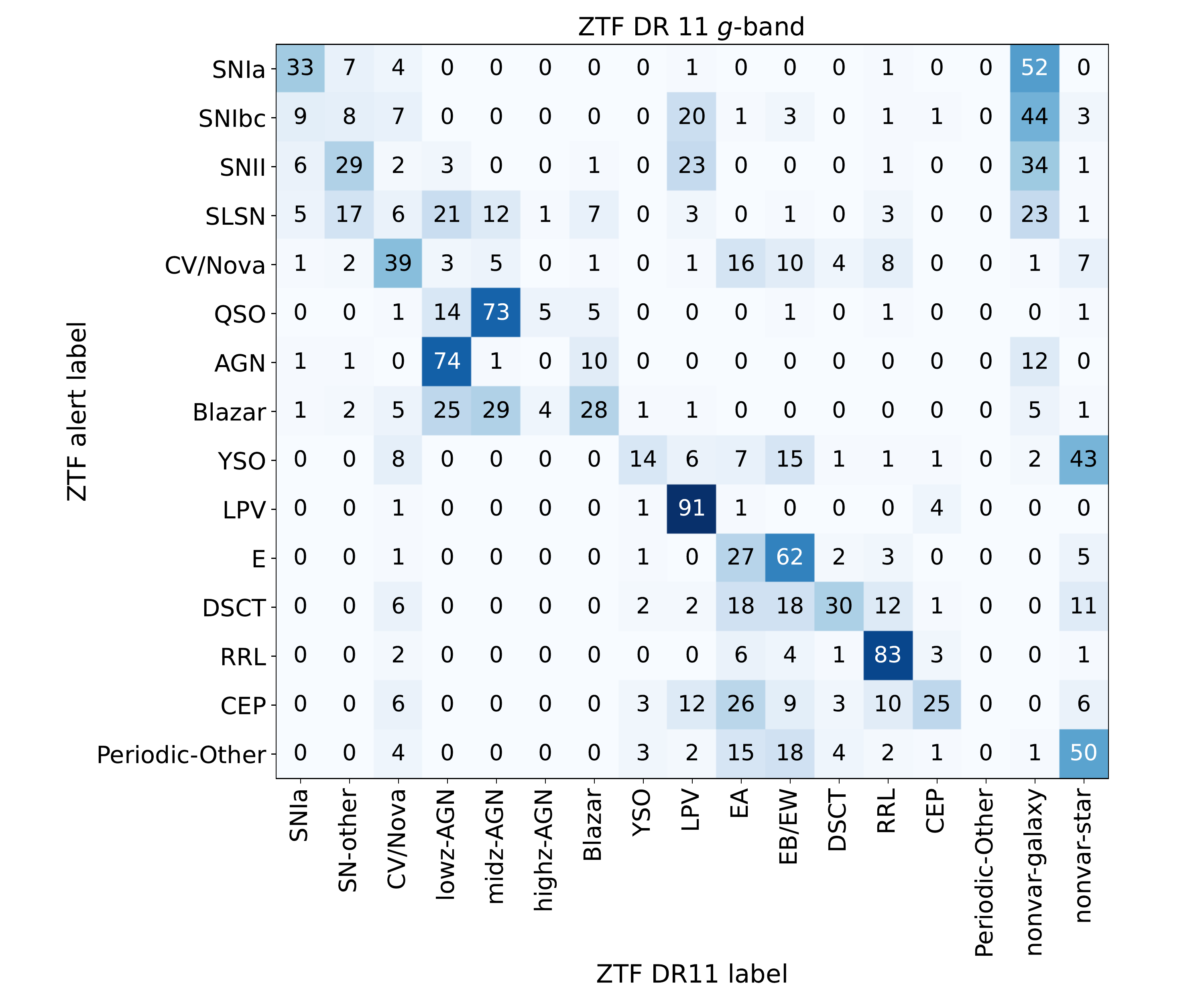} &
  \includegraphics[scale=0.28]{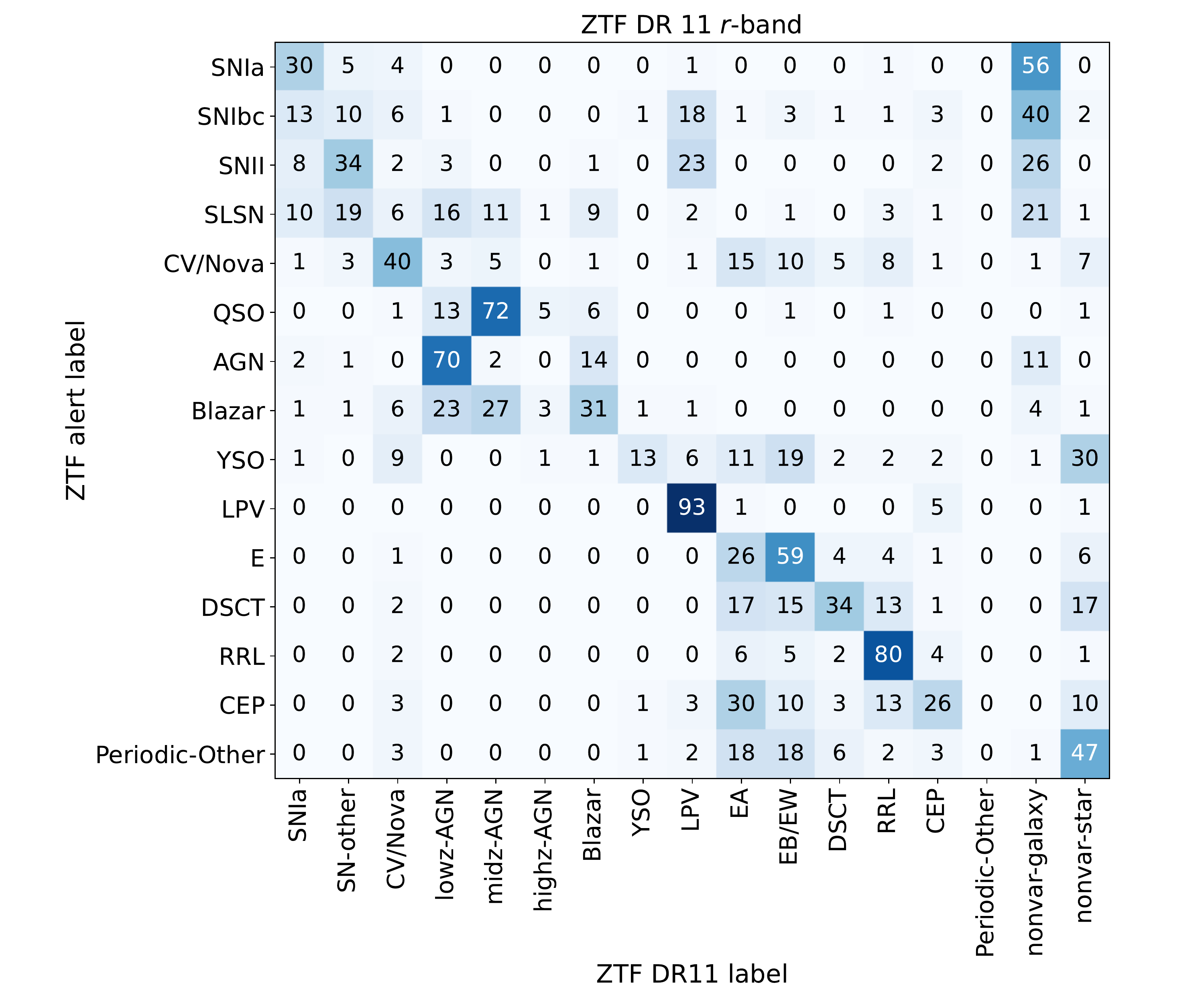} \\
\includegraphics[scale=0.28]{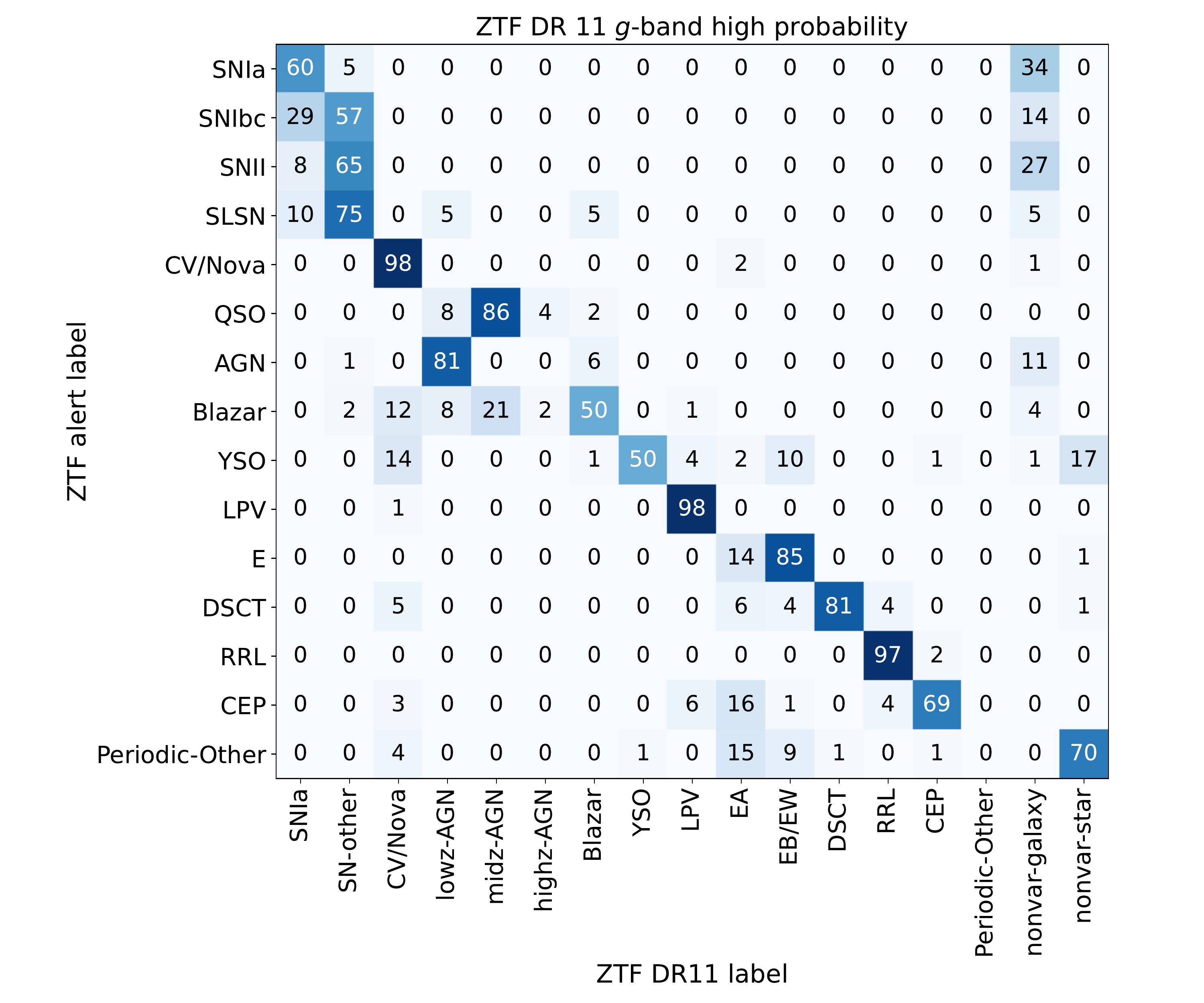} &
  \includegraphics[scale=0.28]{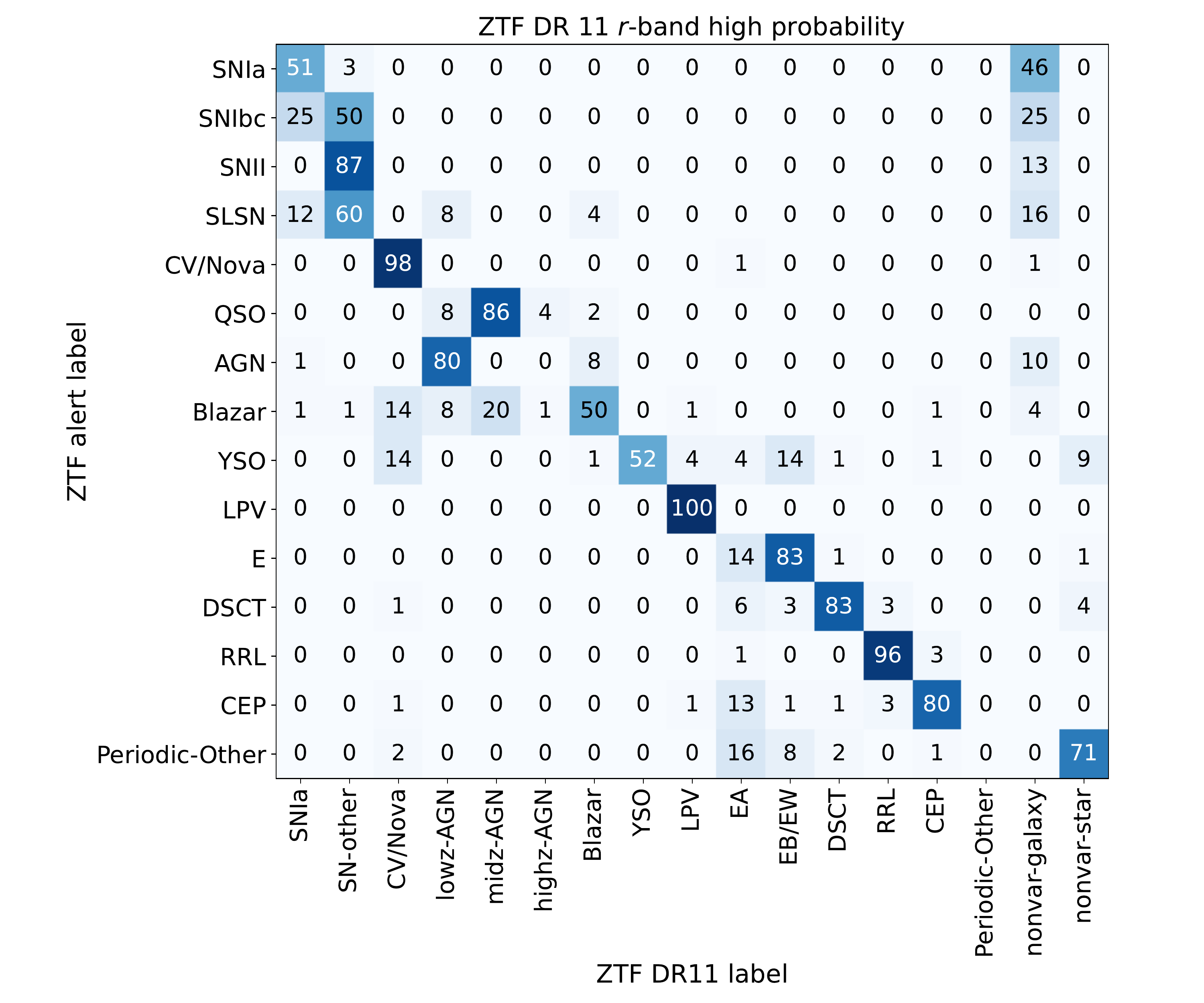} \\
\end{tabular}
\caption{Comparison of the classifications obtained by the ALeRCE broker ZTF alert light curve classifier, and the classifications obtained by our model for the ZTF DR11 light curves. The left panel shows the results when the $g$-band features are used for the DR classifier, the right panel shows the results when the $r$-band features are used for the DR classifier. The bottom panels show the same results, but keeping only those sources with high classification probabilities ($P\geq0.5$) in both the alert and DR samples. We divided each row by the total number of objects per class with alert labels, and we rounded these percentages to integer values.
\label{figure:comp_DR_alerts}}
\end{center}
\end{figure*}

In Figure \ref{figure:comp_DR_alerts} we compare the classifications provided by the ALeRCE light curve classifier with our results. The top panels show the comparison for all the sources in common in both bands (372,125 in $g$ and 360,322 in $r$), while the bottom panels show the comparison when only sources with a final predicted probability $P\geq0.5$; in both ZTF DR11 and ZTF alert models; are considered (123,829 in $g$ and 116,827 in $r$). Clearly, there is more confusion when all sources in common are included in the analysis, but for those with probabilities larger than 0.5, the results agree well. We can see that a high fraction of the SNe candidates from the ZTF alert stream are classified as non-variable galaxies in the ZTF DR11; this is expected since the DR light curves do not necessarily contain the flux from the transient, as the PSF photometry is centered in the position of the host galaxy and not at the position of the transient itself. Moreover, other classes from the alert classification are classified as non-variable in the ZTF DR model. This can be due to problems in the ZTF image subtraction that produces alerts for non-variable objects (like the bogus class from \citealt{Carrasco-Davis21}), or by the difficulty in detecting low-amplitude variations when DR light curves are used.

\begin{figure}[tb]
    \centering
    \includegraphics[width=0.8\linewidth]{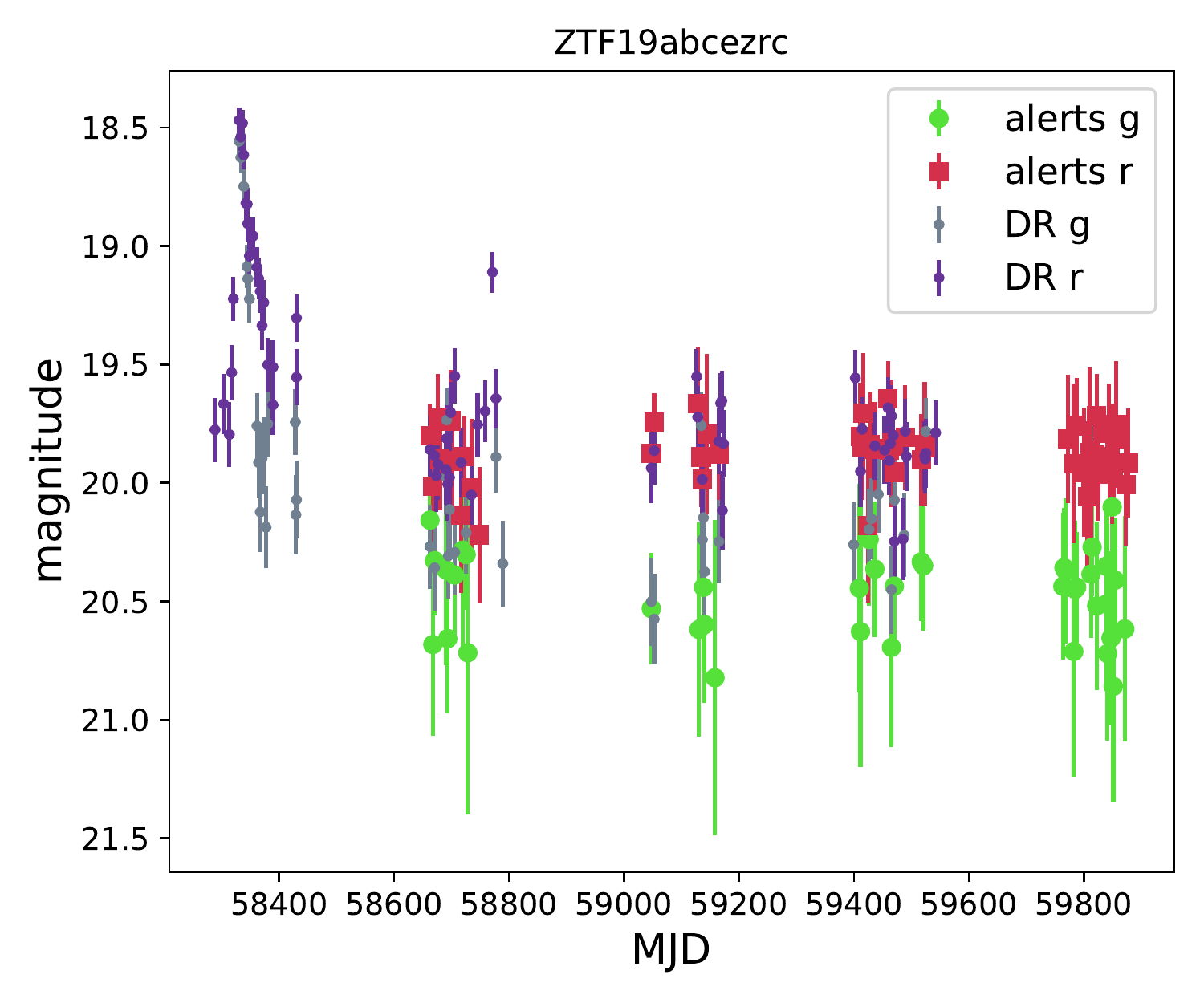}
    \caption{ZTF alert and ZTF DR11 light curve for the source ZTF19abcezrc. \label{figure:sn_template}}
\end{figure}

When comparing the classifications provided by the ALeRCE light curve classifier and those obtained using the ZTF DR11, we notice that there are 49 sources classified as SNIa and 113 as SN-other, in both $g$ and $r$ bands, that are classified as AGN, QSO or Blazar based on their alerts.  After inspecting some examples of these objects, we noticed that most of them were transients present in ZTF that produced alerts when the transient event ended. An example of this is presented in Figure \ref{figure:sn_template}. The figure shows the ZTF DR11 and the alert light curve of the source ZTF19abcezrc. This source was reported to TNS by the ATLAS team \citep{TNSreport}, and it is classified as SNIa in both bands from its ZTF DR11 light curves. From the figure, we can see that the alerts are generated after the SN event finished. The source was classified by the ALeRCE broker as AGN. This confusion can be explained by the lack of a transient event in the alerts, and WISE and optical colors similar to those of AGNs (the color of the host galaxy). These results demonstrate the advantage of combining alert and DR light curves when dealing with ZTF data.

\begin{figure*}[htbp!]
\begin{center}
\begin{tabular}{cc}
\includegraphics[scale=0.3]{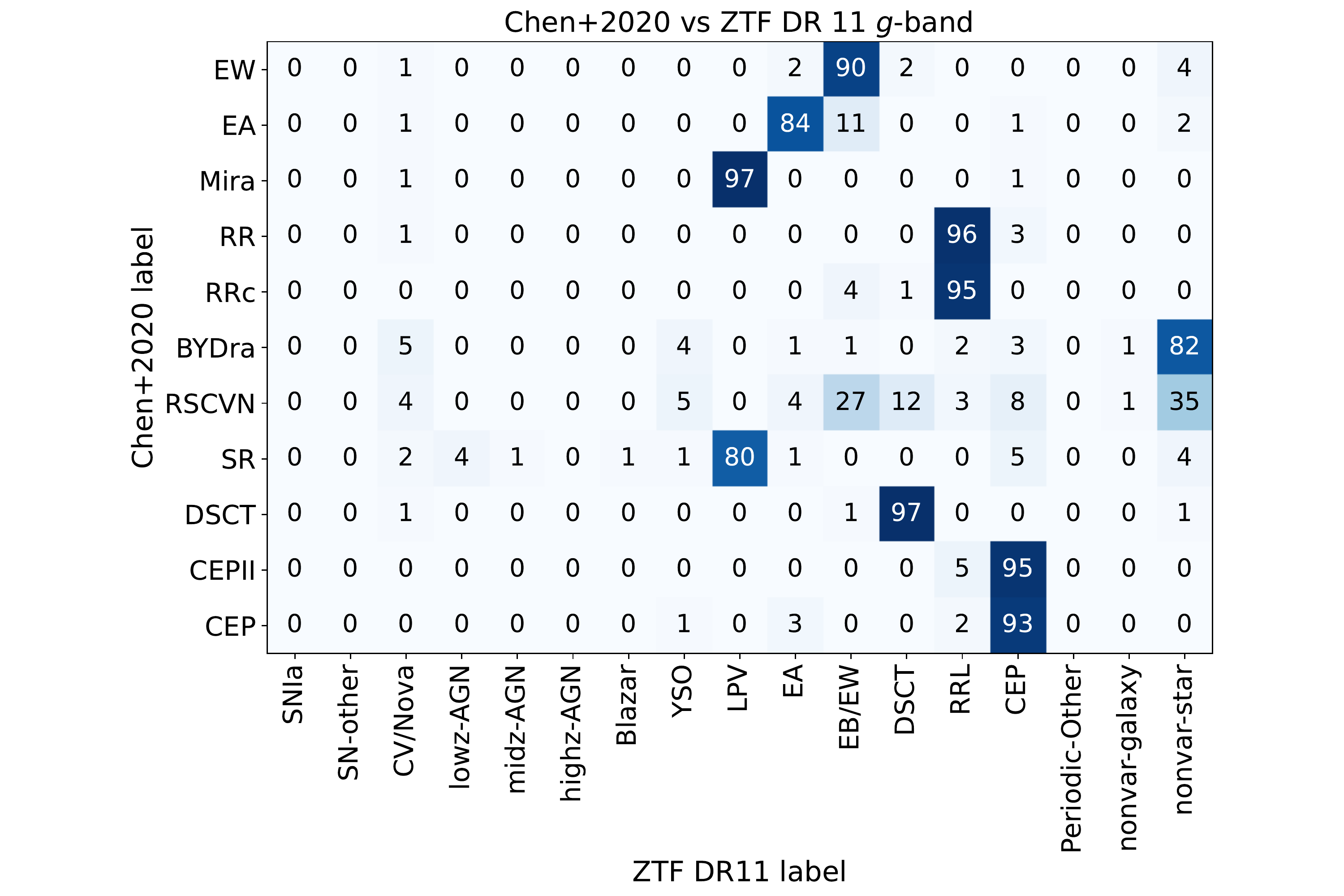} &
  \includegraphics[scale=0.3]{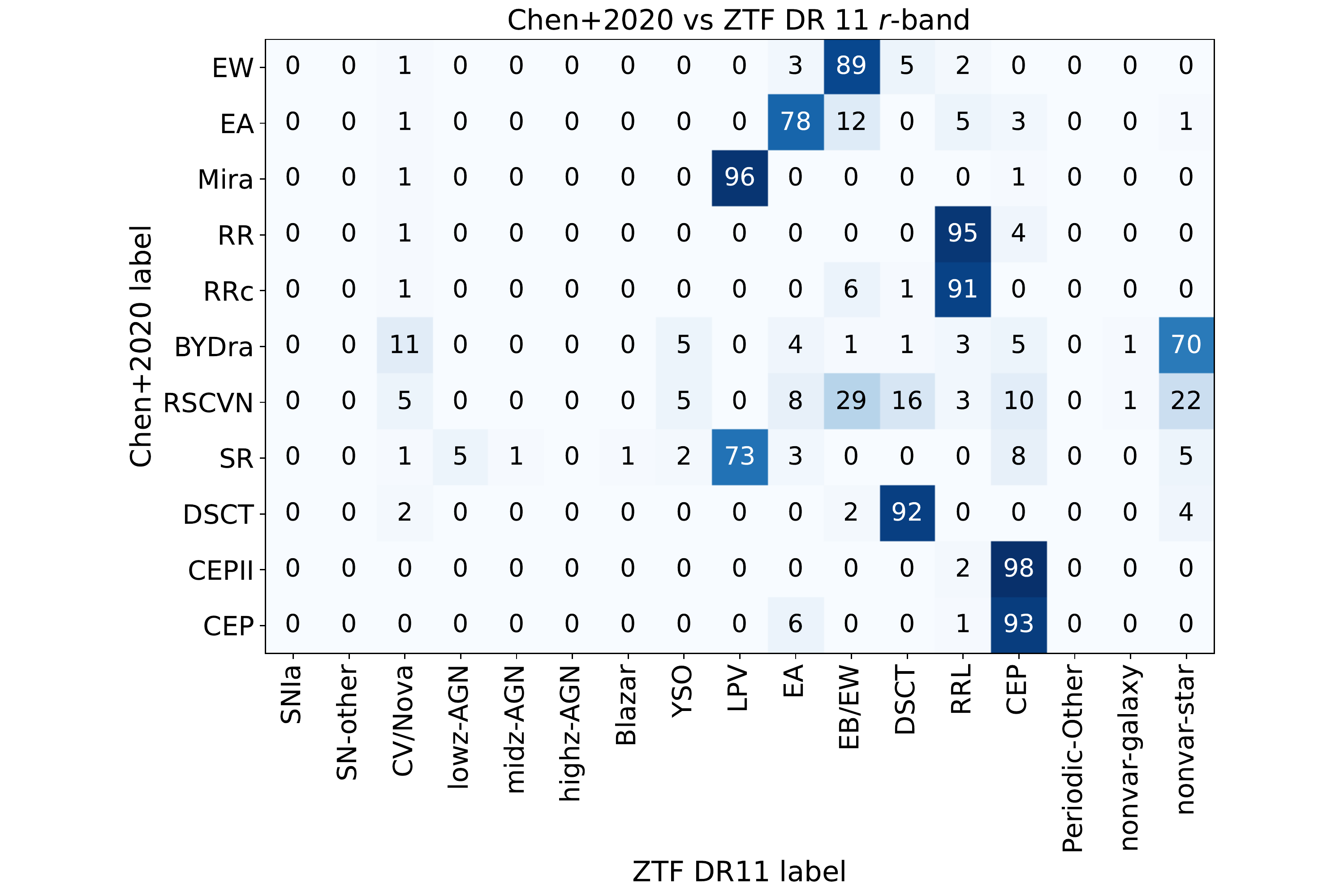} \\

\end{tabular}
\caption{Comparison of the classifications obtained by \cite{Chen20}, using ZTF DR2, and the classifications obtained by our model for the ZTF DR11 light curves. The left panel shows the results when the $g$-band features are used for the DR classifier, the right panel shows the results when the $r$-band features are used for the DR classifier. We divided each row by the total number of objects per class with \cite{Chen20} labels, and we rounded these percentages to integer values. \label{figure:comp_DR_Chen20}}
\end{center}
\end{figure*}

\begin{figure*}[htbp!]
\begin{center}
\begin{tabular}{cc}
\includegraphics[scale=0.5]{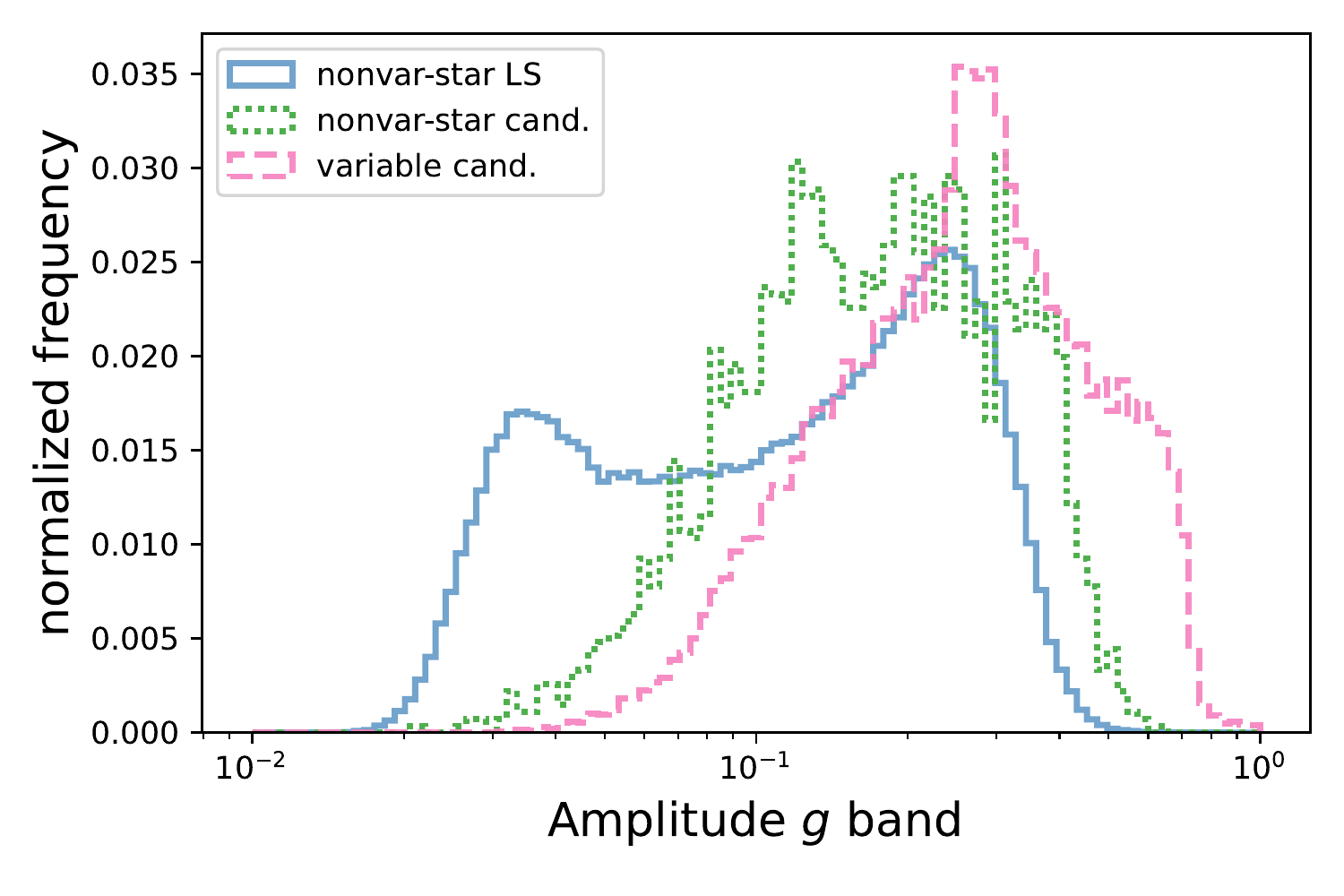} &
  \includegraphics[scale=0.5]{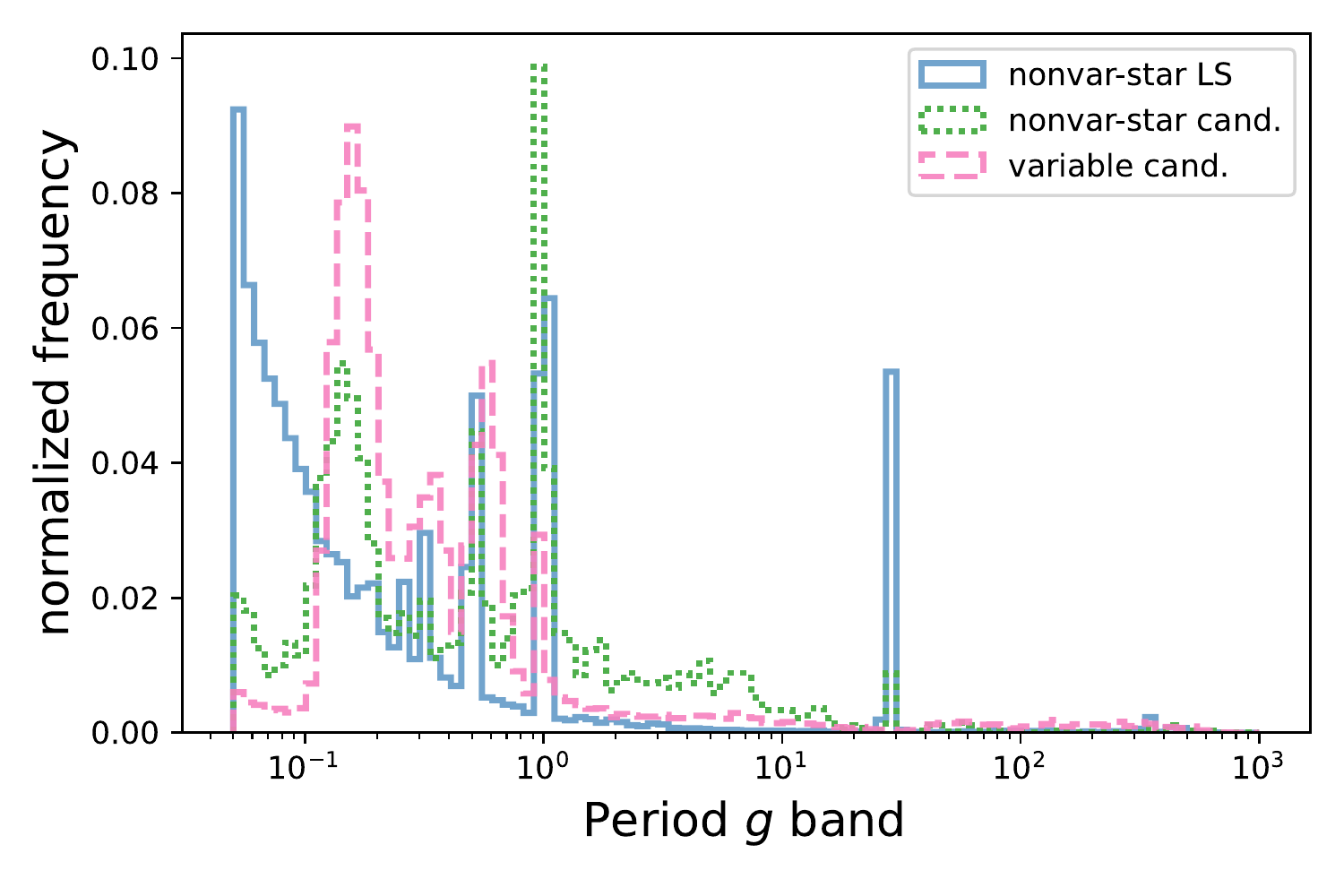} \\

\end{tabular}
\caption{Variability properties of variable star candidates from \cite{Chen20}, classified as Variable (pink) and non-variable (green) by our model. The left panel shows the \texttt{Amplitude} distribution, and the right panel the \texttt{Period} distribution. Sources from the LS classified as nonvar-star are included for reference (blue). 
\label{figure:comp_DR_Chen20_feat}}
\end{center}
\end{figure*}

\cite{Chen20} classified ZTF DR2 light curves into 11 classes of periodic variable stars, and provided a catalog with classifications for 781,602 objects. There are 57,128 and 57,086 sources in their catalog located in the ZTF/4MOST sky which also have classifications from our model in the $g$ and $r$ bands, respectively. In Figure \ref{figure:comp_DR_Chen20} we compare the results obtained by \cite{Chen20} with ours. We can see that there is broad agreement in the classification of the sources classified as variable by our model, when the classes considered are present in both models (i.e., EA, EB/EW, LPV, RRL, DSCT, and CEP). We can also see that very few of their objects are classified as transient or stochastic by our model,  as expected since their catalog does not include this kind of variability. Although around 6\% of their semi-regular variables (SR) are classified as AGN by our model (65 of them are known AGNs from the literature). The largest discrepancy by far, however, is the number of objects that are classified as variable stars by them but non-variable stars by us (including several of their BY Dra and RS CVn variables in this field). In order to understand the origin of this discrepancy, we show in Figure \ref{figure:comp_DR_Chen20_feat} the \texttt{Amplitude} and the \texttt{Period} of the common sources in the $g$-band, separating them according to our classification into nonvar-star and variable. Sources from the LS, classified as nonvar-star are also included for reference. A similar distribution is observed in the $r$-band. We can see that sources classified as nonvar-star have lower \texttt{Amplitude} values compared to the ones classified as variable, which could explain why we do not classify them as variables with our model. We can also see that we are missing sources with \texttt{Amplitude} larger than 0.1, probably due to the noise of the ZTF DR11 light curves, since known nonvar-stars from the LS also show amplitudes larger than 0.1. The \texttt{Period} distribution of the nonvar-star and the variable candidates is also different, with many of the sources classified as non-variable by our model, having periods of around one day. \cite{Chen20} did a very good job in ensuring that their periods were correct, including masking out obvious aliases (0.5 day, 1 day, 29.5 days, etc.), so their classification of low-amplitude sources should be better than ours.

Finally, we decided to exclude from this comparison the works presented by \cite{vanRoestel21} and \cite{Aleo22}, as they do not provide full sky catalogs. We also decided to exclude from the analysis a comparison with works that used other surveys (like \textit{Gaia} or CRTS), since several of their candidates were included in our training set (see Section \ref{section:var_classes_lb}).

\section{Conclusions}\label{section:sumary}

In this work, we presented a hierarchical classifier designed to process ZTF DR11 PSF light curves (measured over the science images for objects detected in the reference images) of both extended and point sources. To the best of our knowledge, this corresponds to the first attempt to classify light curves of extended sources using ZTF DR light curves, which are not optimal for the study of the variability of extended objects. In this work, however, we demonstrate that these light curves can still be used to identify variable extended objects, when a suitable set of features is available.

The main purpose of our classifier is to select AGNs and Blazars at different redshifts from their variability, which will be spectroscopically followed up by the 4MOST survey ChANGES. The main goal of ChANGES is to target an unrivaled legacy sample of AGN with 4MOST, selected from several existing surveys (including ZTF). Although this was the main motivation to develop this classifier, inspired by our previous experience with the ALeRCE broker light curve classifier (\citetalias{Sanchez-Saez21a}), we decided to make a more general classifier that considers different classes of transients, persistently variable, and non-variable objects. The model considers a total of 17 classes, with two non-variable, three transient, five stochastic, and seven periodic classes. 

The model was designed in a hierarchical fashion, following the nature of the classes considered. We used a local classifier per parent node approach, where each node was composed of a balanced random forest classifier. For this reason, we call our model the Hierarchical Balanced Random Forest (HBRF) classifier. We trained two versions of this model, one that uses 61 variability features computed using the ZTF $g$-band, and another one that uses the same number of features but measured from the $r$-band light curves. We also included 10 additional features, which correspond to nine CatWISE and PS1 colors, as well as a PS1 morphology score. We decided not to include as features the coordinates of the sources, to avoid position-dependent biases present in the LS. The 61 variability features used in this work are taken from the detection features used by \citetalias{Sanchez-Saez21a}, excluding multi-band features. We also modified some of the features to properly deal with DR light curves. The LS was constructed using catalogs taken from \citetalias{Sanchez-Saez21a}, but additional catalogs were included, especially for the non-variable classes. 

We notice that, in general, better results are obtained when using the ZTF $g$-band. We propose that this is due to the intensive cadence campaigns conducted by ZTF in the $r$-band that skew the feature distributions and produce a large number of candidates in some classes, particularly for CV/Nova, highz-AGN, and YSO. Therefore, we recommend that any user of this model gives priority to the results obtained using the $g$-band. When using the classifications obtained in the $r$-band, we recommend to filter the catalogs by the classification probability in the node\_init ($P_{init}\geq0.9$), and/or to avoid the use of the model in regions with rapid cadence (using sources with number of epochs lower than 400 in DR11 light curves). Finally, when filtering the classifications by probability, we suggest taking into account the probability distributions per class presented in Section \ref{section:results} and the reliability diagrams discussed in Section \ref{section:prob_cal}.

Using this model, we were able to identify 384,242 unique AGN and Blazar candidates (within a radius of 1.5$''$) in the $g$-band and 4,048,299 candidates in the $r$-band, with 356,631 of these candidates being classified as AGN or Blazar in both bands. Considering the issues in the selection done with the $r$-band, for the selection of sources to be observed by ChANGES, we decided to give priority to the candidates selected in the $g$-band.

We expect to include more classes in a future version of this classifier. In particular, with the growing sample of known TDEs (e.g., \citealt{vanVelzen21,Hammerstein23}), we can start testing new models that include them as a class in the transients node. We also plan to expand the periodic node to new classes and sub-classes of periodic variable stars, including, for instance, rotational classes. Furthermore, future versions of this classifier can take advantage of new models that use light curves directly, without the need of a feature extraction procedure (e.g., \citealt{Becker20,Donoso-Oliva23,Pimentel23, Astorga23}).

\begin{acknowledgements}

The authors acknowledge support from the National Agency for Research and Development (ANID) grants: Millennium Science Initiative Program ICN12\_12009 (PSS,FEB,GCV,MC,PAE,FF,PH,LHG,RK,AMMA,GP), and NCN$19\_058$ (PA, PL); BASAL Center of Mathematical Modelling Grant PAI AFB-170001 (FF,AMMA); Basal CATA FB210003 (FEB,MC); FONDECYT Regular 1190818 (FEB), 1200495 (FEB), 1220829 (PAE), 1200710 (FF), 1211374 (PH), 1201748 (PL); FONDECYT Initiation 11191130 (GCV); FONDECYT Postdoctorado 3200250 (PSS); and from the Max-Planck Society through a Partner Group grant (PA).

This work has been possible thanks to the use of AWS-U.Chile-NLHPC credits.

Powered@NLHPC: This research was partially supported by the supercomputing
infrastructure of the NLHPC (ECM-02).

Based on observations obtained with the Samuel Oschin 48-inch Telescope at the Palomar Observatory as part of the Zwicky Transient Facility project. ZTF is supported by the National Science Foundation under Grant No. AST-1440341 and a collaboration including Caltech, IPAC, the Weizmann Institute for Science, the Oskar Klein Center at Stockholm University, the University of Maryland, the University of Washington, Deutsches Elektronen-Synchrotron and Humboldt University, Los Alamos National Laboratories, the TANGO Consortium of Taiwan, the University of Wisconsin at Milwaukee, and Lawrence Berkeley National Laboratories. Operations are conducted by COO, IPAC, and UW.

Based on observations obtained with the Samuel Oschin Telescope 48-inch and the 60-inch Telescope at the Palomar Observatory as part of the Zwicky Transient Facility project. ZTF is supported by the National Science Foundation under Grants No. AST-1440341 and AST-2034437 and a collaboration including current partners Caltech, IPAC, the Weizmann Institute for Science, the Oskar Klein Center at Stockholm University, the University of Maryland, Deutsches Elektronen-Synchrotron and Humboldt University, the TANGO Consortium of Taiwan, the University of Wisconsin at Milwaukee, Trinity College Dublin, Lawrence Livermore National Laboratories, IN2P3, University of Warwick, Ruhr University Bochum, Northwestern University and former partners the University of Washington, Los Alamos National Laboratories, and Lawrence Berkeley National Laboratories. Operations are conducted by COO, IPAC, and UW.

This work has made use of data from the European Space Agency (ESA) mission
{\it Gaia} (\url{https://www.cosmos.esa.int/gaia}), processed by the {\it Gaia}
Data Processing and Analysis Consortium (DPAC,
\url{https://www.cosmos.esa.int/web/gaia/dpac/consortium}). Funding for the DPAC
has been provided by national institutions, in particular the institutions
participating in the {\it Gaia} Multilateral Agreement.

We acknowledge with thanks the variable star observations from the AAVSO International Database contributed by observers worldwide and used in this research.

\end{acknowledgements}

\bibliographystyle{aa}
\bibliography{bibliography.bib}

\onecolumn
\begin{appendix}

\section{Sky densities per each class}\label{app:sky_densities}

The following figures (from \ref{figure:density_SNIa} to \ref{figure:density_nonvar-star}) show the target sky densities per class, in the Galactic coordinate space. 

\FloatBarrier

\begin{figure*}[hptb!]
\begin{center}
   \includegraphics[width=0.69\linewidth]{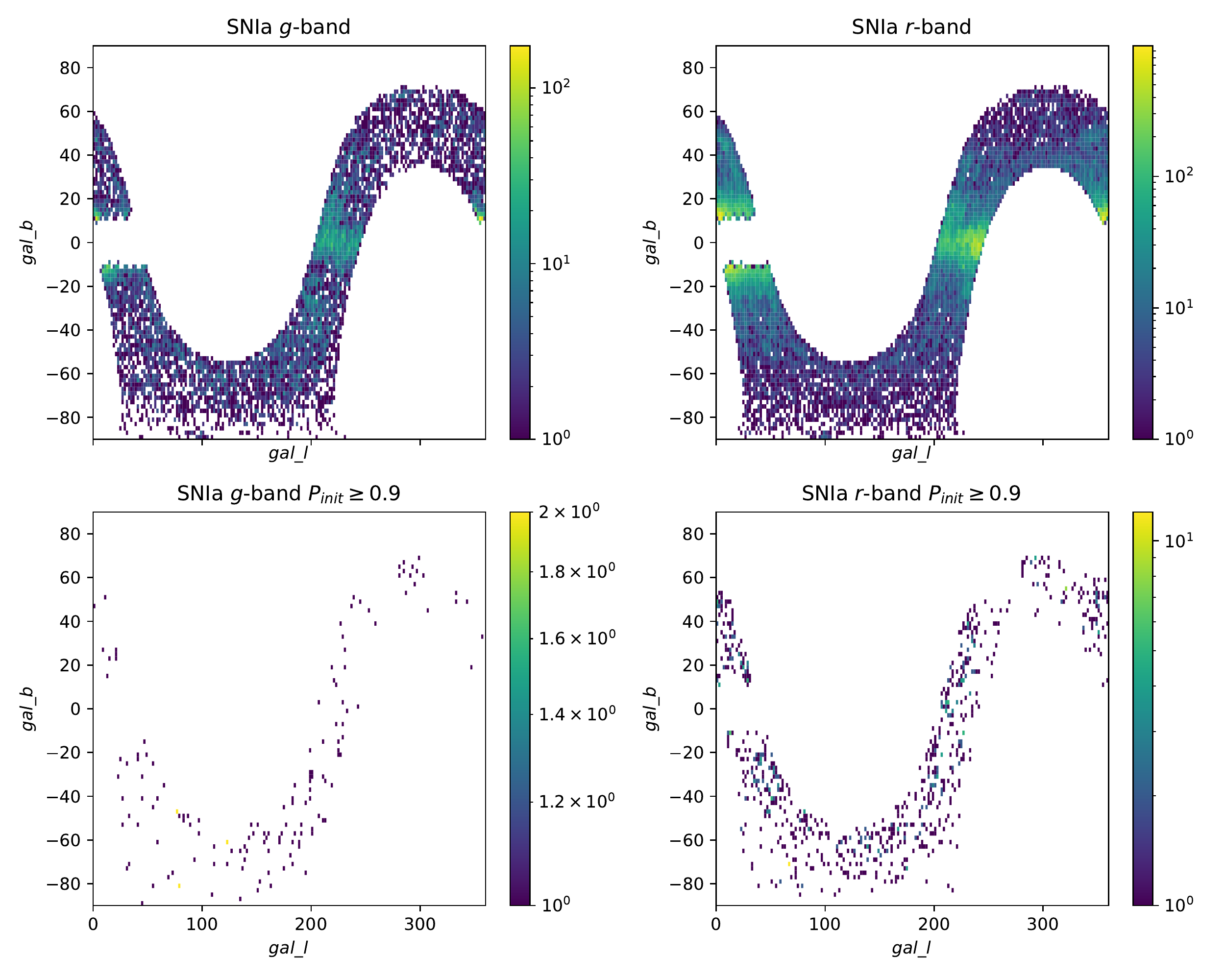} 

\caption{As in Figure \ref{figure:density_midz-AGN} but for the SNIa class.
\label{figure:density_SNIa}}
\end{center}
\end{figure*}

\begin{figure*}[hptb!]
\begin{center}
   \includegraphics[width=0.69\linewidth]{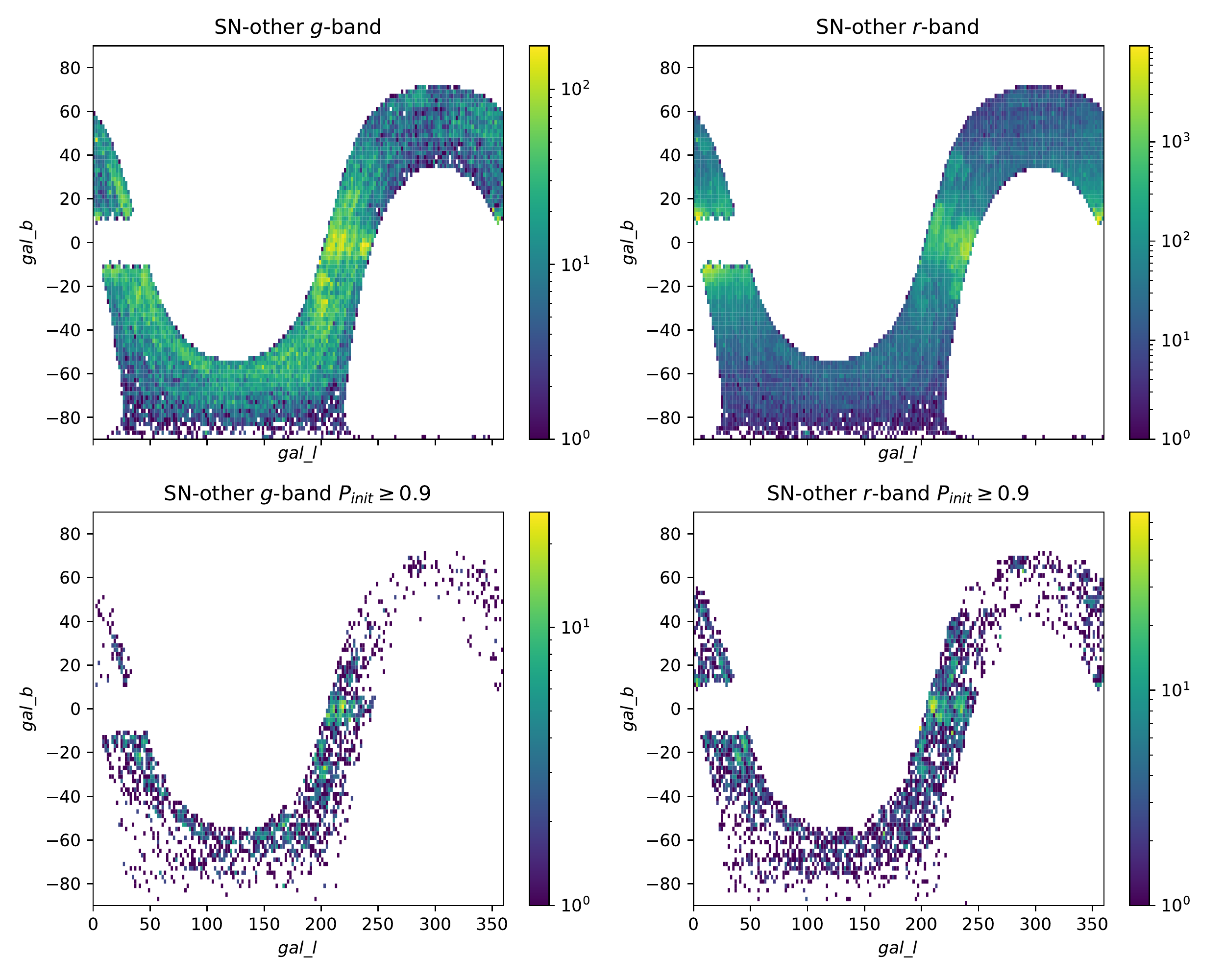} 

\caption{As in Figure \ref{figure:density_midz-AGN} but for the SN-other class.
\label{figure:density_SN-other}}
\end{center}
\end{figure*}

\begin{figure*}[hptb!]
\begin{center}
   \includegraphics[width=0.69\linewidth]{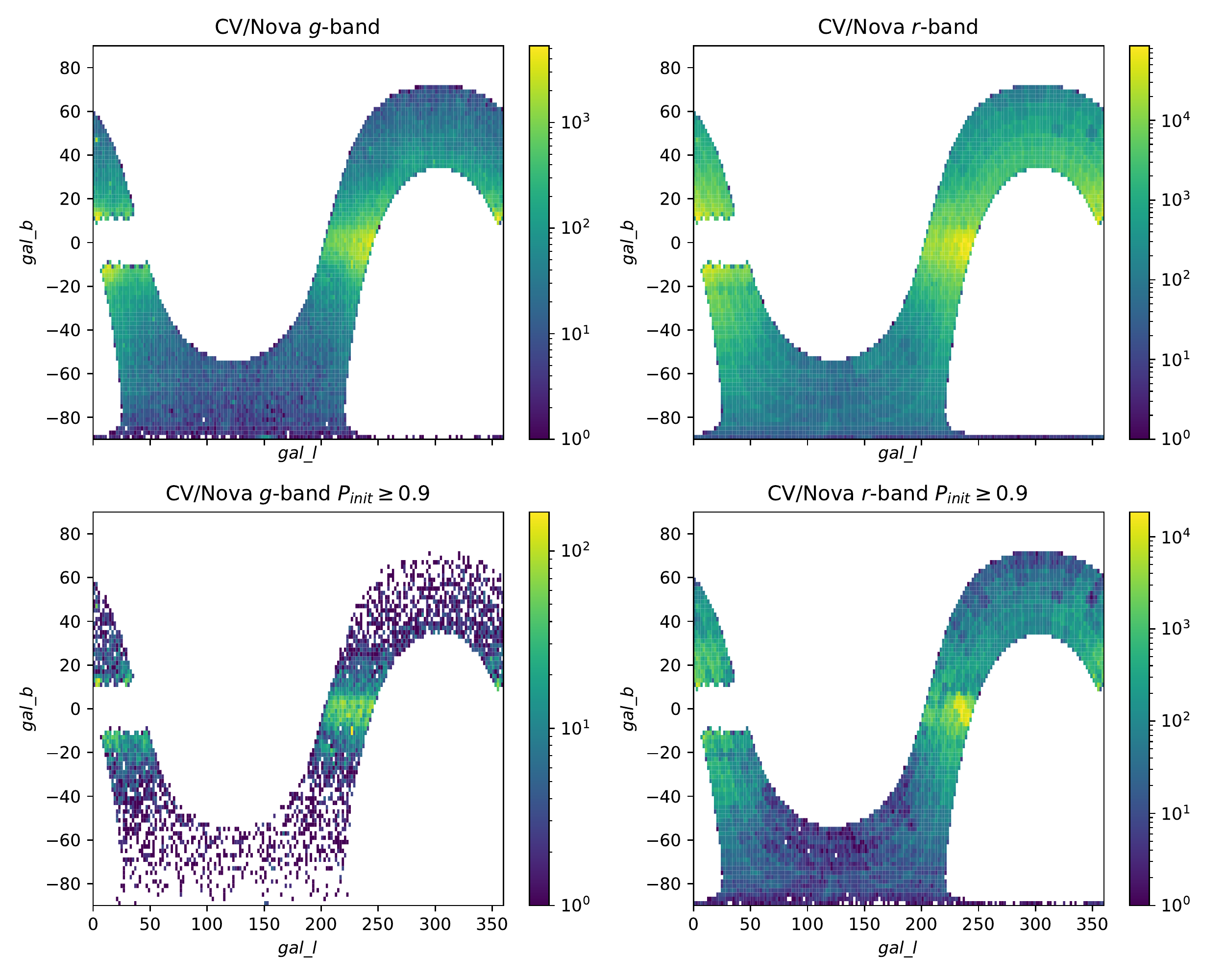} 

\caption{As in Figure \ref{figure:density_midz-AGN} but for the CV/Nova class.
\label{figure:density_CV/Nova}}
\end{center}
\end{figure*}

\begin{figure*}[hptb!]
\begin{center}
   \includegraphics[width=0.69\linewidth]{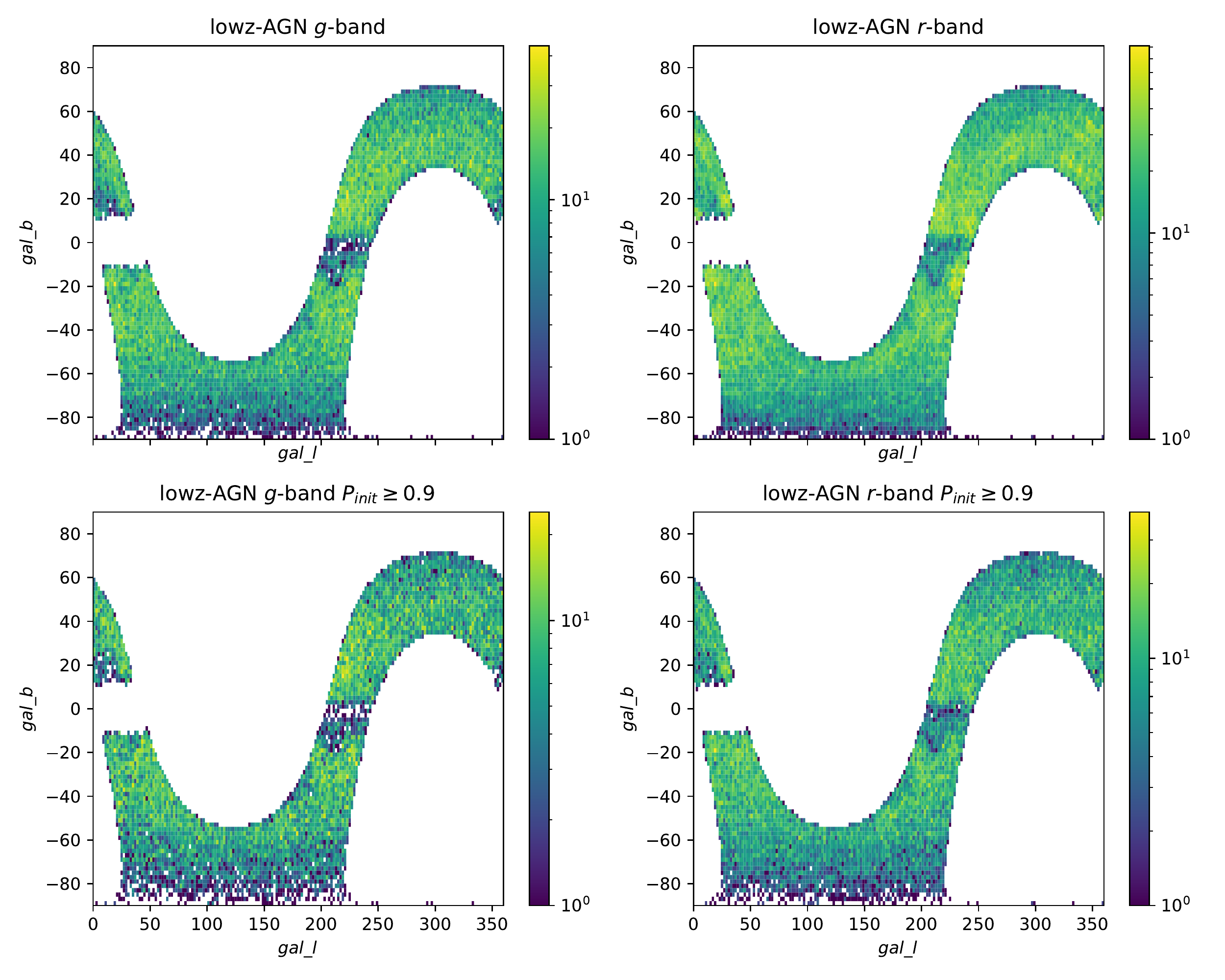} 

\caption{As in Figure \ref{figure:density_midz-AGN} but for the lowz-AGN class.
\label{figure:density_lowz-AGN}}
\end{center}
\end{figure*}

\begin{figure*}[hptb!]
\begin{center}
   \includegraphics[width=0.69\linewidth]{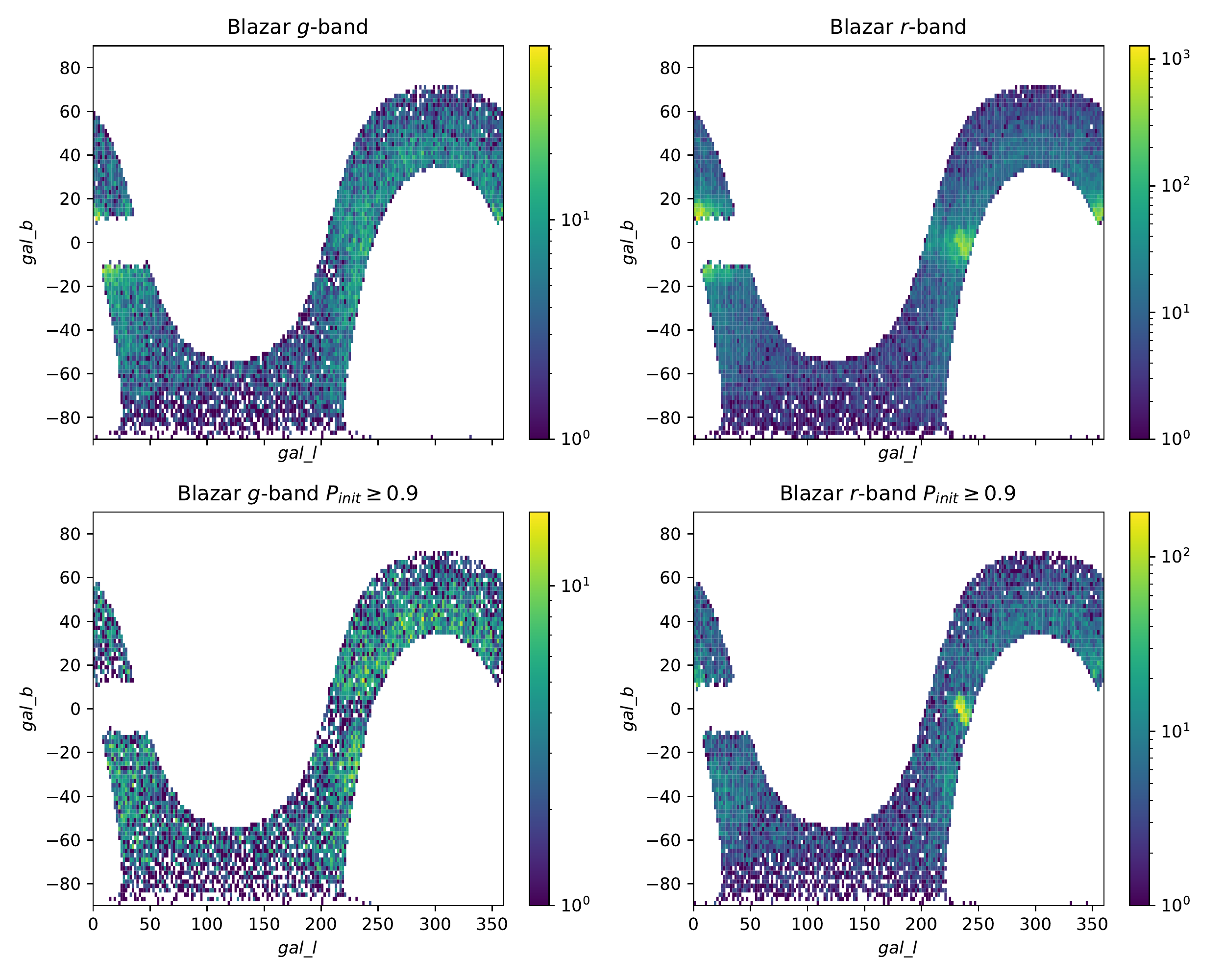} 

\caption{As in Figure \ref{figure:density_midz-AGN} but for the Blazar class.
\label{figure:density_Blazar}}
\end{center}
\end{figure*}

\begin{figure*}[hptb!]
\begin{center}
   \includegraphics[width=0.69\linewidth]{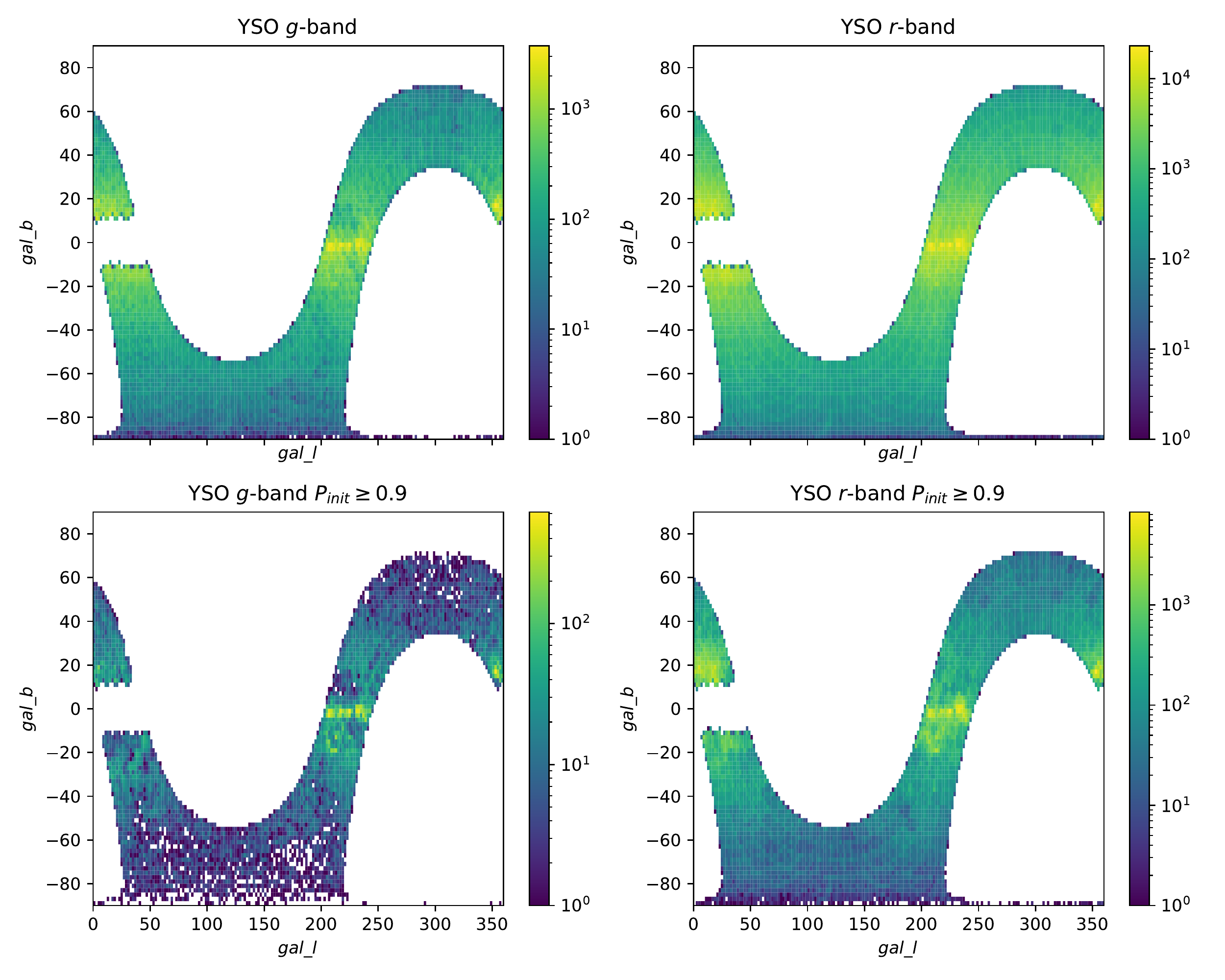} 

\caption{As in Figure \ref{figure:density_midz-AGN} but for the YSO class.
\label{figure:density_YSO}}
\end{center}
\end{figure*}

\begin{figure*}[hptb!]
\begin{center}
   \includegraphics[width=0.69\linewidth]{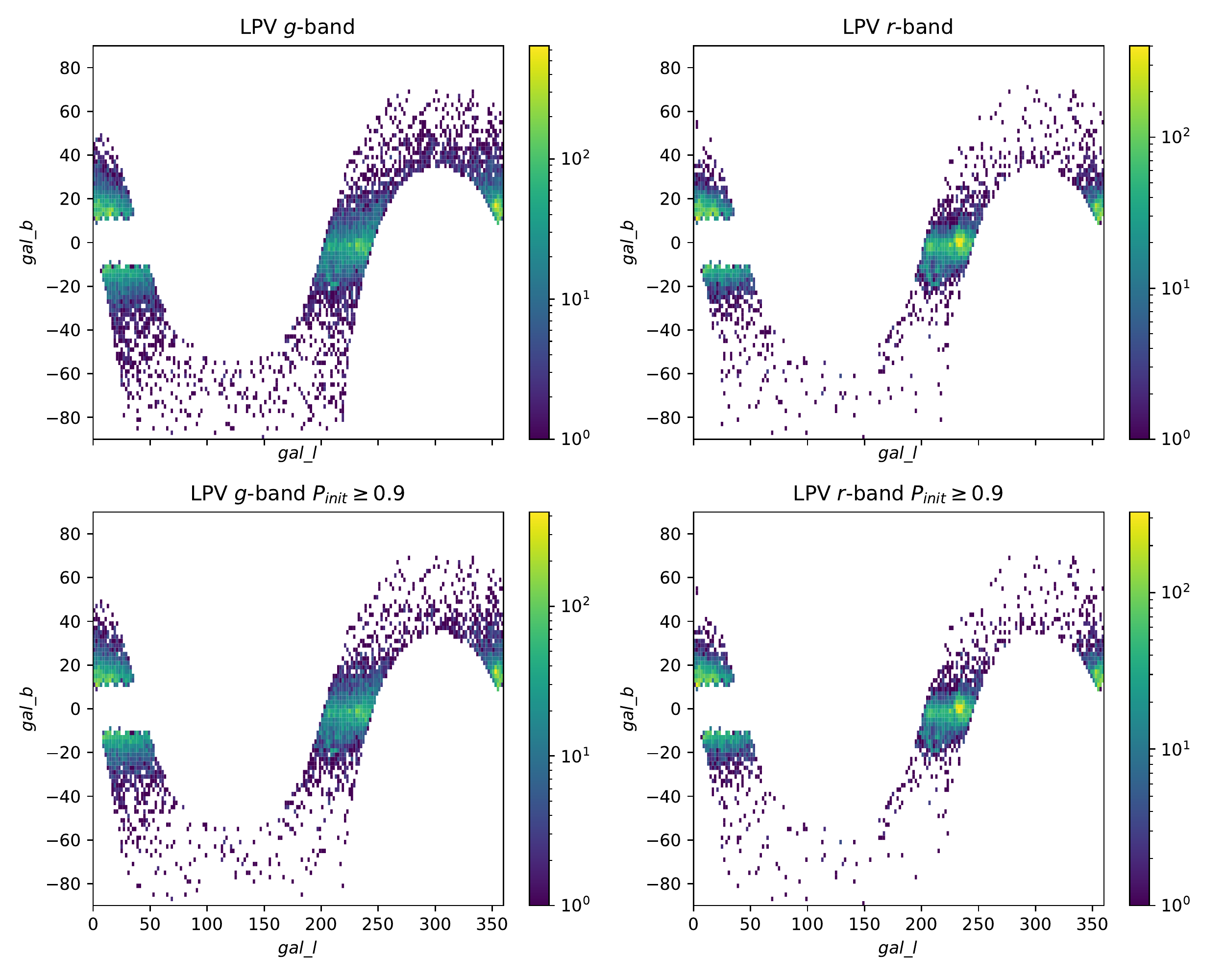} 

\caption{As in Figure \ref{figure:density_midz-AGN} but for the LPV class.
\label{figure:density_LPV}}
\end{center}
\end{figure*}

\begin{figure*}[hptb!]
\begin{center}
   \includegraphics[width=0.69\linewidth]{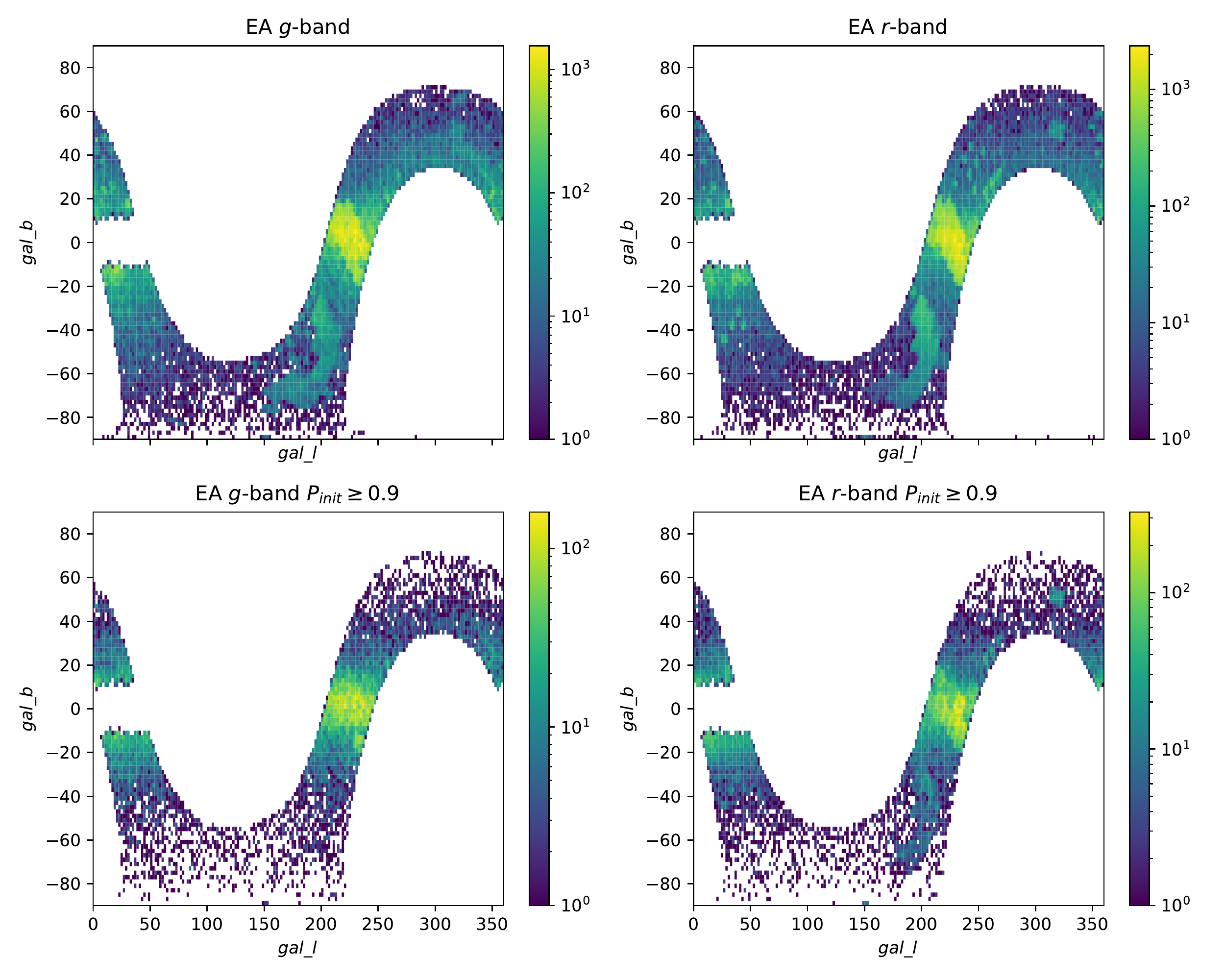} 

\caption{As in Figure \ref{figure:density_midz-AGN} but for the EA class.
\label{figure:density_EA}}
\end{center}
\end{figure*}

\begin{figure*}[hptb!]
\begin{center}
   \includegraphics[width=0.69\linewidth]{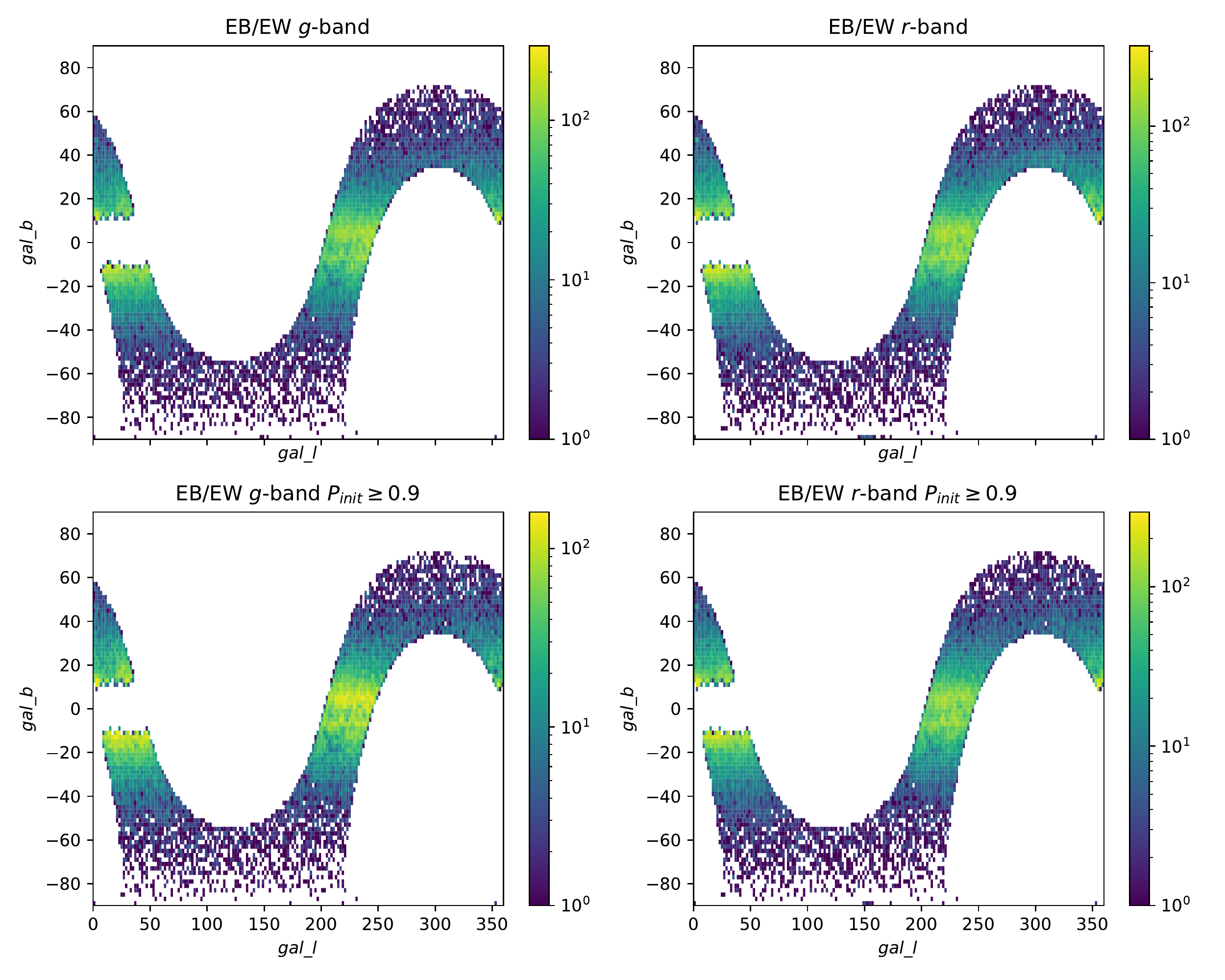} 

\caption{As in Figure \ref{figure:density_midz-AGN} but for the EB/EW class.
\label{figure:density_EB/EW}}
\end{center}
\end{figure*}

\begin{figure*}[hptb!]
\begin{center}
   \includegraphics[width=0.69\linewidth]{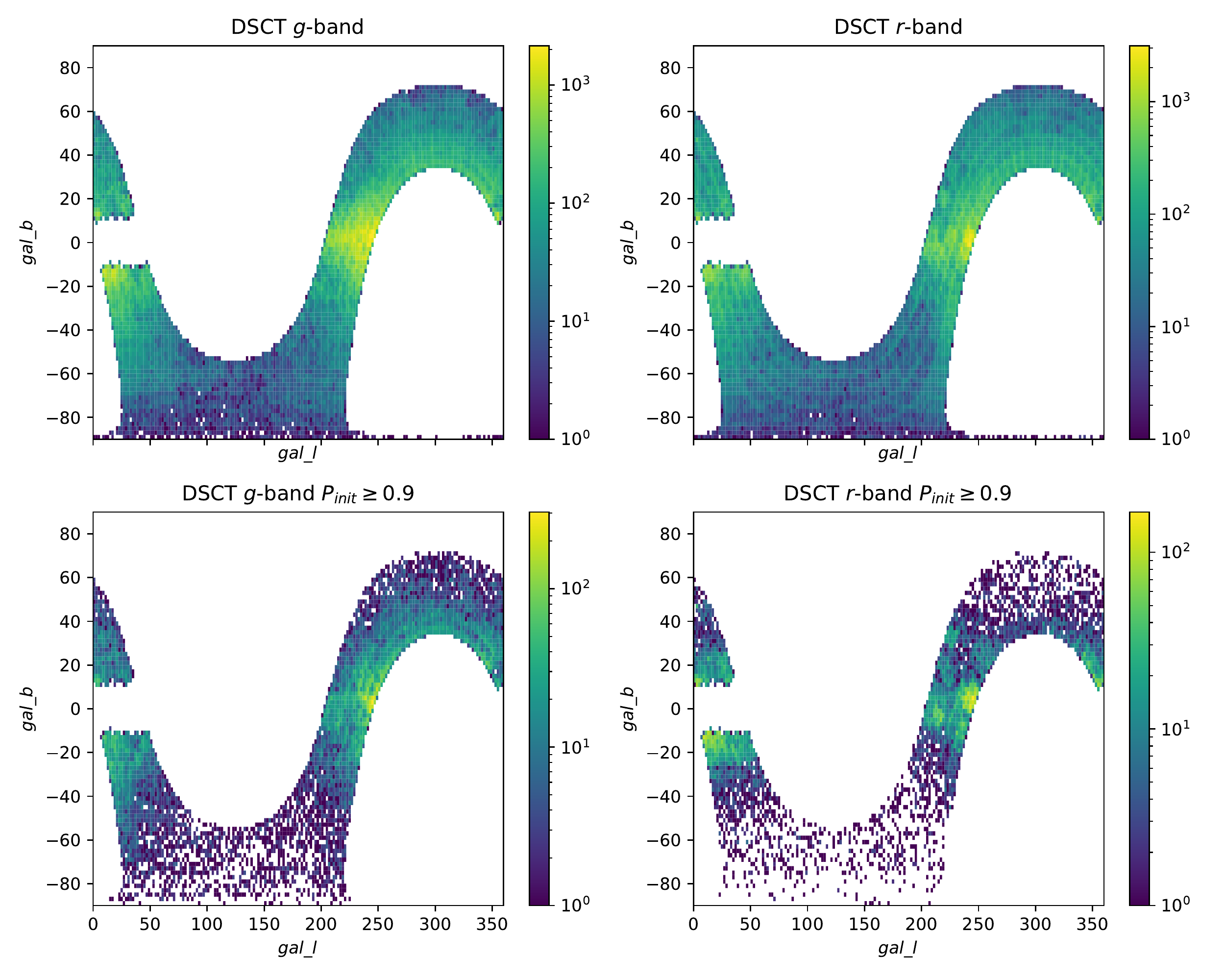} 

\caption{As in Figure \ref{figure:density_midz-AGN} but for the DSCT class.
\label{figure:density_DSCT}}
\end{center}
\end{figure*} 

\begin{figure*}[hptb!]
\begin{center}
   \includegraphics[width=0.69\linewidth]{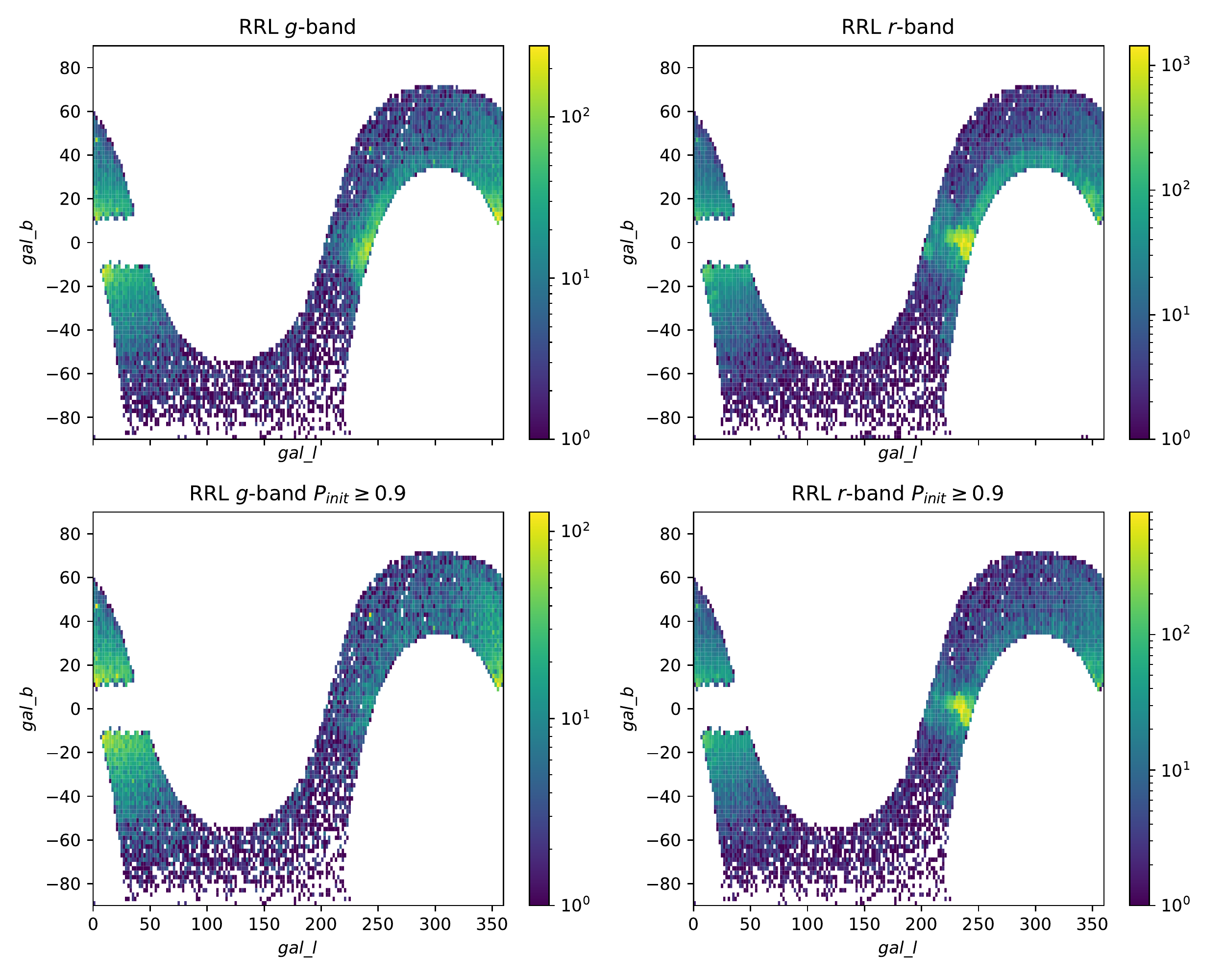} 

\caption{As in Figure \ref{figure:density_midz-AGN} but for the RRL class.
\label{figure:density_RRL}}
\end{center}
\end{figure*} 

\begin{figure*}[hptb!]
\begin{center}
   \includegraphics[width=0.69\linewidth]{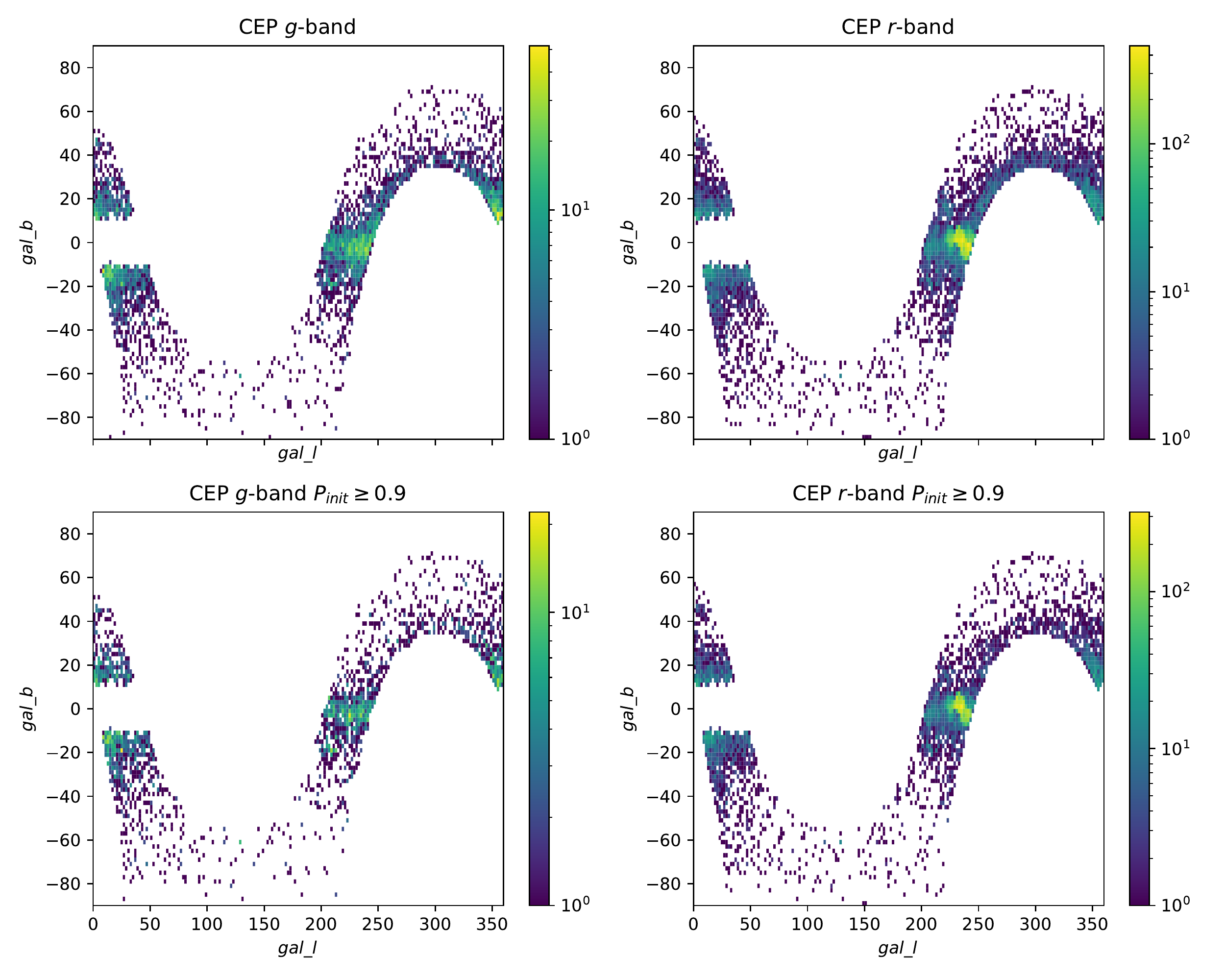} 

\caption{As in Figure \ref{figure:density_midz-AGN} but for the CEP class.
\label{figure:density_CEP}}
\end{center}
\end{figure*}

\begin{figure*}[hptb!]
\begin{center}
   \includegraphics[width=0.69\linewidth]{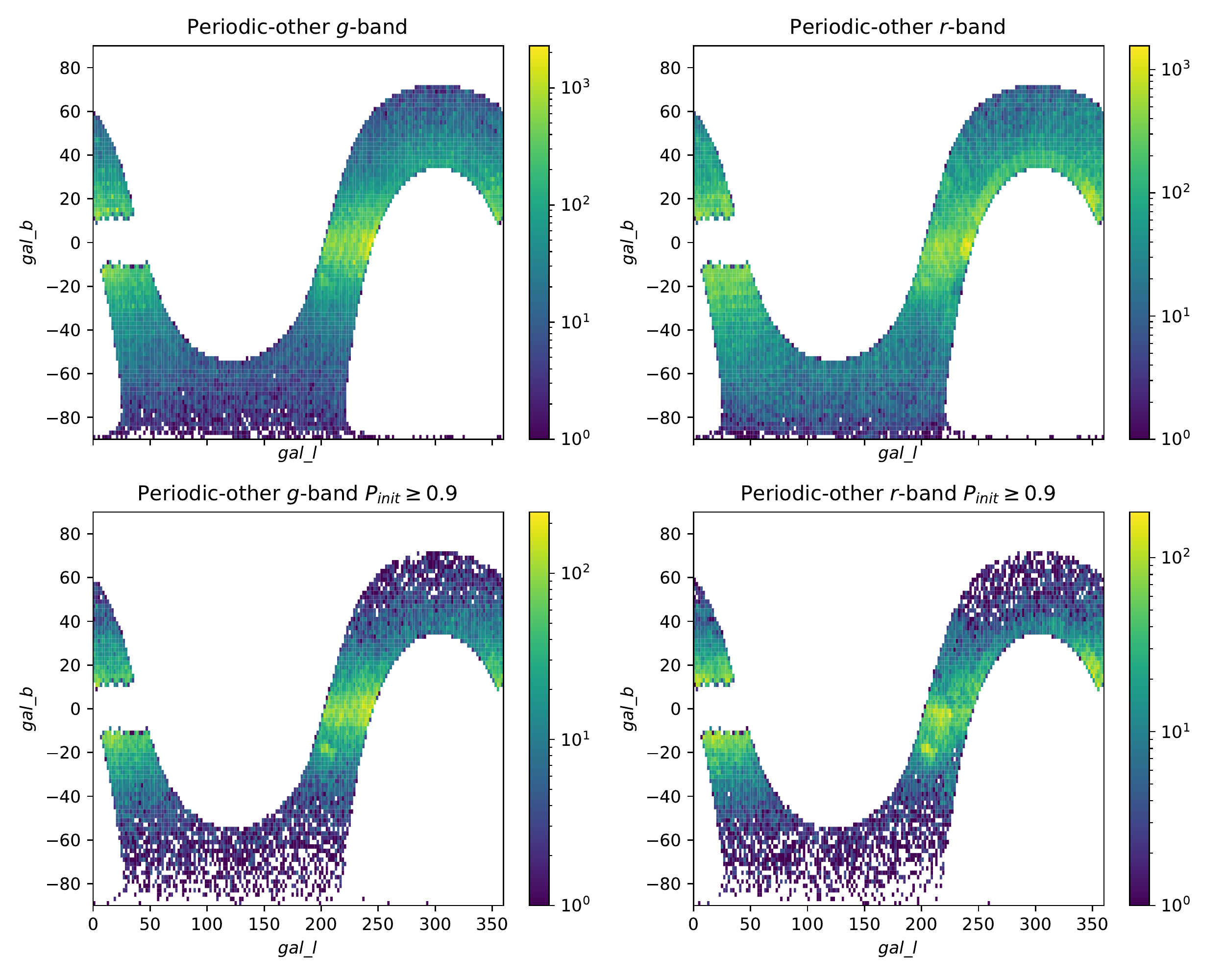} 

\caption{As in Figure \ref{figure:density_midz-AGN} but for the Periodic-other class.
\label{figure:density_Periodic-other}}
\end{center}
\end{figure*} 

\begin{figure*}[hptb!]
\begin{center}
   \includegraphics[width=0.69\linewidth]{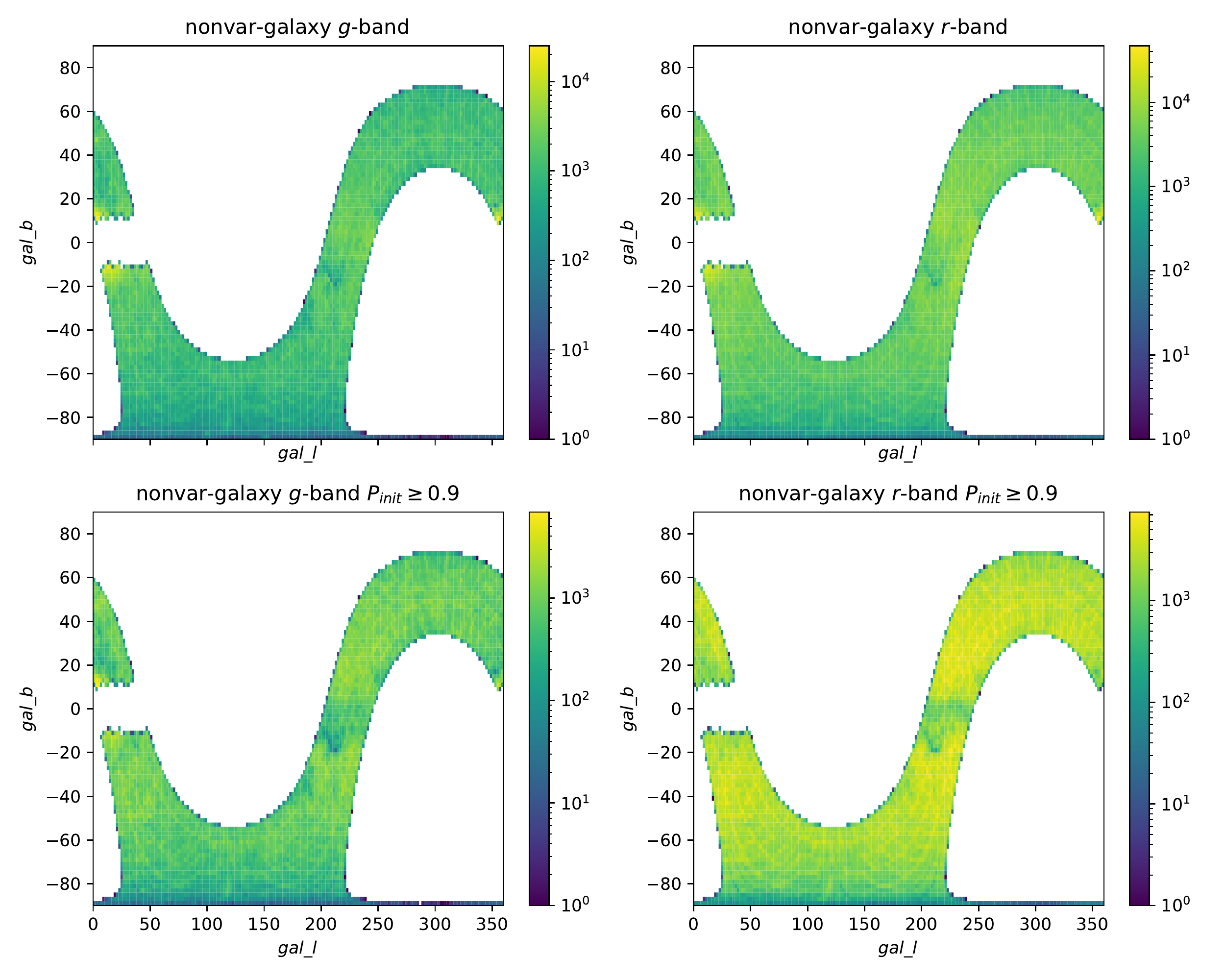} 

\caption{As in Figure \ref{figure:density_midz-AGN} but for the nonvar-galaxy class.
\label{figure:density_nonvar-galaxy}}
\end{center}
\end{figure*} 

\begin{figure*}[hptb!]
\begin{center}
   \includegraphics[width=0.69\linewidth]{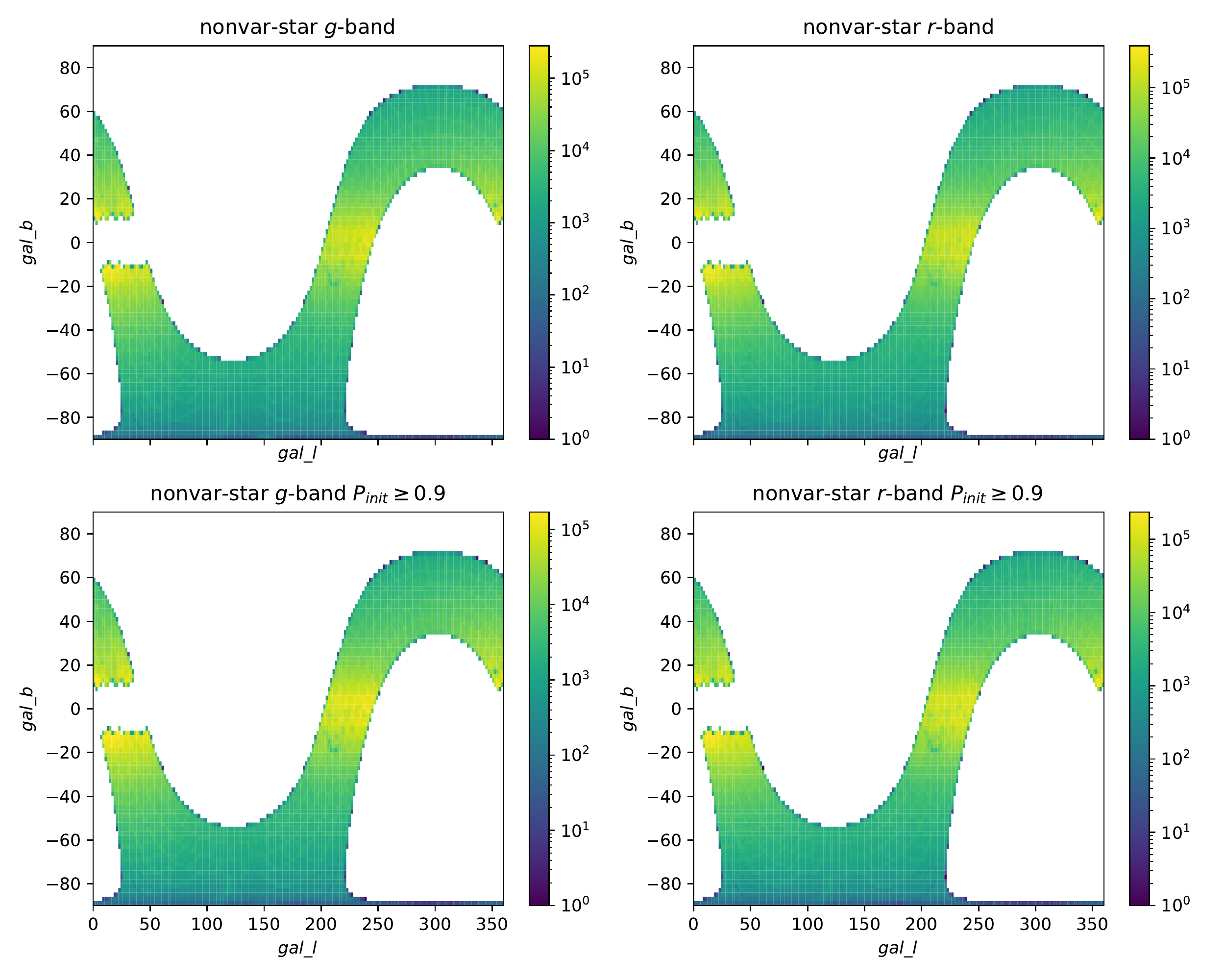} 

\caption{As in Figure \ref{figure:density_midz-AGN} but for the nonvar-star class.
\label{figure:density_nonvar-star}}
\end{center}
\end{figure*} 

\FloatBarrier

\end{appendix}

\end{document}